\documentclass[twocolumn,table]{aastex631}

\newcommand{\oiii}{[\hbox{{\rm O}\kern 0.1em{\sc iii}}]\,5007}
\newcommand{\oii}{[\hbox{{\rm O}\kern 0.1em{\sc ii}}]\,3727}
\newcommand{\ha}{\hbox{{\rm H}\kern 0.1em{\sc $\alpha$}}}
\newcommand{\hb}{\hbox{{\rm H}\kern 0.1em{\sc $\beta$}}}
\newcommand{\cione}{\hbox{[{\rm C}\kern 0.1em{\sc i}]\,($^3P_1\to{}^3P_0$)}}
\newcommand\hidden[1]\null
\def\fdeg{\hbox{$.\!\!^\circ$}}
\def\farcs{\hbox{$.\!\!^{\prime\prime}$}}
\def\arcsec{\hbox{$^{\prime\prime}$}}
\def\arcmin{\hbox{$^{\prime}$}}

\usepackage{textcomp}
\usepackage{amssymb,amsmath}
\usepackage{hyperref}
\usepackage{multirow,tabularx}
\usepackage{longtable}
\usepackage{threeparttable}
\usepackage[caption=false]{subfig}
\usepackage{enumerate,enumitem}
\usepackage{soul}
\graphicspath{ {./figures/}{./} }

\begin{document}

\title{The Centre of Attention: a Powerful Radio Galaxy Pinpoints a NIR-Dark Protocluster at $z\sim 3.9$}

\author[0009-0007-5594-6476]{A. J. Hedge}
\affiliation{International Centre for Radio Astronomy Research, Curtin University, GPO Box U1987, Bentley, WA 6845, Australia}
\author[0000-0003-3506-5536]{N. Seymour}
\affiliation{International Centre for Radio Astronomy Research, Curtin University, GPO Box U1987, Bentley, WA 6845, Australia}
\author[0000-0002-2239-6099]{J. W. Broderick}
\affiliation{SKA Observatory, Science Operations Centre, CSIRO ARRC, 26 Dick Perry Avenue, Kensington, WA 6151, Australia}
\affiliation{International Centre for Radio Astronomy Research, Curtin University, GPO Box U1987, Bentley, WA 6845, Australia}
\author[0000-0002-8984-3666]{A. Gupta}
\affiliation{International Centre for Radio Astronomy Research, Curtin University, GPO Box U1987, Bentley, WA 6845, Australia}
\affiliation{ARC Centre of Excellence for All Sky Astrophysics in 3 Dimensions (ASTRO 3D), Australia}
\author[0000-0002-9149-2973]{J. Afonso}
\affiliation{Instituto de Astrof\'{i}s\'{i}ca e Ci\^{e}ncias do Espa\c{c}o, Faculdade de Ci\^{e}ncias, Universidade de Lisboa, OAL, Tapada da Ajuda, PT1349-018
Lisboa, Portugal}
\author[0000-0003-1516-9450]{L. Ighina}
\affiliation{Center for Astrophysics \textbar{} Harvard \& Smithsonian, 60 Garden St., Cambridge, MA 02138, USA}
\affiliation{INAF, Osservatorio Astronomico di Brera, via Brera 28, 20121, Milano, Italy}
\author[0000-0003-1939-5885]{M. Lehnert}
\affiliation{Universit\'{e} Lyon 1, ENS de Lyon, CNRS UMR5574, Centre de Recherche Astrophysique de Lyon, 69230 Saint-Genis-Laval, France}
\author{G. Noirot}
\affiliation{Department of Astronomy and Physics and Institute for Computational Astrophysics, Saint Mary's University, 923 Robie Street, Halifax, NS B3H 3C3, Canada}
\affiliation{Space Telescope Science Institute, 3700 San Martin Drive, Baltimore, Maryland 21218, USA}
\author[0000-0001-5064-0493]{S. Shabala}
\affiliation{School of Natural Sciences, University of Tasmania, Private Bag 37, Hobart, TAS 7001, Australia}
\author[0000-0003-2686-9241]{D. Stern}
\affiliation{Jet Propulsion Laboratory, California Institute of Technology, 4800 Oak Grove Drive, Pasadena, CA 91109, USA}
\author[0000-0002-4376-5455]{R. J. Turner}
\affiliation{School of Natural Sciences, University of Tasmania, Private Bag 37, Hobart, TAS 7001, Australia}







\begin{abstract}
We report the discovery of a $z\sim3.9$ protocluster identified from Atacama Large Millimetre/sub-millimetre Array Band 3 spectral scans of a bright radio source selected from the GaLactic and Extra-galactic All-sky Murchison Widefield Array (GLEAM) survey. Extended CO(4-3) and \cione~line emission was detected in GLEAM~J005332$-$325630 confirming it to be a $z=3.879$ powerful radio galaxy with luminosity, $L_{\rm 500\;MHz}=1.3\times10^{28}\;{\rm W\,Hz}^{-1}$. This source is part of a sample of candidate high redshift radio galaxies with bright radio fluxes, $S_{\rm 150MHz}>0.1\,$Jy, but host galaxies with $K_s({\rm AB})\gtrsim23$ mag. The molecular gas associated with the radio galaxy host has two kinematically separate components, likely in-falling and indicative of a recent interaction or merger with another galaxy. One 100-GHz continuum source $\sim120\,$pkpc away is found to have both CO(4-3) and \cione~emission lines and a further five protocluster members are identified from CO(4-3) emission alone, all at similar redshift ($\Delta v<700\,$km s$^{-1}$) and within a radius of $1.1\arcmin$. Using photometry from the High Acuity Widefield K-band Imager $K_s$-band and the Dark Energy Survey $g,\,r,\,i,\,z$ and $Y$ bands, we find this protocluster harbours a rare, optically-dark, very massive $M_*\sim10^{12}\,{\rm M}_\odot$ galaxy. Comparisons with the TNG300 cosmological simulation puts this galaxy in a dark matter halo of $M_{\rm DM}\sim3\times10^{13}\,{\rm M}_\odot$ which will evolve into a Coma-like DM halo ($M_{\rm DM}\sim10^{15}\,{\rm M}_\odot$) by the present day.
\end{abstract}



\section{Introduction} \label{sec:intro}
Our understanding of how the most extreme supermassive black holes (SMBHs; $M_{\rm BH}>10^8\,{\rm M}_\odot$) in the early Universe managed to form and evolve, particularly at $z>7$, remains one of the most challenging topics for current theoretical models \citep[][]{SMBH_formation_evolution_Volonteri_2012,First_SMBHs_Smith_2017,SMBHs_in_early_universe_Smith_2019}. Observational constraints posed by the discovery of powerful active galactic nuclei (AGN) hosting SMBHs at high redshift, that are more massive than permitted by formation via small seeds or constant Eddington accretion, are not yet reconciled with theory. The recent discovery of ultra high redshift (UHz; $z>5$) quasars, such as J0313$-$1806 at $z=7.64$ and even UHZ1 at $z\approx10.1$, located well within the epoch of reionisation \citep[EoR; $6\lesssim z\lesssim25$; see review by][]{Cosmic_Dawn_EoR_SKA_Koopmans_2015} highlights the extreme nature of these SMBHs. The SMBH in J0313$-$1806, with a mass of $M_{\rm BH}=(1.6\pm0.4)\times10^9\,{\rm M}_\odot$, was able to grow to such a size in less than $700\,$Myr after the Big Bang, possibly from a direct-collapse black hole seed of mass $M_{\rm seed}=10^4$--$10^5\,{\rm M}_\odot$ \citep[][]{QSO_z_7.642_Wang_2021}. Similarly, the X-ray luminous and Compton-thick AGN in UHZ1 favours a formation via direct-collapse black hole from a massive seed \citep[][]{UNCOVER_SMBH_growth_UHZ1_JWST_spectroscopy_Goulding_2023,Heavy_seed_SMBH_z_10_x-ray_quasar_Bogdan_2024}. Finding additional AGN at very high redshifts is crucial for understanding the efficiency of accretion, the formation method (i.e. massive seeds, hierarchal mergers or sustained super-Eddington accretion) and the properties of SMBHs during the early stages of cosmic evolution.

Powerful radio emission serves as an unambiguous tracer of accretion processes onto central SMBHs which, unlike emission shortward of the near infrared (NIR), is virtually unaffected by dust and the dense interstellar medium (ISM) of the host galaxy. As such, a robust search for SMBHs can be made by hunting for radio-powerful $\left(L_{\rm 500 MHz}>10^{27}\,{\rm W\,Hz}^{-1}\right)$ AGN, some of these likely at high redshift and with extreme mass. Radio-loud quasars, such as VIK J2318$-$3113 at $z=6.44$ \citep[][]{RL_QSO_z_6.44_Ighina_2021,RL_QSO_z_6.44_properties_Ighina_2022} and P172$+$18 at $z=6.82$ \citep[][]{RL_QSO_z_6.82_Banados_2021,RL_QSO_z_6.82_resolved_radio_Momjian_2021} as well as others at $z>5$ \citep[e.g.][]{24_RL_quasars_z_4.9_to_6.6_Gloudemans_2022,Pan-STARRS1_z_5.6_quasar_survey_Banados_2023,RL_QSO_z_5.3_Belladitta_2023,EoR_RL_QSOs_Ighina_2023,RL_QSO_z_6.5_multi-wl_Ighina_2024}, have been identified near the end of the EoR. Moreover, the discovery of J0410$-$0139, the most distant known blazar at $z=6.996$ \citep[][]{Blazar_z_7_Banados_2024}, underscores the presence of powerful radio sources and the reach of their emission across cosmic time. Yet the most powerful radio emission known in the distant Universe continues to be that produced by high redshift radio galaxies \citep[HzRGs; $z>2$; see review by][]{HzRG_review_paper_Miley_DeBreuck_2008}. The Hubble $K$--$z$ relation ($K$ represents the $2.2$-\textmu{}m apparent magnitude) indicates that HzRGs are among the most massive galaxies at any redshift \citep[][]{K_z_relation_for_RG_Rocca-Volmerange_2004}. Since the nuclear emission in HzRGs tends to be obscured by the dusty torus, they are not only efficiently identified through their high radio luminosity, but they also allow for more detailed studies of their host galaxies, unimpeded by AGN emission at rest-frame ultraviolet through to NIR wavelengths, useful for observing the host galaxy \citep[e.g.][]{RG_Hosts_across_cosmic_time_Seymour_2007,Spitzer_HzRG_survey_DeBreuck_2010,Disentangling_SFR_AGN_in_HzRG_ULIRGS_z_1-4_Drouart_2016}. HzRGs play a vital role in understanding the co-evolution between SMBHs and massive host galaxies \citep[e.g.][]{MRC1138-262_HzRG_z_2.2_Spiderweb_galaxy_coeval_growth_Seymour_2012}, especially in the early Universe.

Another recurring characteristic of powerful HzRGs is their association with galaxy overdensities and clusters \citep[e.g.,][]{HzRG_protocluster_BCG_ancestor_Venemans_2002,HzRG_protocluster_BCG_population_2007,Clusters_around_radio_loud_AGN_z_1_3_Wylezalek_2013,Cluster_MIR_luminosity_function_z_1_3_Wylezalek_2014,CARLA_clusters_around_radio_loud_AGN_z_2_structures_Noirot_2016,CARLA_clusters_around_radio_loud_AGN_z_1_3_structures_Noirot_2018}. Many of the most extreme galaxy clusters at their respective redshifts house a HzRG, often near the centre of the cluster's dark matter halo, a well-studied case being the Spiderweb galaxy \citep[][]{Spiderweb_galaxy_z_2.2_Lyman_alpha_emitters_Kurk_2000,Spiderweb_galaxy_Miley_2006}. A more recent case that overlaps with sub-millimetre galaxies (SMGs) being SPT2349$-$56 \citep[][]{SPT2349-56_galaxy_cluster_core_z_4.3_Miller_2018,SPT2349-56_z_4.3_radio_loud_AGN_Chapman_2024}. In order to feed all the galaxies in such systems, an abundance of molecular and atomic gas is often found at millimetre and sub-millimetre wavelengths in bright clouds (usually identified by emission of ro-vibrational CO or neutral/ionised carbon transitions) around the cluster members and at low surface brightness in the cold intracluster medium (ICM). Such observations allow for spectroscopic redshifts of many sources to be determined, mapping of the fuel being used to feed SMBHs or galaxies to form stars, and estimates of the cluster's projected evolution into the equivalent of the largest clusters seen in the local Universe \citep[Spiderweb as an example:][]{Spiderweb_galaxy_CO_Emonts_2013,Molecular_gas_fuels_Spiderweb_galaxy_Emonts_2016,Spiderweb_HzRG_recyled_gas_CGM_CI_Emonts_2018,Spiderweb_galaxy_excess_dusty_starbursts_Dannerbauer_2014}.

The information we present in this paper is in relation to a series of studies \citep[][henceforth referred to as D20, B22 and B24, respectively]{GLEAMing_HzRGs_1_Drouart_2020,GLEAMing_HzRGs_2_Broderick_2022,GLEAMing_HzRGs_3_Broderick_2024} pursuing radio-powerful, ultra high redshift radio galaxies (UHzRGs; $z>5$).

In short, our sample of candidate UHzRGs contains compact, isolated radio sources with low-frequency spectra that are steep or turning over, yet they have faint or no counterpart in shallow $K$-band imaging. For more details regarding selection, see the previous studies.

14 priority targets with faint $K$-band host galaxy detections or limits were included in 100-GHz spectroscopic follow-up with the Atacama Large Millimetre/sub-millimetre Array \citep[ALMA;][]{ALMA_Wootten_Thompson_2009} to target redshifted common molecular emission lines from ro-vibrational CO and neutral/ionised carbon transitions.
For radio galaxies, ALMA has been used to observe the molecular gas of the Spiderweb \citep[][]{Spiderweb_HzRG_recyled_gas_CGM_CI_Emonts_2018} and the Dragonfly \citep[$z=1.92$,][]{Dragonfly_2_z_2_ALMA_CO_6-5_merger_ULIRG_HzRG_Emonts_2015,Dragonfly_3_z_2_HzRG_jet_brightening_disk_merger_ULIRG_Lebowitz_2023}, and for SMGs, ALMA has notably confirmed the redshifts for the protoclusters SPT2349$-$56 \citep[][]{SPT2349-56_galaxy_cluster_core_z_4.3_Miller_2018} and the Distant Red Core \citep[DRC;][]{DRC_z_4_proto-cluster_DSFG_Oteo_2018}. ALMA probes some common properties between extreme HzRGs and high redshift SMG protoclusters such as vigorous star formation rates (SFRs), large quantities of molecular gas to fuel star formation, and massive stellar masses (at their respective redshifts).

This paper focuses on one UHzRG candidate from the ALMA subset (GLEAM~J005332$-$325630, henceforth J0053), which we spectroscopically confirm to lie at $z=3.879$ within a protocluster of mostly optically/NIR-dark low-mass galaxies. In Section \ref{sec:data} we describe the ALMA observations and their data reduction, as well as new SED modelling of aperture photometry using the Dark Energy Survey (DES). In Section \ref{sec:results} we present the results of continuum detections and spectral line finding, and finally in Section \ref{sec:discussion} we explore the interpretations for each detected source, properties of the protocluster and how it compares to both HzRGs in galaxy overdensities and other SMG protoclusters at comparable redshift.

We assume a flat $\Lambda$ cold dark matter ($\Lambda$CDM) cosmology with $\Omega_M=0.3$, $\Omega_\Lambda=0.7$ and $H_0=70\;{\rm km\,s}^{-1}\,{\rm Mpc}^{-1}$. We also give uncertainties as $\pm1\sigma$, magnitudes in the AB system \citep[][]{AB_magnitude_system_absolute_SED_white_dwarfs_Oke_1974}, and logarithms as the decimal logarithm (base 10). We adopt a labelling convention for any source or emission line detections by appending a letter to the shorthand for this field, e.g. J0053x.

\section{Observations and Data Reduction} \label{sec:data}
\subsection{ALMA observations and data products} \label{sec:obs}
The Band 3 (100\,GHz) observations of J0053 took place from 2023 January 2 to 2023 January 9 (Cycle 9; programme ID 2022.1.01657.S) using the 12-m array in the C-3 configuration (with baselines of 15--500\,m), detailed in Table \ref{tab:obs_log}. The effective field of view at the central frequency of the band was $58\arcsec{}$ in diameter with an estimated best angular resolution from the used baselines of $1.4\arcsec{}$ and a spectral resolution of $7.813$\,MHz per channel.
The spectral setup used is identical in design to that in D20, which itself is based on the setup detailed in \citet{ALMA_redshifts_SPT_survey_Weiss_2013}, enabling almost full coverage of the ALMA Band 3 receiver ($84$--$116$ GHz).
To achieve this frequency coverage, the observing strategy we implemented separates observations into five tunings with staggered frequencies, enabling a large bandwidth. To make the calibrator observations more efficient each tuning was observed separately which resulted in observations scheduled at different times, sometimes weeks apart with different weather and $u,v$-coverage. Each tuning had at least 12.6 minutes of on-source time, accumulating at least one hour of on-source time across Band 3.
For each tuning, 7.5-GHz is split evenly into two sidebands with symmetric frequency offset from a local oscillator frequency, each of which is also split into two 1.875-GHz spectral windows by the correlator. The actual frequency coverage obtained with this strategy is $84.2$--$114.9$ GHz, with a small fraction of the bandwidth ($96.2$--$102.8$ GHz) observed twice by overlapping sidebands.

\begin{table*}[ht]
    \centering
    \caption{ALMA observing log for J0053, with most details available on the ALMA Science Archive (programme ID 2022.1.01657.S). When multiple observations for a tuning exist, the statistics are of the combined observations for total integration time and 80th percentile projected baseline. We detail, for each tuning observation, the date(s) of observation, the tuning's four spectral windows, angular resolution, number of usable antennas, integration time, 80th percentile projected baseline (L80 BL) and the precipitable water vapour (PWV) level. All observations were made with the 12m array in the C-3 configuration.}
    \label{tab:obs_log}
    \begin{tabular*}{\textwidth}{@{\extracolsep{\fill}}ccccccccccc}
        \hline\hline
        Tuning & Date & SPWs & Ang. res. & N$_{\rm ant}$ & Int. time & L80 BL & PWV \\
         &  & [GHz] & [\arcsec] &  & [s] & [m] & [mm] \\
        \hline
        \multirow{4}{*}{1} & \multirow{4}{*}{2023-01-08} & 84.17--86.04 & \multirow{4}{*}{1.2} & \multirow{4}{*}{41} & \multirow{4}{*}{756.000} & \multirow{4}{*}{335.416} & \multirow{4}{*}{1.2} \\
         &  & 86.05--87.92 &  &  &  &  & \\
         &  & 96.17--98.04 &  &  &  &  & \\
         &  & 98.05--99.92 &  &  &  &  & \\
        \hline
        \multirow{4}{*}{2} & \multirow{4}{*}{2023-01-08} & 87.92--89.79 & \multirow{4}{*}{1.3} & \multirow{4}{*}{41} & \multirow{4}{*}{756.000} & \multirow{4}{*}{270.911} & \multirow{4}{*}{2.1} \\
         &  & 89.80--91.67 &  &  &  &  & \\
         &  & 99.92--101.79 &  &  &  &  & \\
         &  & 101.80--103.67 &  &  &  &  & \\
        \hline
        \multirow{4}{*}{3} & \multirow{2}{*}{2023-01-08} & 91.67--93.54 & \multirow{2}{*}{1.1} & \multirow{2}{*}{23} & \multirow{4}{*}{756.000} & \multirow{4}{*}{214.535} & \multirow{2}{*}{1.9} \\
         &  & 93.55--95.42 &  &  &  &  & \\
         & \multirow{2}{*}{2023-01-09} & 103.67--105.54 & \multirow{2}{*}{1.1} & \multirow{2}{*}{41} &  &  & \multirow{2}{*}{1.6} \\
         &  & 105.55--107.42 &  &  &  &  & \\
        \hline
        \multirow{4}{*}{4} & \multirow{2}{*}{2023-01-06} & 95.42--97.29 & \multirow{2}{*}{1.5} & \multirow{2}{*}{41} & \multirow{4}{*}{1572.480} & \multirow{4}{*}{298.700} & \multirow{2}{*}{4.5} \\
         &  & 97.30--99.17 &  &  &  &  & \\
         & \multirow{2}{*}{2023-03-03} & 107.42--109.29 & \multirow{2}{*}{1.0} & \multirow{2}{*}{36} &  &  & \multirow{2}{*}{0.9} \\
         &  & 109.30--111.17 &  &  &  &  & \\
        \hline
        \multirow{4}{*}{5} & \multirow{4}{*}{2023-01-02} & 99.17--101.04 & \multirow{4}{*}{1.5} & \multirow{4}{*}{43} & \multirow{4}{*}{816.480} & \multirow{4}{*}{232.213} & \multirow{4}{*}{1.8} \\
         &  & 101.05--102.92 &  &  &  &  & \\
         &  & 111.17--113.04 &  &  &  &  & \\
         &  & 113.05--114.92 &  &  &  &  & \\
        \hline\hline
    \end{tabular*}
\end{table*}

The quality assurance report (QA0) for the first observing block of tuning 3 indicates the execution fraction for that tuning was not fulfilled due to a large number of antennas exceeding the observatory's phase calibration RMS limit of 1 radian, hence a second observing block was continued the following day. Despite this, both observations pass their quality assurance tests and are used in our reduction.
It is evident from the QA2 and QA0 reports for tuning 4 that it was re-observed on March 3 due to marginal weather conditions (high PWV) which resulted in a high bandpass decorrelation fraction. During re-observation, 11 antennas exceeded the observatory's phase calibration RMS limit, although the QA0 report's summary indicates 36 antennas were still usable and yielded data of sufficient quality. We found that combining both observations led to a marginal improvement in resolution and sensitivity that outweighed the marginal data quality of the first observation.

Calibrated data and quality assurance reports were obtained from the ALMA Science Archive\footnote{\url{https://almascience.nrao.edu/aq}} after observations had been processed with the ALMA pipeline whilst further reduction including imaging was performed locally. The uncertainty in calibrated flux density is expected to be around 5\%.

\subsubsection{Imaging} \label{sec:imaging}
Imaging was carried out using the Common Astronomy Software Applications \citep[CASA;][]{CASA_McMullin_2007} package, primarily making use of the following routines: \texttt{concat} for joining different tunings' calibrated measurement sets together, \texttt{tclean} to create the cleaned image and cube with primary beam corrections, and \texttt{imsmooth} to smooth the resolution of image cubes to the smallest common resolution of the joined tunings, or to a specified target resolution.

Using a close to natural Briggs weighting scheme \citep[robustness $=2$;][]{Briggs_deconvolution_scheme_resolved_sources_Briggs_1995}, the resulting continuum image has a $1\farcs51\times1\farcs37$ synthesised beam with position angle $-85\fdeg5$ (measured north through east) and a mean background RMS noise level of $6.2$\,\textmu{}Jy\,beam$^{-1}$.
After imaging and cleaning the full cube, we smooth each channel to a common beam, which is $2\farcs11\times1\farcs73$ with position angle $82\fdeg8$. Frequency-dependent noise variations in the cube are due to differing observing conditions for each tuning and a small fraction of channels being observed twice by overlapping tunings. Keeping this in mind, the RMS noise across the entire cube is approximately $400$\,\textmu{}Jy\,beam$^{-1}$\,channel$^{-1}$.

\subsubsection{3-D emission line finding} \label{sec:line_finding}
The most prominent emission lines were initially found manually by extracting the spectrum of the only bright continuum source in the ALMA field-of-view (not the radio galaxy; see Section \ref{sec:continuum}), whereby associated CO emission of an unknown transition was found. Following this, CO emission was also identified and associated with the radio galaxy at the same frequency and of similar intensity in the moment-0 linemap.
After identifying the most obvious spectral features by eye, we opted to use an application to automate spatial and spectral searches and to more robustly detect features that would otherwise be missed. To achieve this we used version 2 of the HI Source Finding Application \citep[SoFiA-2;][]{SoFiA_description_Serra_2015,SoFiA2_description_Westmeier_2021} which is capable of detecting and characterising emission lines distributed in 3-D data cubes, not limited to HI.


\subsection{Literature Data} \label{sec:lit_data}
\subsubsection{ATCA}
The Australian Telescope Compact Array \citep[ATCA;][]{ATCA_overview_Frater_1992} was used by B22 (programme ID C3377) to measure the compactness of the radio emission at 5.5- and 9-GHz for the initially selected sample of radio galaxies. We only use the 5.5-GHz ATCA data of J0053, which has higher S/N, primarily for determining astrometry relative to features at other wavelengths.
\subsubsection{HAWK-I}
B24 obtained deep $K_s$-band ($2.146$-\textmu{}m; $3\sigma\gtrsim23.3$ AB) observations with the High Acuity Widefield K-band Imager \citep[HAWK-I;][]{HAWK-I_Kissler-Patig_2008} at the Very Large Telescope \citep[VLT;][]{VLT_white_book_ESO_1998} (programme ID 108.22HY.001) identify sources with fainter NIR counterparts which are thus more likely be at ultra-high redshift. We use HAWK-I data in this work to provide photometry in the NIR.
\subsubsection{DES}
We use Data Release 2 of the Dark Energy Survey \citep[DES;][]{DES_design_Sanchez_2010,DES_DR2_Abbott_2021}, observed with the Dark Energy Camera \citep[DECam;][]{DECam_Flaugher_2015}. DES releases deep wide-area coverage of the sky in five bands from the visible to the near-infrared: $g,\,r,\,i,\,z$ and $Y$, with central wavelengths of $473,\,642,\,784,\,926$ and $1009$ nm, respectively. These reach median $10\sigma$ (AB) depths of $24.7,\,24.4,\,23.8,\,23.1$ and $21.7$, respectively. Combined with our similarly deep HAWK-I data, DES provides good wavelength coverage for photometric modelling.

\subsection{SED modelling} \label{sec:sed_modelling}
We utilise the BayEsian Analysis of GaLaxy sEds \citep[BEAGLE;][]{Beagle_Chevallard_2016} code to fit SEDs and galaxy properties to photometry that we extract from the Data Release 2 of DES, via cutouts produced by the Data Central Data Aggregation Service \citep[][]{Data_central_data_aggregation_service_Miszalski_2022}. Combining the DES bands ($g$, $r$, $i$, $z$, $Y$) with our HAWK-I $K_s$-band data, we have up to six photometric points that BEAGLE can use for fitting (see Table \ref{tab:photometry}).
\subsubsection{Photometry} \label{sec:photometry}
We inspect the region around each candidate protocluster member in all of the aforementioned bands to choose a location to place a 2\arcsec{} aperture to measure the flux, prioritising the placement on signatures of a host galaxy if there are any. Otherwise, we use the centroid of the CO(4-3) emission (see Figure \ref{fig:photometry_cutouts} for the cutouts used in extracting photometry). We adopt the method used in B24, measuring the flux in randomly placed 2\arcsec{} apertures over the images, to estimate the uncertainty. Our $1\sigma$ uncertainties for the DES bands are within $10\%$ of the noise level inferred from the DES DR2 median coadded depths in $1.95\arcsec{}$ diameter apertures.

\begin{table*}[htbp]
    \centering
    \caption{Photometry of candidate protocluster members in 2\arcsec{} diameter apertures. All flux densities are given in \textmu{}Jy. Invalid values (i.e. the summed flux and error is negative) are discarded by BEAGLE and are marked with a {\bf \dag}.}
    \label{tab:photometry}
    \begin{tabular*}{\textwidth}{@{\extracolsep{\fill}}ccccccc}
        \hline\hline
        Source & $g$ & $r$ & $i$ & $z$ & $Y$ & $K_s$ \\
        \hline
        J0053a & $-0.05\pm0.06$ & $0.21\pm0.08$ & $0.26\pm0.15$ & $0.42\pm0.28$ & $0.53\pm0.78$ & $1.11\pm0.58$\\
        J0053b & $-0.09\pm0.06^{\boldsymbol{\dag}}$ & $-0.05\pm0.08$ & $0.01\pm0.14$ & $-0.33\pm0.26^{\boldsymbol{\dag}}$ & $-0.34\pm0.77$ & $-0.33\pm0.57$\\
        J0053c & $0.09\pm0.07$ & $0.00\pm0.08$ & $-0.08\pm0.15$ & $0.24\pm0.27$ & $0.50\pm0.77$ & $0.65\pm0.54$\\
        J0053d & $-0.04\pm0.06$ & $0.27\pm0.08$ & $0.38\pm0.15$ & $0.29\pm0.27$ & $0.55\pm0.73$ & $6.12\pm0.57$\\
        J0053e & $0.11\pm0.06$ & $0.32\pm0.08$ & $0.58\pm0.14$ & $0.52\pm0.26$ & $-0.66\pm0.79$ & $-0.25\pm0.58$\\
        J0053f & $0.36\pm0.06$ & $0.86\pm0.09$ & $1.31\pm0.15$ & $1.00\pm0.27$ & $1.09\pm0.75$ & $2.07\pm0.56$\\
        J0053g & $-0.04\pm0.06$ & $-0.17\pm0.09^{\boldsymbol{\dag}}$ & $0.00\pm0.15$ & $0.35\pm0.27$ & $0.37\pm0.74$ & $-0.68\pm0.57^{\boldsymbol{\dag}}$\\
        J0053h & $-0.05\pm0.07$ & $-0.05\pm0.09$ & $0.23\pm0.15$ & $-0.36\pm0.28^{\boldsymbol{\dag}}$ & $-0.52\pm0.75$ & $-0.22\pm0.59$\\
        \hline\hline
    \end{tabular*}
    \footnotesize{The flux densities measured for source e are likely a lower redshift interloper closest to the molecular CO(4-3), appearing brighter at the shorter wavelengths. Further discussion follows in \ref{sec:source_properties}}
\end{table*}

\subsubsection{BEAGLE} \label{sec:beagle_params}
We summarise the parameters and priors provided to BEAGLE in Table \ref{tab:beagle_params} and we report the fit results in Section \ref{sec:beagle_fitting}.
We have divided Table \ref{tab:beagle_params} into cosmological, nebular and star formation parameters. The outputs we obtain from the BEAGLE fitting are: the log-likelihood of the maximum-a-posteriori (MAP) walker ($l(\theta)$), stellar mass ($M_*$), $V$-band attenuation optical depth ($\tau_{V\mathrm{, eff}}$), star formation timescale ($\tau$) and the star formation rate ($\psi$). Setting nebular parameters to be `dependent' simply fixes their value to that of the galaxy's fitted stellar parameter, which we do for simplicity. Additionally, the nebular dust-to-metal mass ratio ($\xi$) is kept at its default value and we fix the metallicity to solar metallicity. Our chosen prior ranges for the $V$-band optical depth, stellar mass and star formation rate are based on the assumptions that the millimetre continuum detections and protocluster environment imply high star formation rates (see Section \ref{sec:aless_sfr}), and that the lack of host galaxy detections favour either unobscured low-mass galaxies or high-mass, high $V$-band optical depth galaxies.

We use BEAGLE with the provided libraries. This includes the \citet[][]{stellar_population_synthesis_BC03_Bruzual_2003} stellar population synthesis (SPS) code updated with stellar evolutionary tracks from \citet[][]{PARSEC_stellar_evolutionary_track_Bressan_2012} and \citet[][]{COLIBRI_stellar_evolutionary_track_Marigo_2013}, the MILES library of spectral templates \citep[][]{MILES_spectral_templates_Sanchez-Blazquez_2006} and a fixed \citet[][]{IMF_Chabrier_2003} Galactic-disc initial mass function (IMF). As the \citet[][]{IMF_Chabrier_2003} IMF's distribution of stellar light contributions turns over at stellar masses $M_*<1\,{\rm M}_\odot$, BEAGLE's stellar mass estimates using this IMF are lower compared to using a classical \citet[][]{IMF_Salpeter_1955} IMF. With this, more mass is tied up in massive stars that contribute more stellar light to the integrated galaxy SED, requiring fewer stars (lower stellar mass) to achieve a fixed brightness than in an IMF which has more mass tied up in low-mass stars that contribute less stellar light. There is speculation of early Universe galaxies requiring even more top-heavy IMFs to reconcile with recent {\it James Webb Space Telescope} ({\it JWST}) photometry that reveals large numbers of high redshift galaxies that seem too bright for their theorised mass assembly (e.g. \citet[][]{CEERS_1_early_galaxy_formation_JWST_Finkelstein_2023,Nebular_dominated_galaxies_hot_stars_top-heavy_IMF_Cameron_2024} from observations and \citet[][]{Simulating_top-heavy_IMF_high_SFE_enhanced_UV_luminosity_Jeong_2025} from simulations). We do not use any top-heavy IMFs as current discussions revolve around higher redshifts than we consider here, but acknowledge that their usage would result in lower mass estimates from SPS and could provide an alternative explanation for improbably high masses.

\begin{table}[htbp]
    \centering
    \caption{Setup of parameters and priors for the BEAGLE runs used to fit candidate protocluster members. All prior distributions are uniform.}
    \label{tab:beagle_params}
    \begin{tabular}{ccc}
        \hline\hline
        Parameter & Value & Prior range \\
        \hline
        \multicolumn{3}{c}{Cosmological parameters}\\
        \hline
        $H_0$ & 70 km s$^{-1}$ Mpc$^{-1}$ & - \\
        $\Omega_M$ & 0.3 & - \\
        $\Omega_\Lambda$ & 0.7 & - \\
        Redshift & 3.879 & - \\
        \hline
        \multicolumn{3}{c}{Nebular and dust parameters}\\
        \hline
        IGM absorption & \citet[][]{IGM_absorption_model_Inoue_2014} & - \\
        Nebular $\log U$ & dependent & - \\
        Nebular $\xi$ & 0.3 & - \\
        Nebular $Z$ & dependent & - \\
        Attenuation & \citet[][]{Dust_attenuation_model_Calzetti_2000} & - \\
        $\tau_{V\mathrm{, eff}}$ & - & $-1, 7$ \\
        \hline
        \multicolumn{3}{c}{Star formation parameters}\\
        \hline
        SFH & constant & - \\
        $\log\tau/\mathrm{yr}$ & - & $8, 9.15$ \\
        $\log Z/{\rm Z}_\odot$ & 0 & - \\
        $\log M_*/{\rm M}_\odot$ & - & $9, 13$ \\
        $\log\psi$ & - & $1, 3.5$ \\
        \hline\hline
    \end{tabular}
\end{table}

In order for BEAGLE to derive a plausible value for the star formation timescale, $\tau$, we manually set the upper limit of the prior range to be $200\,$Myr less than the Universe's age at the assumed redshift, forcing formation to begin after the Universe was $200\,$Myr old.

\section{Results} \label{sec:results}
\subsection{Continuum}
\label{sec:continuum}
\begin{figure*}[ht]
    \centering
    \includegraphics[width=\linewidth,trim={0 0 0.2cm 0},clip]{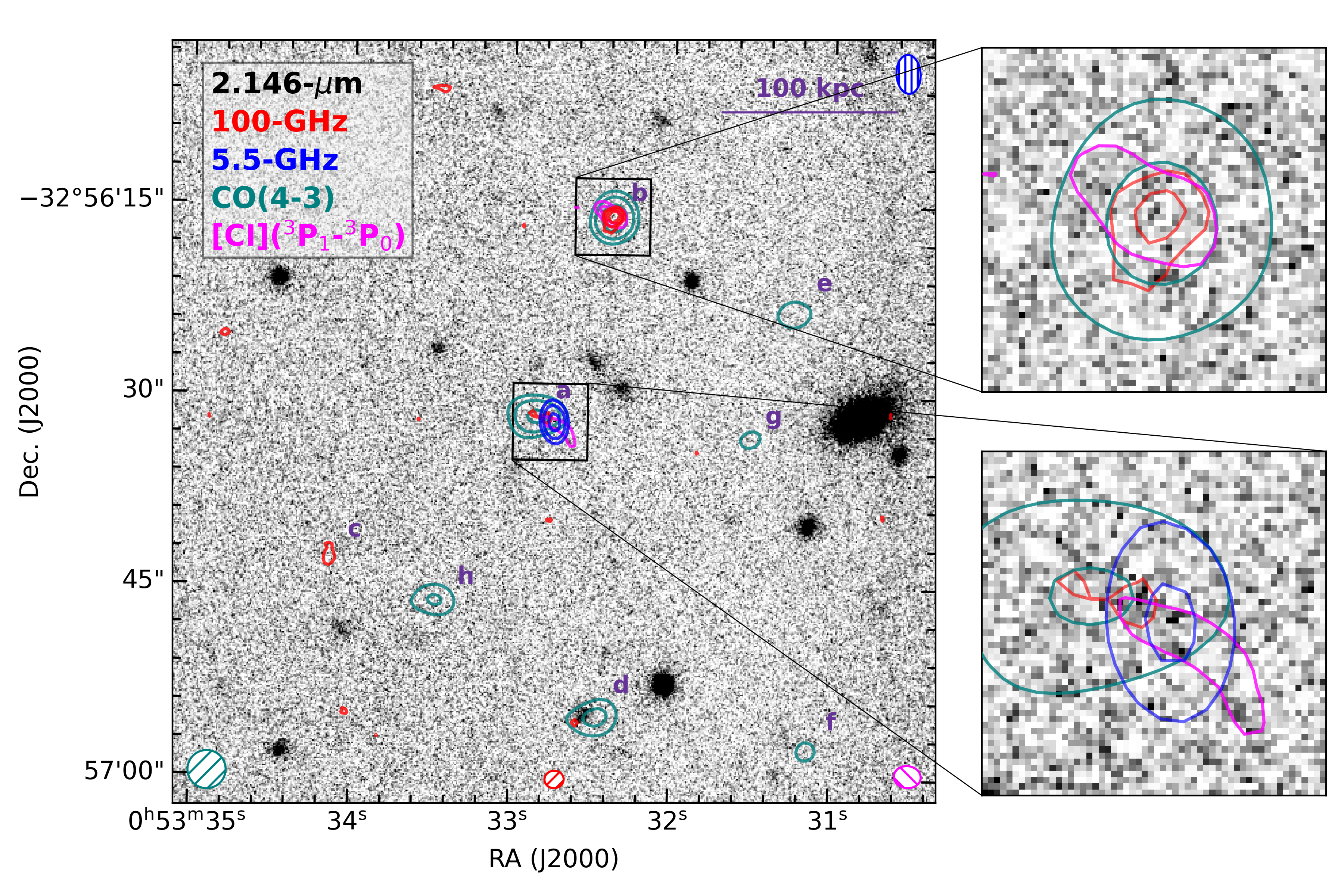}
    \caption{Full ALMA field-of-view ($58\arcsec{}$ on a side) of the radio galaxy (J0053a) and its environment. The underlying image is from HAWK-I $K_s$-band, with ALMA 100-GHz continuum overlaid in red (beam at bottom of figure) and ATCA 5.5-GHz continuum overlaid in blue (beam in upper-right corner). Additionally, CO(4-3) line emission is overlaid in teal using a $3\arcsec{}\times3\arcsec{}$ smoothed beam resolution (beam shown in lower-left corner) and \cione{} line emission is overlaid in fuchsia using the smallest common resolution (beam shown in lower-right). All contours (except \cione{}) start at $3\sigma_{\rm RMS}$ ($2.5\sigma_{\rm RMS}$) and increase by factors of $\sqrt{2}$. In the $6\arcsec{}\times6\arcsec{}$ insets the contours increase by factors of $2$ instead so that the NIR image is less obstructed. The two most prominent sources with insets are separated by approximately $17\arcsec{}$ corresponding to a linear separation of $120\,$pkpc at $z=3.879$ (redshift confirmation in Section \ref{sec:spec_lines_analysis}). {\it Top inset}: J0053b, {\it Bottom inset}: J0053a.}
    \label{fig:overlays}
\end{figure*}

We combine our 100-GHz ALMA continuum data with the 5.5-GHz ATCA and $K_s$-band data into one comprehensive map of the continuum emission, overlaid with line emission in Figure \ref{fig:overlays}. At 100-GHz we identify three significant continuum sources by eye, which are at least the size of the synthesised beam, at S/N $>3$. The target radio galaxy (J0053a) has a weak 100-GHz continuum detection which is possibly resolved into two components. The brightest continuum detection (J0053b) is undetected at other wavelengths. A third continuum detection (J0053c) is also undetected at other wavelengths. The characteristics for continuum sources are obtained with CASA's \texttt{imfit} and summarised in Table \ref{tab:source_stats}.

\begin{table*}[ht]
    \centering
    \begin{threeparttable}
        \caption{Continuum source properties. Integrated flux density uncertainties are calculated by \texttt{imfit} from the RMS in residual images around the sources. The sources, in order, are: the radio galaxy, the bright continuum source (companion), and the other in-field S/N $>3$ detection.}
        \label{tab:source_stats}
        \begin{tabular*}{\textwidth}{@{\extracolsep{\fill}}cccccc}
            \hline\hline
            Source & RA & Dec. & Deconvolved size & $S_{\rm 100 GHz}^{\rm peak}$ & $S_{\rm 100 GHz}^{\rm int}$ \\
             & [J2000] & [J2000] & [\arcsec$\times$\arcsec; $\deg$] & [\textmu{}Jy beam$^{-1}$] & [\textmu{}Jy] \\
            \hline
            J0053a & 00h53m32.806s & -32d56m31.578s & $(2.4\pm1.2)\times(1.3\pm1.2);~85\pm29$ & $22.4\pm6.5$ & $58.2\pm22.8$\\
            J0053b & 00h53m32.382s & -32d56m16.092s & $(1.6\pm0.5)\times(0.4\pm0.6);~155\pm27$ & $42.5\pm6.5$ & $65.7\pm15.4$\\
            J0053c & 00h53m34.144s & -32d56m42.614s & $2.4\times0.9;~0^{*}$ & $26.4\pm5.6$ & $41.2\pm13.3$\\
             \hline\hline
        \end{tabular*}
        \begin{tablenotes}
            \item [$*$] Upper limits to the deconvolved size.
        \end{tablenotes}
    \end{threeparttable}
\end{table*}

\subsection{Spectra and line analysis} \label{sec:spec_lines_analysis}
Once the two high signal-to-noise (S/N) detections previously identified manually had been confirmed (using SoFiA parameter \texttt{scfind.threshold}\,$=6.0$), we extracted spectra from the cube using the spatial masks that comprise the SoFiA detections. With manual review and inspection, SoFiA did not appear to produce any detections below S/N $\sim6\sigma$ that we were confident in, picking up an increasing amount of noise as `sources' at higher frequencies ($\gtrsim107\,$GHz) where the sensitivity of the ALMA Band 3 begins to decline. Hence any additional detections were obtained manually by searching for lower S/N features either spatially coincident with a detected line but at a different frequency, or spectrally coincident with a detected line but at a different location spatially.

\begin{figure}[ht]
    \centering
    \includegraphics[width=0.8\linewidth]{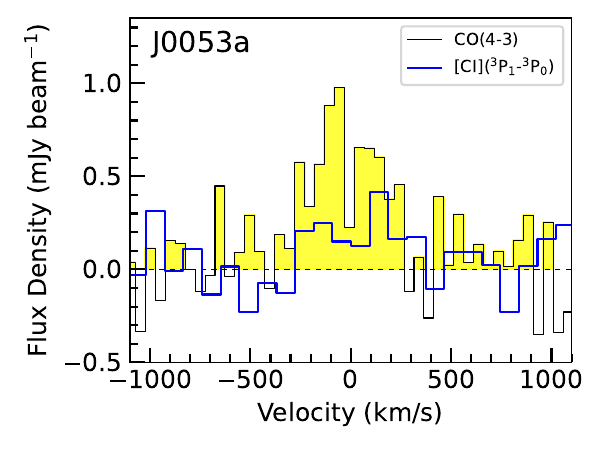}
    \\
    \includegraphics[width=0.8\linewidth]{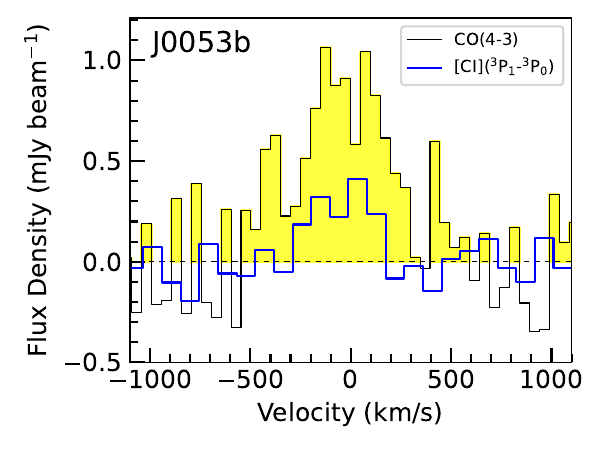}
    \caption{Emission line profiles for the spectroscopically confirmed sources ({\it Top}: J0053a, {\it Bottom}: J0053b) in velocity, relative to the systematic redshifts. Black outline with yellow filling shows the CO(4-3) line binned by a factor of 2. Blue outline shows the weaker \cione{} line binned by a factor of 4.}
    \label{fig:line_velocities}
\end{figure}

The high-S/N SoFiA catalogue contained the two previously identified most obvious spectral features. Manually searching for lower-S/N features as described above yielded seven additional emission lines, two of which were identified as neutral \cione{} transitions (S/N $\sim3$) associated with the two most obvious CO features, hence confirming each line as CO(4-3) and their redshift. At the fitted CO(4-3) frequencies, this constrains the redshift of J0053a as $z=3.879\pm0.002$ and J0053b as $3.878\pm0.002$. We show the alignment of the line profiles in velocity relative to the quoted redshifts in Figure \ref{fig:line_velocities}.

The remaining five emission lines (S/N $4-6$) are assumed to be CO(4-3) due to their similar frequency to the two confirmed CO(4-3) sources and hence lie at very similar redshift (justification for this is outlined in Section \ref{sec:purity}).
We use the lines' spatial and spectral positions to make moment-0 linemaps spanning approximately the full-width zero-intensity (FWZI) of the line, maximising its signal, from which we employ \texttt{photutils} \citep[][]{photutils_1.6.0_Bradley_2022} to create a 2-D segmentation map of the line emission at the $3\sigma$ level. We create segments for each identified line from which we extract the spectra for each source, ensuring we are not missing a significant number pixels containing line emission and recover a total flux for each line. We fit a model Gaussian to each line using \texttt{lmfit} \citep[][]{lmfit_1.2.2_Newville_2023} after subtracting a first order polynomial approximation of any underlying continuum. We also calculate a signal to noise ratio for line(s) in the extracted spectrum.
For a simple S/N metric, we bin the spectrum by the number of channels contained within the fitted line's full-width half-maximum (FWHM) and divide the (un-binned) fitted line's peak flux by the binned RMS noise in the spectra local to the line (typically spanning $\sim5\,$GHz split either side of the line). We estimate the RMS noise near the line since the noise changes across the ALMA Band 3 due to deeper integration on some frequencies by the overlapping tunings.

\begin{figure*}[hbpt]
    \centering
    \includegraphics[trim={0.9cm 0.4cm 1.5cm 1.5cm},clip,width=\linewidth]{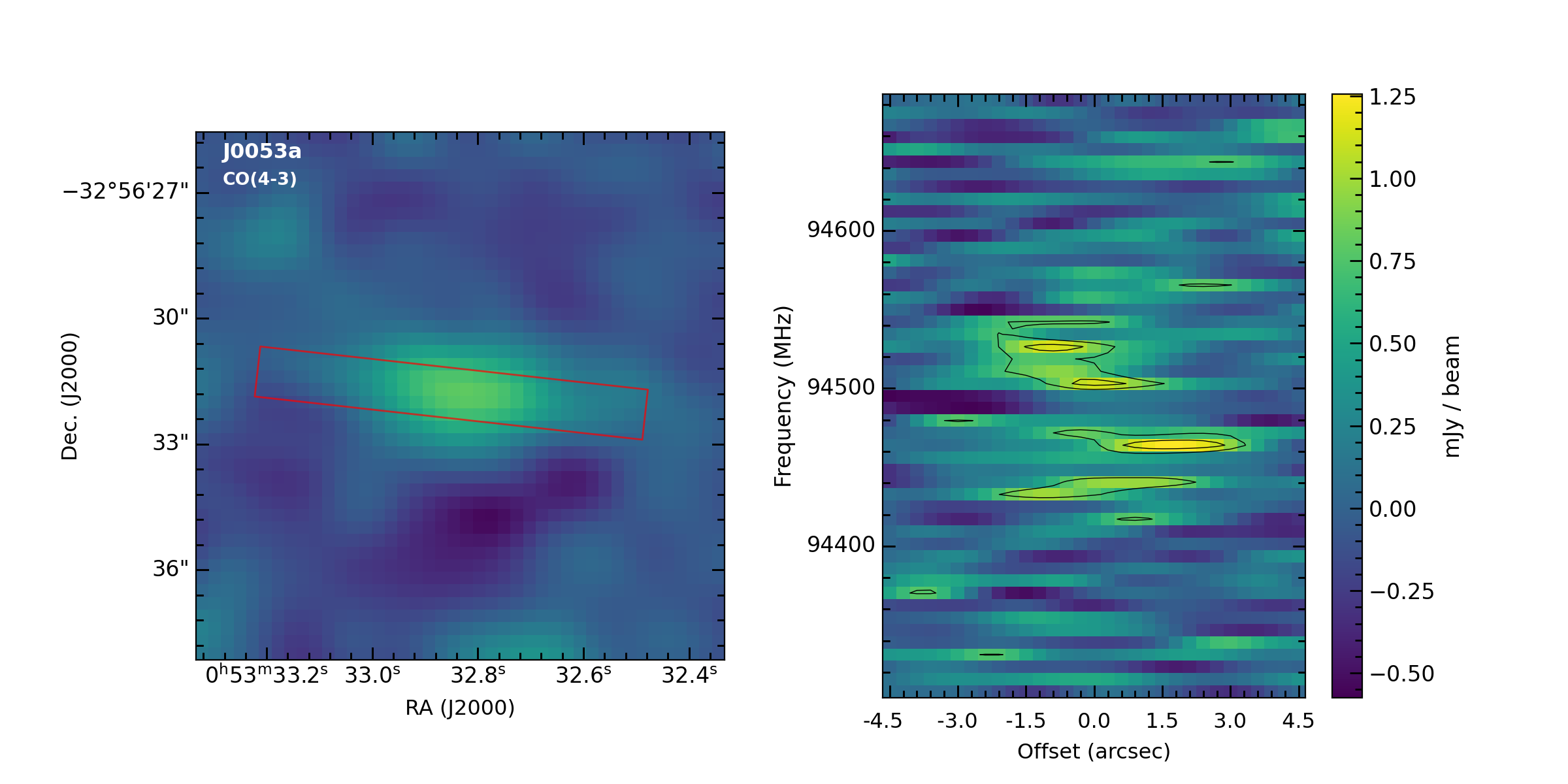}
    \caption{Position-velocity diagram (PVD) of the CO(4-3) associated with J0053a in the cube, after smoothing to the smallest common resolution across the five tunings. The PVD ({\it right panel}) is made using \texttt{pvextractor} \citep[][]{radio_astronomy_tools_python_Ginsburg_2015,pvextractor_Ginsburg_2016} with a 9\arcsec{} length slit of 1.2\arcsec{} width centred on the centroid of the CO(4-3) segment and along the position angle of the segment, both estimated by \texttt{photutils} on the FWZI linemap ({\it left panel}).}
    \label{fig:pvd_J0053a}
\end{figure*}

We show in Figure \ref{fig:pvd_J0053a} a position-velocity diagram (PVD) of the CO(4-3) associated with J0053a. The same line's profile in Figure \ref{fig:line_velocities} appears to have two peaks, which become separable into two distinct components in the PVD, with a noticeable decrease in flux density in the few channels between the emission lines. The overlap of the components along the spatial offset axis makes discerning and separating them more difficult in a 1-D spectrum. Figure \ref{fig:pvd_J0053b} shows a similar PVD for the CO(4-3) associated with J0053b.

\begin{figure*}[ht]
    \centering
    \includegraphics[width=0.83\textwidth]{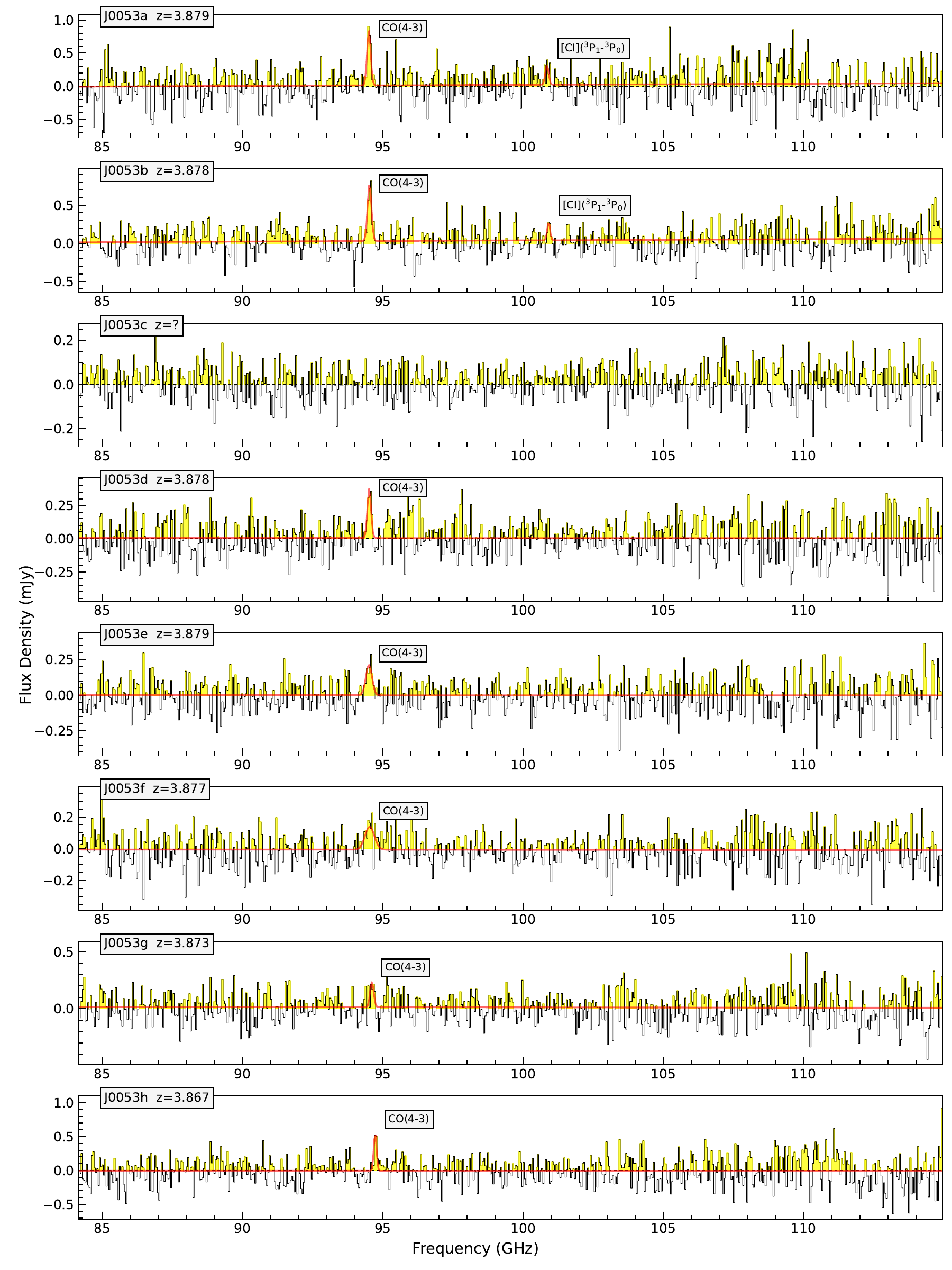}
    \caption{Full spectra for each of the candidate protocluster members, binned by a factor of six so that each channel spans approximately $47\;$MHz. These have been extracted from source masks encompassing the CO(4-3) emission (presumed for candidates without spectroscopic confirmation by the additional \cione{} line) or the 100-GHz continuum emission. All candidates with a redshift indicated are based on the assumption that the fitted line is CO(4-3). Small differences in their redshifts correspond to velocity offsets of less than 1000 km/s, indicating these are likely part of the same structure. The red line indicates the model fit, which includes the identified line(s) and underlying continuum.}
    \label{fig:spectra}
\end{figure*}

We show the full spectra for each of the candidate protocluster members in Figure \ref{fig:spectra} and additionally characterise the emission lines by fitting a Gaussian to the spectra, summarised in Table \ref{tab:line_stats}.

\begin{table*}[ht]
    \centering
    \begin{threeparttable}
        \caption{Line properties of significant detections attributed to the CO emitters. The listed properties are determined from fitting a single Gaussian to the line profile above the continuum.}
        \label{tab:line_stats}
        \begin{tabular*}{\textwidth}{@{\extracolsep{\fill}}lcccccccc}
            \hline\hline
            Source & Molecular line & S/N & $\nu_{\rm obs}$ & Redshift & $\Delta v_{\rm J0053a}$ & Peak flux & FWHM & Total flux \\
             &  &  & [GHz] &  & [km s$^{-1}$] & [mJy beam$^{-1}$] & [km s$^{-1}$] & [mJy km s$^{-1}$] \\
            \hline
            \multirow{2}{0.6em}{J0053a} & CO(4-3)\tnote{1} & 6.8 & 94.50 & 3.879$\pm$0.002 & 0 & 0.834$\pm$0.150 & 473$\pm$98 & 420$\pm$115\\
             & \cione{} & 3.8 & 100.85 & 3.880$\pm$0.005 & 93 & 0.205$\pm$0.096 & 404$\pm$217 & 88.0$\pm$62.8 \vspace{0.2cm} \\
            \multirow{2}{0.6em}{J0053b} & CO(4-3)\tnote{1} & 7.9 & 94.522 & 3.878$\pm$0.002 & -71 & 0.754$\pm$0.113 & 508$\pm$88 & 407$\pm$94\\
             & \cione{} & 2.8 & 100.906 & 3.877$\pm$0.004 & -80 & 0.139$\pm$0.065 & 327$\pm$178 & 48.3$\pm$34.8 \vspace{0.2cm} \\
            J0053d & CO(4-3)$^{\dag}$ & 7.7 & 94.512 & 3.878$\pm$0.003 & -37 & 0.381$\pm$0.096 & 472$\pm$137 & 191$\pm$74\\
            J0053e & CO(4-3)$^{\dag}$ & 5.7 & 94.502 & 3.879$\pm$0.006 & -4 & 0.215$\pm$0.061 & 800$\pm$261 & 183$\pm$79\\
            J0053f$^*$ & CO(4-3)$^{\dag}$ & 5.0 & 94.528 & 3.877$\pm$0.009 & -89 & 0.145$\pm$0.042 & 1232$\pm$411 & 191$\pm$84\\
            J0053g$^*$ & CO(4-3)$^{\dag}$ & 4.3 & 94.608 & 3.873$\pm$0.004 & -343 & 0.228$\pm$0.086 & 474$\pm$206 & 115$\pm$66\\
            J0053h & CO(4-3)$^{\dag}$ & 6.1 & 94.724 & 3.867$\pm$0.002 & -709 & 0.588$\pm$0.159 & 345$\pm$108 & 216$\pm$89\\
             \hline\hline
        \end{tabular*}
        \begin{tablenotes}
            \item [1] Lines that were detected with S/N in excess of 6 by SoFiA.
            \item [$\dag$] Transition is prescribed from the similarity to the spectroscopically confirmed sources' (a, b) CO lines.
            \item [*] These line identifications are considered tentative due to lower S/N of the line or the much larger dispersion in their FWHM.
        \end{tablenotes}
    \end{threeparttable}
\end{table*}

\subsubsection{Statistical analysis concerning the weak detection of the carbon line} \label{sec:low_snr_stat_analysis}
In order to investigate the robustness of manually identified carbon lines at S/N $\sim3$, we simulate a noise-only spectrum representative of our data with identical spectral dimension and noise sampled from a normal distribution with zero mean and standard deviation equal to the RMS noise across the entire cube (given in Section \ref{sec:imaging}). We then attempt to fit a Gaussian of similar characteristics as our detections to the anticipated location in the noise spectrum. Simulating a large number ($N=10^4$) of these randomly generated noise spectra, the false-positive rate of detecting a S/N $\ge3$ Gaussian line is $\approx2.5$\%.

For just one of our identified carbon lines to be false, the probability of 2.5\% is quite high. However we have two definite CO detections with very small spectral offset, which leads us to infer that for whichever CO line it may be, it must be the same for both sources.
Whilst the two CO detections could be random alignment in frequency of two different transitions, the differing transition -- if not CO(4-3) -- would have to be a lower transition, otherwise a second CO transition would appear at a higher frequency in the observed bandwidth. This leaves three discrete, approximate ($|\delta z|\le0.01$) redshifts that satisfy the random alignment. For a simple test we consider two spectra, where one is the same as one of our known sources at the canonical redshift, and the other is of a random source of unknown redshift. Permitting $\le1$ CO line detections, sampling a redshift from a uniform distribution such that $z\in(0,4)$ results in a probability $\le0.75\%$ for a differing CO transition to match our observations. Given there are at least five detections of CO emission within approximately $1\,{\rm arcmin}^2$ at similar frequency, the probability of physical association in an overdensity outweighs the probability of random alignment of all the detections.
Hence these identical lines become an external prior. In order to reject our proposed redshift, we must consider the probability for both of the carbon detections to be false, which is only 0.0625\%.
Furthermore, this already low false-positive rate does not account for other observed properties such as the correlation between the identified carbon lines and the CO lines in velocity/redshift. We argue that pure coincidence yielding two false-positive detections which are aligned in velocity with the known CO is less likely than what the reported low S/N ratios on their own imply.

\subsubsection{Robustness and purity of single-line CO emitters} \label{sec:purity}
We justify our claims of the lower S/N candidate protocluster members with only the single CO line detection as corresponding to CO(4-3) in two ways. Firstly, we calculate the purity of the detections, using the same formalism described as the \textit{fidelity} in \citet[][]{ALMA_CII_spectroscopic_survey_Hubble_UDF_Aravena_2016}, where the purity is $P(>S/N)=1-N_{\rm positive}/N_{\rm negative}$, and the positive and negative detection counts are obtained using the \texttt{detect\_sources} function of \texttt{photutils}' \texttt{segmentation} module at varying thresholds of S/N. The purity works as a proxy for realising the fraction of real sources by assuming negative sources are unphysical and represent the detection noise, assuming the noise in the cube is Gaussian with a mean flux density of zero. We choose to use \texttt{photutils} for this as it is the closest programmatic method to our manual line identifications, which were performed using linemaps of various spectral widths and central frequencies close to the CO(4-3) identified for J0053a/b.

We assume a line with a FWHM of $449\;$km s$^{-1}$ (the mean of all the detected lines) and test this method on both a nominal linemap (centred at $\approx94.52\;$GHz to capture as many candidate protocluster lines as possible) and linemaps with linearly spaced central frequencies across the entire cube. We find that the purity in the nominal linemap reaches $100\%$ with five positive detections at S/N $\ge3.8$, whereas the mean purity across the cube tapers off close to $100\%$ at S/N $\ge4.7$. This results in an elevated source count and purity in the nominal linemap compared to a typical linemap created anywhere else in the cube with an mean excess of four positive detections, relative to a typical linemap, down to S/N $\ge3.0$. Additionally, the six candidate protocluster lines that fall in the nominal linemap (J0053h is too offset in velocity to be included) are detected at S/N $\ge3.2$, for which there are seven positive detections and the purity is $\approx71\%$. With the caveat of small number statistics in this linemap, the purity would suggest two out of the seven positive detections are spurious, which could be either one or two of the detected candidate protocluster lines. This happens to correspond well with the tentative lines listed in Table \ref{tab:line_stats}, whereby J0053f has a very large FWHM so its signal would not be maximised in the nominal linemap and J0053g is offset enough in velocity that only part of the line would be contained in the nominal linemap. Nevertheless, the majority of the candidate protocluster line detections appear reliable.

Secondly, we make conservative estimates for the probabilities that the single-line candidate protocluster members are not emitting CO(4-3) but instead a different transition. We use the $50$th percentile fitted Schechter function parameters to the CO(1-0) luminosity function over $1.95<z<2.85$ from \citet[][]{COLDz_CO_luminosity_function_high_redshift_Riechers_2019},

\begin{align} \label{eqn:CO_luminosity_function}
    \log\Phi_{\rm CO}=\,&\log\Phi^*_{\rm CO} + \alpha\left(\log L'_{\rm CO}-\log L'^*_{\rm CO}\right) \\\nonumber
    &- \frac{L'_{\rm CO}}{L'^*_{\rm CO}\ln10} +\log(\ln10)
\end{align}

where the intrinsic luminosities, $L'$, are in units of K km s$^{-1}$ pc$^{2}$ and the luminosity density term, $\Phi$ is in units of Mpc$^{-3}$ dex$^{1}$. The fitted parameters are $\log L'^*_{\rm CO}=10.70$, $\log\Phi^*_{\rm CO}=-3.87$ and $\alpha=0.08$. For simplicity we assume this luminosity function does not evolve over the redshift range of interest ($1\lesssim z\lesssim 7$, corresponding to the redshifts of the CO transitions we will consider). We also assume a fully thermalised CO spectral line energy distribution (SLED) so that the characteristic intrinsic luminosity, $\log L'^*_{\rm CO}$, is the same for all transitions we consider. Both these assumptions result in over-estimates of the luminosity density for $J>1$ and higher redshift.

Once again, we assume a line with a FWHM of $449\;$km s$^{-1}$ and bin the cube in frequency so that each channel width is approximately this FWHM (i.e. the cube is averaged by a factor of 22). We then assume the noise in this binned cube to be the mean RMS per channel, ${\rm RMS_{binned}}\approx98\;$\textmu{}Jy beam$^{-1}$ channel$^{-1}$, although the RMS does vary over the cube due to overlaps in the tuning setup and varying atmospheric transmission. At this noise level and assumed FWHM, a pixel of total flux $132\;$mJy km s$^{-1}$ is the threshold for a $3\sigma$ detection. Considering transitions from CO(2-1) up to CO(6-5), we calculate the mean redshift that each line would be present in the cube and use this redshift in calculating the luminosity limit for each line or equivalent redshift bin. Integrating from this luminosity limit in the Schechter function, we obtain conservative (over-)estimates for the expected CO luminosity densities. Multiplying these by the respective co-moving volumes that the ALMA cube spans in each transition's redshift range, we obtain the expected number of detectable lines for each transition in the binned cube, which we use as a proxy for chance interlopers emitting each of the considered CO transitions. Over the \textit{full} cube, the expected number of blindly detected CO lines is $<1$ for each transition. Furthermore, to estimate the expected number of chance interlopers that affect our protocluster candidates, we limit the co-moving volumes for each transition to correspond to the apparent frequency range of the protocluster, e.g. for CO(4-3) we take this as the co-moving volume in $3.865<z<3.885$. The expected number of blindly detected CO lines in this narrow range is $<0.01$ for each transition. Not only is the probability of chance alignment of non-CO(4-3) low, but the number of protocluster candidates is more than an order of magnitude higher in density than expected for blind detections in a random pointing spanning the \textit{entire} ALMA Band 3 bandwidth.

The line detections appear far more likely to be physically related in a protocluster structure rather than chance alignment of CO(4-3) or other transitions, and the purity of the lines is within expectation considering the visual identification, with at least four of the lines able to be considered as `pure'.

\subsection{BEAGLE fitting} \label{sec:beagle_fitting}
We present a summary of the fitted galaxy properties from the SED modelling in Table \ref{tab:beagle_results}, quoting the mean and 68\% credible intervals as uncertainties. As J0053c does not have a line detection nor significant optical/NIR detections to estimate a photometric redshift, and may not be part of the same structure as the other sources, we do not assume a redshift or fit it using BEAGLE. Out of our seven sources that we fit with BEAGLE, only three (IDs J0053a/d/f) have reliable photometry in all six of the considered bands. Sources with ID/s J0053b/g/h have predominantly noise sampled in their apertures and are less informative to BEAGLE. For these sources, we still provide the aperture flux and flux uncertainty of non-detections to BEAGLE as the fluxes are still weighted by their uncertainties and exclude some resulting SED solutions, but BEAGLE will ignore any photometry that has no positive value within its uncertainty.

\begin{table*}[htbp]
    \centering
    \begin{threeparttable}
        \caption{BEAGLE fit results for galaxy properties. Parameter values are given as the mean with 68\% credible interval uncertainties. We report the mean rather than the maximum-a-posteriori (MAP) value as in some cases the parameter is poorly constrained and the MAP value does not necessarily represent the marginal probability distribution well (indicated by an asterisk). We do not include the results for J0053e as the photometry does not agree with any galaxy templates at the expected redshift.}
        \label{tab:beagle_results}
        \begin{tabular*}{\textwidth}{@{\extracolsep{\fill}}lccllll}
            \hline\hline
            Source & MAP $l(\theta)$ & \# usable & $M_*$ & $\tau_{V\mathrm{, eff}}$ & $\tau$ & $\psi$ \\
             &  & data & [$\log {\rm M}_\odot$] &  & [$\log\mathrm{yr}$] & [$\log {\rm M}_\odot\,\mathrm{yr}^{-1}$] \\
            \hline
            J0053a & $-6.2$ & $6$ & $10.64\,_{-0.66}^{+0.53}$ & $1.78\,_{-0.64}^{+0.49}$ & $8.62\,_{-0.35}^{+0.35}$ & $2.38\,_{-1.16}^{+0.88}$ * \\
            J0053b & $-4.4$ & $4$ & $10.84\,_{-1.08}^{+1.11}$ * & $4.90\,_{-1.49}^{+1.43}$ * & $8.61\,_{-0.40}^{+0.38}$ & $2.01\,_{-0.71}^{+0.80}$ * \\
            J0053d & $-7.8$ & $6$ & $12.12\,_{-0.20}^{+0.17}$ & $1.92\,_{-0.40}^{+0.38}$ & $8.58\,_{-0.38}^{+0.36}$ & $2.13\,_{-0.75}^{+0.79}$ * \\
            J0053f & $-9.9$ & $6$ & $10.56\,_{-0.31}^{+0.27}$ & $0.38\,_{-0.27}^{+0.30}$ & $8.54\,_{-0.33}^{+0.35}$ & $1.78\,_{-0.39}^{+0.42}$ \\
            J0053g & $-4.9$ & $4$ & $11.13\,_{-1.19}^{+1.30}$ * & $4.24\,_{-1.72}^{+1.79}$ * & $8.60\,_{-0.39}^{+0.37}$ & $2.11\,_{-0.73}^{+0.86}$ * \\
            J0053h & $-6.6$ & $5$ & $10.79\,_{-1.11}^{+1.15}$ * & $4.75\,_{-1.45}^{+1.50}$ * & $8.59\,_{-0.39}^{+0.37}$ & $2.03\,_{-0.73}^{+0.78}$ * \\
            \hline\hline
        \end{tabular*}
        \begin{tablenotes}
            \item[*] Poorly constrained parameters. These tend to have a wide marginal probability distribution, with the 68\% credible intervals spanning $>1\,$dex.
        \end{tablenotes}
    \end{threeparttable}
\end{table*}

\section{Discussion} \label{sec:discussion}
\subsection{Fitted galaxy properties}  
The Bayesian fits produced by BEAGLE are limited by the number of photometric data points available.
Constraints on star formation rate suffer from the lack of rest-frame UV detections. However, our sources with all six photometric points yield realistic credible intervals. The star formation timescale, which depends on our chosen $200\,$Myr `constant' star formation history, does not seem to differ greatly in the credible intervals for each source (indeed, the marginal probability distributions are wide for each source) and does not necessarily yield any new information about the candidate members in the protocluster. The most useful parameters are the stellar mass and the $V$-band attenuation optical depth, both of which are constrained by the available photometry.

We note some features of the BEAGLE output for the sources with good photometry. J0053a/d/f all have high stellar mass with the range in their 68\% credible intervals $\lesssim1\,$dex, especially J0053d which is the only candidate protocluster member with an obvious $K$-band detection of the host galaxy. The $V$-band optical depths for J0053a/d suggest significant dust obscuration, which for J0053a is supported by the observed rise in continuum across ALMA Band 3, probing the grey-body cold dust emission.

To verify if BEAGLE's mass estimates are plausible, we consider J0053d and compare to the mass estimated from the mass-to-light ratio of an evolving dusty starburst template from P\'egase \citep[][]{Pegase_spectral_evolution_galaxies_with_dust_Fioc_2019}. We find that for J0053d's observed $K_s$-band flux, a dusty starburst in the nearest redshift bin, $z=3.779$, would be $M_*\sim4\times10^{11}\,{\rm M}_\odot$. This is within $0.5\,$dex of BEAGLE's estimate and an acceptable lower limit given that BEAGLE is allowing for posteriors with higher dust and higher mass. The true mass could be closer to either estimate, but confirmation would require deeper observations.


\begin{figure*}[htbp]
    \centering
    \includegraphics[width=0.45\linewidth]{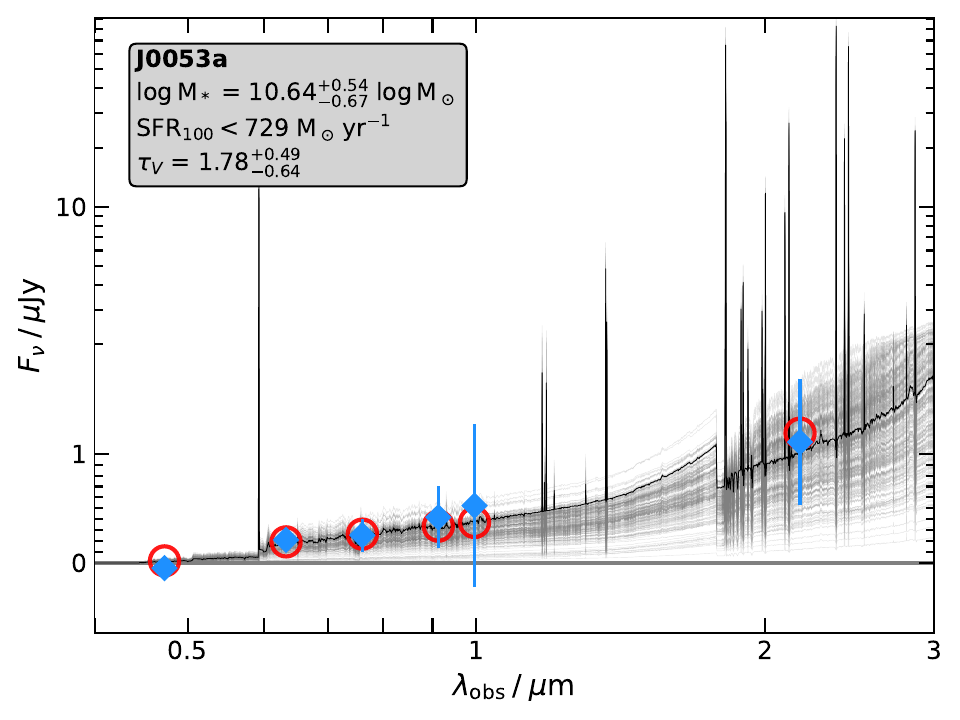}
    \quad
    \includegraphics[width=0.45\linewidth]{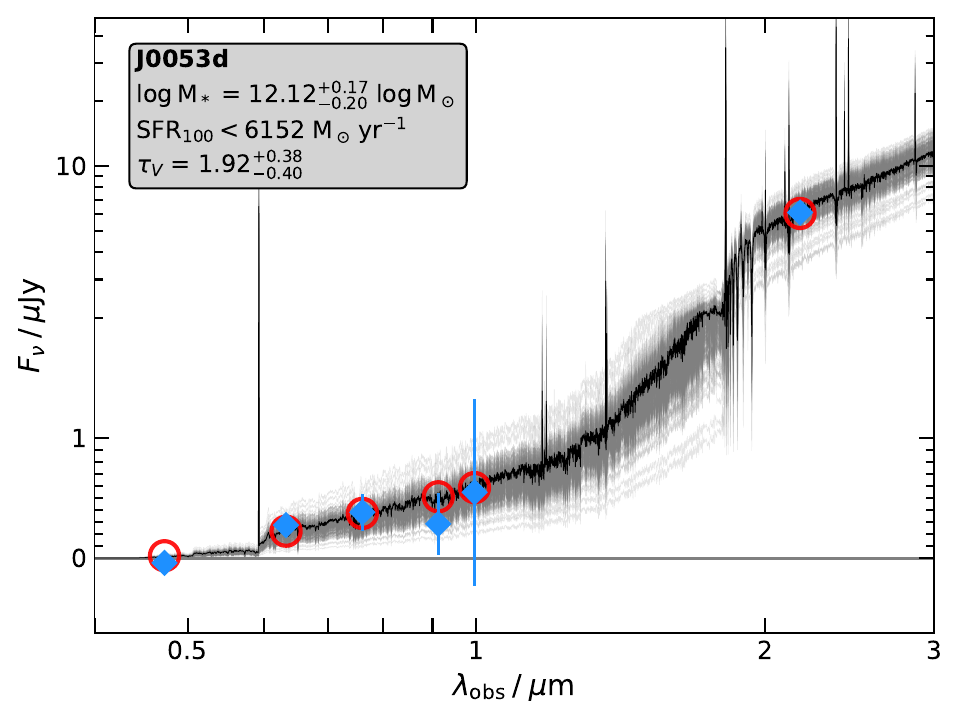}
    \caption{Fitted SEDs of J0053a ({\it left}) and J0053d ({\it right}) with BEAGLE. Black SEDs are the MAP solutions and grey SEDs are alternative solutions for walkers within the 68\% credible intervals of the fitted parameters, representative of the scatter in the fitted SED. Blue diamonds with error bars mark the input (observed) photometry and their $1\sigma$ uncertainties. Open red circles mark the predicted MAP photometry with the same filters. A summary of the primary constraints are inset in the upper left of each panel including the mean and 68\% credible intervals for both the mass and $V$-band optical depth, and the upper limit on the SFR over the past $100\,$Myr from the star formation history.}
    \label{fig:beagle_seds}
\end{figure*}

Generally, we disregard the posteriors for all sources but J0053a/d/f. We utilise the mass' upper 95\% credible interval as a limit for the other sources, as non-detections of flux do provide a constraint on the stellar mass with BEAGLE (albeit a very weak one).

\subsection{Interpretation of sources} \label{sec:source_properties}
Many of these sources are detected (or inferred from their association to molecular gas emission) for the first time in these ALMA Band 3 observations. Unless otherwise stated, we assume these sources are similar to sub-millimetre galaxies (SMGs) given their faintness at optical and NIR wavelengths as well as their clustering in space, similar to SMG protoclusters \citep[e.g.][]{DRC_z_4_proto-cluster_DSFG_Oteo_2018}. With the available data (optical, NIR, millimetre and centimetre wavelengths), we discuss the likely nature of each source with the primary interpretations being either SMGs that are heavily dust-obscured with rest UV-optical emission reprocessed to longer wavelengths, or less dusty, low-mass galaxies.

\begin{enumerate}[label=\textbf{J0053\alph*}),leftmargin=*,wide=0pt]
    \item The target radio galaxy at the centre of the pointing has an extremely marginal ($\sim1.9\sigma$) detection of a host galaxy in the HAWK-I image from B24 which appears to coincide with the western component of the faint ALMA continuum, also closest to the 5.5-GHz ATCA centroid (see Figure \ref{fig:overlays}). Using the power-law model of J0053a's radio synchrotron emission in Table 7 of B22, it has a predicted 100-GHz synchrotron contribution of $4.4\,$\textmu{}Jy, below the peak flux of $22.4\,$\textmu{}Jy, suggesting that this up-turn in the radio SED has a low synchrotron contribution at this frequency. J0053a has a rising continuum across the $84.2$--$114.9\,$GHz bandwidth covered by the ALMA tunings (approximated by a first-order polynomial which gives $\alpha_{100\;\mathrm{GHz}}=3.0\pm1.0$). This is consistent with the $100$-GHz continuum being driven mainly by thermal emission of cold dust. In the DES cutouts, the marginal $\sim1\,$\textmu{}Jy HAWK-I source has upper limits in the $r,\,i$ and $z$ bands of $\sim0.30,\,0.41$ and $0.70\,$\textmu{}Jy respectively (see Table \ref{tab:photometry}). The intriguing case with J0053a is that the CO emission appears to extend a significant distance away from the faint host galaxy. The unresolved ATCA emission is offset slightly from the potential host galaxy and even moreso from the extended CO emission, so the radio detection is likely a one-sided jet or the potential host is incorrect. With the spatially resolved kinematics of the CO shown in Figure \ref{fig:pvd_J0053a}, it appears there are two distinct clouds of gas that overlap spatially, but are separated in velocity. This is indicated by the sharp drop in both the PVD and binned spectrum in Figure \ref{fig:line_velocities}, where the flux density drops to less than a third of the peak line flux for three consecutive channels in the PVD. Whilst it is possible that the dip in flux is spurious, the dip occurs across three consecutive channels and the standard deviation of the flux densities in this dip is lower than the standard deviation of the flux densities away from the dip and emission line, appearing to vary more smoothly despite the sharp drop. Assuming that both clouds are bound to a galaxy and are not isolated in the ICM, this suggests that the host of J0053a is participating in a major merger with a radio-quiet, NIR-dark galaxy to the east. Powerful and massive radio galaxies have long been associated with mergers and merger-induced starbursts \citep[][]{Gas-rich_mergers_starbursting_radio_galaxies_Ivison_2012}. These likely fuel accretion onto the SMBH, enabling the powerful radio emission and leading to a high stellar mass. For a young radio galaxy, this may be at an early stage of the merger or interaction between galaxies. With this interpretation, we understand the distinct gas clouds traced by CO to be in-falling on the merger somewhat along the line of sight, resulting in the velocity separation. This is distinct from a scenario whereby the molecular gas is outflowing via mechanical jet feedback, in which we would expect the blue-shifted gas cloud to be aligned closer to the ATCA centroid (if a one-sided jet). The carbon line for this source has very low S/N and appears to sit in close proximity to the faint host galaxy and radio emission.
    
    \item The brightest continuum source at $100$-GHz in the field of view, also with a rising continuum within ALMA Band 3 ($\alpha_{100\;\mathrm{GHz}}=3.2\pm0.6$). It has no counterpart in $K$-band nor in the radio at or below 9-GHz. This ALMA-discovered companion galaxy is almost certainly an SMG with emission peaking between the sub-millimetre and FIR. The large amount of available cold gas inferred from the CO line ($M_{\rm H_2}>1.28\times10^{10}\,{\rm M}_\odot$; Table \ref{tab:cluster_stats}) suggests that the galaxy is capable of copious star formation. Since it is not detected at lower radio frequencies, we determine that it has weaker synchrotron emission and that the 100-GHz emission is dominated by thermal emission -- primarily by cold dust, as is seen in SMGs with steep ($\alpha\sim3$) greybody emission on the optically thin side of the FIR `bump'. The weaker synchrotron from the inferred SFR would result in a $5.5$-GHz flux density below our ATCA detection limit. 
    
    \item The third continuum source reported in Table \ref{tab:source_stats}. We do not find any associated emission in the DES or HAWK-I images, nor significant emission lines around the source, so we cannot determine a redshift or whether it is part of the protocluster. A protocluster member with weak CO emission is one possibility for the source, whereby a past interaction stripped away the molecular gas or the gas may have been recently depleted, leaving behind a starved SMG.
    
    \item The associated CO(4-3) has a similar FWHM to the lines of J0053a/b, but is fainter. The emission is likely related to the galaxy seen filling the left side of the contours in the HAWK-I image (see Figure \ref{fig:overlays}). Fitting this galaxy with BEAGLE suggests it is very dust-reddened and has an extreme $M_*\sim10^{12}\,{\rm M}_\odot$ stellar mass. We verify that the AGN-free galaxy templates used are an appropriate match by obtaining a flux density limit with the {\it Widefield Infrared Survey Explorer} \citep[{\it WISE};][]{WISE_Wright_2010} in unWISE \citep[][]{unWISE_coadds_Lang_2014} $W1$ ($3.4\,$\textmu{}m) and $W2$ ($4.5\,$\textmu{}m) of $\sim10\,$\textmu{}Jy, which indicates that the SED flattens longward of $K_s$-band and shows no evidence of hot dust due to an AGN torus. Similarly, there are no prominent emission lines that would fall in the $K_s$-band that would warrant using an extreme emission line galaxy template. The fitted SED of this ``Big Red Galaxy'' is shown in Figure \ref{fig:beagle_seds} alongside that of the radio galaxy, including the MAP walker and walkers within the 68\% credible interval. Considering this source had one of the best fits produced by BEAGLE, the large mass raises questions regarding the process that allowed it to accumulate so much mass, whether it had already interacted with (and stripped) the mass of other members of the protocluster, and whether it hosts an AGN, which could be heavily obscured by the excessive dust or currently be switched-off.
    
    \item The CO(4-3) appears diffuse with a large FWHM. We note that the aperture photometry is extracted on a slightly offset source in the stacked DES image, which may not be the host galaxy due to its photometry appearing to resemble that of a lower redshift galaxy with brighter emission in the shorter wavelength bands. The true host could be undetected and aligned with this other source.
    
    \item The extremely broad line profile of the CO(4-3) is quite noisy as it is closer to the edge of the effective ALMA field-of-view. However, SED fitting of a possible host galaxy slightly North-East of the CO(4-3) -- which is detected in most DES bands and detected, although diffuse, with HAWK-I -- does appear to have a template match at the inferred redshift. The SED modelling of this host galaxy suggests it is the least dusty of the candidate protocluster members and its mass is on the lower end, although this could be an unassociated lower redshift interloper.
    
    \item This is the weakest CO(4-3) detection. The lack of any host galaxy detection does not distinguish between an extremely dusty SMG or a low-mass galaxy, evidenced by the poorly constrained galaxy properties in Table \ref{tab:beagle_results}.
    
    \item Relative to the other sources, this CO(4-3) emission has the largest velocity offset of approximately 700 km/s relative to J0053a. The observed velocity dispersion and fitted width is much smaller than for the others. However, the offset is still plausible for J0053h to belong to the group of sources. The detection of a possible host galaxy is similar to J0053g in that it is ambiguous and poorly constrained by SED fitting.
\end{enumerate}

Given eight currently detected candidate protocluster members within $\sim 1\,{\rm arcmin}^2$ and $3.85<z<3.90$, this field's source density is 20 times that of the Cosmic Assembly Near-infrared Deep Extragalactic Legacy Survey \citep[CANDELS;][]{CANDELS_Grogin_2011} which has $0.4$ sources per arcmin$^2$ per $0.05$ redshift bin, valid over the range $3.8<z<4.0$ in the Great Observatories Origins Deep Survey - South (GOODS-S) field \citep[][]{CANDELS_photz_catalogue_Kodra_2023}.
The radio galaxy (J0053a), which is likely involved in a merger, has pinpointed this over-dense region similar to other powerful HzRGs that act as beacons of rapid galaxy growth \citep[e.g.,][]{HzRG_protocluster_BCG_ancestor_Venemans_2002,HzRG_protocluster_BCG_population_2007,Clusters_around_radio_loud_AGN_z_1_3_Wylezalek_2013,Cluster_MIR_luminosity_function_z_1_3_Wylezalek_2014,CARLA_clusters_around_radio_loud_AGN_z_2_structures_Noirot_2016,CARLA_clusters_around_radio_loud_AGN_z_1_3_structures_Noirot_2018}. Such a HzRG in an over-density at $z>3$ allows for comparison with counterpart systems at lower redshifts, such as the Spiderweb galaxy \citep[$z=2.16$;][]{Spiderweb_galaxy_z_2.2_Lyman_alpha_emitters_Kurk_2000,Spiderweb_galaxy_Miley_2006}, to probe the evolution of radio galaxies and clusters of galaxies.

\subsection{Estimating protocluster properties} \label{sec:protocluster_properties}
We estimate several properties of the candidate protocluster members, prioritising conservative estimates where assumptions are required. Properties such as star formation rate, molecular gas mass and stellar mass for the candidate protocluster members allow us to constrain the same for the protocluster as a whole and compare this over-density to similar sources in the literature (i.e. HzRGs and SMGs at $z\sim4$).

\subsubsection{Star formation rate estimate from ALESS templates} \label{sec:aless_sfr}
ALMA's pathfinder, the Atacama Pathfinder Experiment \citep[APEX;][]{APEX_ALMA_pathfinder_Gusten_2006} performed the Large Bolometric Camera \citep[LABOCA;][]{LABOCA_Large_APEX_Bolometric_Camera_Siringo_2009} Extended {\it Chandra} Deep Field South (ECDFS) survey \citep[LESS;][]{LESS_LABOCA_ECDFS_Survey_South_Weiss_2009}, paving the way for ALMA to observe a number of sub-millimetre selected sources in the ALMA LESS \citep[ALESS;][]{ALESS_SMG_catalogue_multiplicity_Hodge_2013} survey.
Assuming the 100-GHz continuum-detected J0053a/b have their flux dominated by cold dust like SMGs, we compare our fluxes and limits at different wavelengths to the SED templates of the ALESS SMGs \citep[][]{ALESS_SMG_UV_to_radio_modelling_da_Cunha_2015}. We are mainly interested in the median SFR and 16th--84th percentiles which are consistent for both optically bright and optically faint sources, as well as the whole sample. From this, we can scale our observed-frame 3$\,$mm flux densities to the equivalent redshifted average ALESS templates and infer a crude estimate of the SFR. These estimates and their uncertainties are presented in Table \ref{tab:cluster_stats}. We also note that the ALESS templates imply observed $K_s$-band flux densities fainter than J0053a/b, by factors of $2-10$.

\subsubsection{Molecular gas mass estimates from CO(4-3)} \label{sec:molecular_gas}
At high redshift and mm/sub-mm wavelengths, observations of high-J level ($J_\mathrm{upper}=3\to J_\mathrm{upper}=11$) ro-vibrational CO ($\nu=0$) transitions have proven to be efficient to detect and resolve kinematically (and sometimes spatially) with ALMA \citep[][]{ALMA_redshifts_SPT_survey_Weiss_2013,High-z_DSFG_Casey_2014,sub-mm_SED_high_z_DSFG_Spilker_2014,ALMA_NOEMA_molecular_gas_high_z_SFG_Birkin_2021,CO_molecular_gas_state_local_high_z_galaxies_Liu_2021}. As the second most abundant molecule behind molecular hydrogen, studies utilising these CO emission lines provide unique insight into the state of the molecular gas and estimates of properties that may be difficult to obtain otherwise. Many estimates depend on scaling relations, so the uncertainties remain quite high, but for properties such as total molecular gas mass where the detection of inert H$_2$ is unfeasible, any constraint is welcome. A number of studies report the ratio between molecular and atomic gas masses \citep[][]{Molecular_atomic_gas_ratio_in_spiral_galaxies_Young_1989,Molecular_gas_in_spiral_galaxies_Young_1991,Molecular_gas_in_spiral_galaxies_Casoli_1998}, both fuel for star formation, though this ratio seems to vary strongly with the target, typically varying between $0.1$ and $1$.

Due to collisional excitations of CO with other CO and H$_2$ molecules, the observed total flux densities of CO lines make great tracers for H$_2$ and we can estimate the interacting H$_2$ mass from the CO(1-0) luminosity using scale factors that have been calibrated for various types of objects.

We start by calculating the total flux of our CO(4-3) line, $I_\mathrm{CO(4-3)}$ from the Gaussian we had fit spectrally,

\begin{equation} \label{eqn:co_total_flux}
    \left[\frac{I_\mathrm{CO(4-3)}}{\mathrm{Jy\;km\;s}^{-1}}\right]=\sqrt{2\pi}\left[\frac{\int S_\nu d\Omega\mathrm{\;(peak)}}{\mathrm{Jy}}\right]\left[\frac{\sigma}{\mathrm{km\;s}^{-1}}\right],
\end{equation}

\noindent where $\sigma$ is the standard deviation (related to the FWHM by $2\sqrt{2\ln{2}}$) and $\int S_\nu d\Omega\mathrm{\;(peak)}$ is the peak amplitude after integrating spatially (noting that our integrated flux is no longer per beam). We then obtain the intrinsic luminosity of the CO(4-3) line, $L^\prime_\mathrm{CO(4-3)}$ following the process in \citet[][]{high_z_molecular_gas_Solomon_2005} and references therein:

\begin{align} \label{eqn:co_lum}
    \left[\frac{L^\prime_\mathrm{CO(4-3)}}{\mathrm{K\;km\;s}^{-1}\;\mathrm{pc}^2}\right]=\,&3.25\times10^7\left[\frac{I_\mathrm{CO(4-3)}}{\mathrm{Jy\;km\;s}^{-1}}\right]\left[\frac{d_L}{\mathrm{Mpc}}\right]^2\\\nonumber
    &\times\left[\frac{\nu_\mathrm{CO(4-3),rest}}{\mathrm{GHz}}\right]^{-2}\left[1+z\right]^{-1},
\end{align}

\noindent From here we require two key factors to obtain an estimate of the H$_2$ gas mass: $r_{4-1}$, which is the ratio between the CO(4-3) and CO(1-0) intrinsic luminosities, and $\alpha_\mathrm{CO}$, which is the calibrated conversion factor from CO(1-0) line luminosity to H$_2$ gas mass. This yields the H$_2$ gas mass as

\begin{equation} \label{eqn:h2_mass}
    \left[\frac{M_{\rm H_{2}}}{\mathrm{M}_\odot}\right]=\frac{\alpha_\mathrm{CO}}{r_{4-1}}\left[\frac{L^\prime_\mathrm{CO(4-3)}}{\mathrm{K\;km\;s}^{-1}\;\mathrm{pc}^2}\right].
\end{equation}

Both $r_{4-1}$ and $\alpha_\mathrm{CO}$ can vary depending on the state of the gas and the object in question. Ideally, an estimate of the ratio from a J-level transition CO line luminosity to the (1-0) line luminosity can be obtained by re-constructing the CO spectral line energy distribution (SLED) with more than one observed line. however, we only have CO(4-3). The range of $\alpha_\mathrm{CO}$ at high redshift is also difficult to constrain, so we make the conservative assumption, of a fully thermalised CO SLED (i.e. $r_{4-1}=1$) and low $\alpha_{\rm CO}=0.8$ \citep[used in][]{Hot_cold_molecular_gas_outflow_NGC_3256_secondary_Emonts_2014,Dragonfly_2_z_2_ALMA_CO_6-5_merger_ULIRG_HzRG_Emonts_2015,Dragonfly_1_z_2_ATCA_CO_1-0_merger_ULIRG_HzRG_Emonts_2015}, similar to local ULIRGs \citep[see][]{CO_H2_molecular_gas_mass_conversion_nuclear_rings_starbursts_ULIRGs_Downes_1998}. Our H$_{\rm 2}$ gas masses are therefore lower limits.

Note that for fully thermalised, optically thick CO emission, the ratio of the velocity integrated flux of a transition, $I_{\rm CO(J\to J-1)}$, to the first ($1\to0$) transition is equal to $J^2$, whereas for the intrinsic line luminosity, Equation \ref{eqn:co_lum} contains the rest frequency of the line. For rotational CO, this scales with $J$ such that $\nu_{\rm CO(J\to J-1)}\approx J\nu_{\rm CO(1\to0)}$. This is often referred to as the `CO ladder' and the dependence on $J$ means any intrinsic line luminosity ratios for fully thermalised CO gas reduce to 1 as the $J^2$ factor from the velocity integrated flux ratio cancels with the $J^2$ factor in the rest frequency ratio.

\subsection{Comparison to radio galaxies in clusters and SMG clusters at \texorpdfstring{$z\sim4$}{z\~4}} \label{sec:clusters}
We compare the J0053 protocluster, and its members, to various HzRGs and SMGs as listed in Table \ref{tab:cluster_stats}.

\begin{table*}[htbp]
    \centering
    \begin{threeparttable}
        \caption{Estimated SFR, molecular gas mass, stellar mass and gas depletion timescale for J0053a/b, and a number of powerful HzRGs or highly star-forming SMGs in (proto-)clusters. The table is divided into: high redshift sources associated with powerful radio emission (HzRGs), SMG protocluster members, and total protocluster properties. Sources/protocluster from this work are bolded. Unless otherwise stated, the molecular gas mass is derived from CO(1-0) line luminosities (inferred from CO(4-3) assuming a thermalised CO SLED for this work) with a conversion factor $\alpha_\mathrm{CO}=0.8$ and the gas depletion time is the minimum ($t_\mathrm{depl}=M_\mathrm{H_2}/\mathrm{SFR}$, i.e. assuming $100\%$ star formation efficiency; all the molecular gas is available for star formation).}
        \label{tab:cluster_stats}
        \begin{tabular*}{\textwidth}{@{\extracolsep{\fill}}cccccc}
            \hline\hline
            Source & Redshift & SFR & $M_{\mathrm{H}_2}$ & $M_*$ & min($t_\mathrm{depl}$) \\
             &  & [M$_\odot\,$yr$^{-1}$] & [M$_\odot$] & [M$_\odot$] & [Myr] \\
            \hline
            \textbf{J0053a (HzRG)} & $\mathbf{3.879}$ & $\mathbf{270\pm106,\,963\pm377\,^{a}}$ & $\mathbf{>1.31\times10^{10}}$ & $\mathbf{<1.5\times10^{11}\,^{c}}$ & $\mathbf{>48.5,\,>13.6}$\\
            MRC 1338-26 & $2.16$ & $1390\pm150$ & $\sim2\times10^{10}$ & $<1.1\times10^{12}$ & $\sim40$\\
            SPT2349-56 (B) & $\sim4.303$ & $1227\pm409$ & $(11.2\pm2.0)\times10^{10}$ & - & $91$\\
            SPT2349-56 (C) & $\sim4.303$ & $907\pm302$ & $(6.7\pm1.2)\times10^{10}$ & - & $74$\\
            SPT2349-56 (G) & $\sim4.303$ & $409\pm137$ & $(2.9\pm1.3)\times10^{10}$ & - & $71$\\
            MRC 0943-242$\,^{e}$ & $2.92$ & $41^{+26}_{-33}$ & $(4.9\pm2.7)\times10^9$ & $1.66\times10^{11}$ & $\sim120$ \\
            TN J0121+1320$\,^{e}$ & $3.52$ & $626^{+267}_{-243}$ & $(2.60\pm1.46)\times10^{10}$ & $1.05\times10^{11}$ & $\sim42$ \\
            4C + 03.24$\,^{e}$ & $3.57$ & $142^{+240}_{-130}$ & $(6.5\pm2.2)\times10^9$ & $<1.86\times10^{11}$ & $\sim46$ \\
            4C + 19.71$\,^{e}$ & $3.59$ & $84^{+173}_{-62}$ & $(2.54\pm0.88)\times10^{10}$ & $<1.35\times10^{11}$ & $\sim30$ \\
            \hline
            \textbf{J0053b (SMG)} & $\mathbf{3.878}$ & $\mathbf{305\pm71,\,1088\pm254\,^{a}}$ & $\mathbf{>1.28\times10^{10}}$ & $\mathbf{<8.8\times10^{11}\,^{c}}$ & $\mathbf{>42.0,\,>11.8}$\\
            DRC-1 & $\sim4.0$ & $\sim2900\,^{a}$ & $\sim2.62\times10^{11}\,^{b}$ & - & $\sim90$ \\
            DRC-2 & $\sim4.0$ & $\sim990\,^{a}$ & $\sim1.18\times10^{11}\,^{b}$ & - & $\sim120$ \\
            DRC-3 & $\sim4.0$ & $\sim902\,^{a}$ & $\sim1.78\times10^{11}\,^{b}$ & - & $\sim200$ \\
            DRC-1 $^{d}$ & $\sim4.0$ & $1744\pm1162$ & $(8.62\pm2.15)\times10^{11}$ & $(16\pm7)\times10^{10}$ & $494$ \\
            DRC-2 $^{d}$ & $\sim4.0$ & $1132\pm1013$ & $(2.94\pm0.73)\times10^{11}$ & $(8\pm6)\times10^{10}$ & $260$ \\
            DRC-3 $^{d}$ & $\sim4.0$ & $1527\pm1303$ & $(2.68\pm0.67)\times10^{11}$ & $(17\pm12)\times10^{10}$ & $176$ \\
            \hline
            \textbf{J0053 (total in-field)} & $\mathbf{\sim3.9}$ & $\mathbf{\lesssim2000}$ & $\mathbf{>5.5\times10^{10}}$ & $\mathbf{<6\times10^{12}}$ & $\mathbf{\gtrsim27.5}$\\
            SPT2349-56 (total) & $\sim4.303$ & $6000\pm600$ & $6\times10^{11}$ & - & $100$\\
            DRC$_\mathrm{tot}$ \citet{DRC_z_4_proto-cluster_DSFG_Oteo_2018} & $\sim4.0$ & $\sim6500\,^{a}$ & $\sim6.6\times10^{11}\,^{b}$ & - & $\sim100$\\
            DRC$_\mathrm{tot}$ \citet{DRC_z_4_detailed_evolution_Long_2020}$^{d}$ & $\sim4.0$ & $\sim5973$ & $\sim1.8\times10^{12}$ & $\sim9\times10^{11}$ & $\sim300$ \\
            \hline\hline
        \end{tabular*}
        \begin{tablenotes}
            \item [$a$] Estimates of SFR for this work and the Dusty Star-forming Galaxies (DSFGs) in the DRC are obtained by rescaling the ALESS template to the observed wavelength of the respective ALMA data. For this work, the ranges given are indicative of the optically-bright and optically-faint ALESS template scalings, since both could be plausible scenarios given our $K_s$-band limits are greater than both templates at the canonical redshift. Thus a low and high ALESS-scaled SFR is given for J0053a/b as well as two corresponding min($t_\mathrm{depl}$).
            \item [$b$] DRC molecular gas mass(es) instead derived from \cione{}, with Equation 1 of \citet{DRC_z_4_proto-cluster_DSFG_Oteo_2018}.
            \item [$c$] Upper limits on the stellar mass for this work are derived from the 68\% upper credible intervals from BEAGLE.
            \item [$d$] The methods used to estimate these properties in \citet{DRC_z_4_detailed_evolution_Long_2020} deviate from the other examples given, utilising SED fitting code \citep[CIGALE;][]{CIGALE_1_original_Burgarella_2005,CIGALE_2_SINGS_test_sample_Noll_2009,CIGALE_3_python_port_Boquien_2019} and a method from \citet{Mass_and_SFR_law_dust_continuum_ALMA_COSMOS_Scoville_2016} to estimate molecular gas mass which uses the link between the Rayleigh-Jeans tail tracing dust and the molecular gas within the ISM, calibrated by rest-frame $850\,\mu$m luminosity and molecular gas mass (this absorbs variations in dust temperature, opacities and abundances into a single ratio, further calibrated by CO(1-0) measurements in DSFGs making it well-suited for DRC over more common $\alpha_\mathrm{CO}$ conversions). Molecular gas mass uncertainties with this method are $\sim25\%$. See \citet{DRC_z_4_detailed_evolution_Long_2020} and \citet{Mass_and_SFR_law_dust_continuum_ALMA_COSMOS_Scoville_2016} for comprehensive details and discussions of the choices of methods and derivations leading to their values.
            \item [$e$] Four sources taken from a sample of seven HzRGs in \citet[][]{Faint_Ci_z_3.5_HzRG_Kolwa_2023} which exhibit faint \cione{} emission. The molecular gas masses are estimated using \cione{} \citep[][Equation 3]{Faint_Ci_z_3.5_HzRG_Kolwa_2023}. The SFRs are reported from \citet[][]{ALMA_quenching_HzRGs_Falkendal_2019}. The stellar masses are taken from \citet[][]{Spitzer_HzRG_survey_DeBreuck_2010} for MRC 0942-242 and from \citet[][]{Disentangling_SFR_AGN_in_HzRG_ULIRGS_z_1-4_Drouart_2016} for the other three.
        \end{tablenotes}
    \end{threeparttable}
\end{table*}

If we interpret the hosts of our candidate protocluster members, identified by molecular clouds with CO(4-3) emission, as SMGs not yet detected at optical to NIR wavelengths (other than J0053d), then the environment in this protocluster bears similarity to the high redshift cluster discoveries densely populated by SMGs such as the core of SPT2349-56 \citep[14 sources at $z\sim4.3$;][]{SPT2349-56_galaxy_cluster_core_z_4.3_Miller_2018,SPT2349-56_z_4.3_radio_loud_AGN_Chapman_2024} and the central complex of the Distant Red Core \citep[DRC, 11 sources at $z\sim4$;][]{DRC_z_4_proto-cluster_DSFG_Oteo_2018,DRC_z_4_detailed_evolution_Long_2020}. If J0053's currently known members form the protocluster core, like with the 14 sources in SPT2349-56 and the 11 sources in the DRC, then not only is this core significantly larger ($\sim430\,$kpc) in comparison, but could be more widespread and populated than the current field of view and depth with ALMA reveals.

Powerful radio galaxies acting as beacons for SMBH growth and galaxy overdensities are common \citep[e.g.,][]{HzRG_protocluster_BCG_ancestor_Venemans_2002,HzRG_protocluster_BCG_population_2007,MRC1138-262_HzRG_z_2.2_Spiderweb_galaxy_coeval_growth_Seymour_2012,Dragonfly_1_z_2_ATCA_CO_1-0_merger_ULIRG_HzRG_Emonts_2015,Clusters_around_radio_loud_AGN_z_1_3_Wylezalek_2013,Cluster_MIR_luminosity_function_z_1_3_Wylezalek_2014,CARLA_clusters_around_radio_loud_AGN_z_2_structures_Noirot_2016,CARLA_clusters_around_radio_loud_AGN_z_1_3_structures_Noirot_2018}. We therefore also compare J0053a to HzRGs. Indeed, SPT2349-56 recently was discovered to have a radio-loud AGN \citep[][]{SPT2349-56_z_4.3_radio_loud_AGN_Chapman_2024} in the central complex of its SMGs (overlapping with its components B, C and G). This perhaps makes SPT2349-56 the closest cousin to the J0053 protocluster (although we only detect J0053a/b in continuum at 100-GHz with ALMA). At lower frequencies, J0053a has a luminosity $L_{\rm 500\;MHz}=1.3\times10^{28}\;{\rm W\,Hz}^{-1}$ and the comparison radio sources in Table \ref{tab:cluster_stats} have luminosities in the range $4.7\times10^{28}<L_{\rm 500\;MHz}<2.7\times10^{29}\;{\rm W\,Hz}^{-1}$ (excluding SPT2349-56 which has 1-2 orders of magnitude lower radio luminosity, $\sim9.8\times10^{26}\;{\rm W\,Hz}^{-1}$). J0053a is less radio powerful than the comparison HzRGs chosen, as it lies at an intermediate redshift compared to the target $z>5$ redshift range in B22.


Compared to the SFRs listed, our estimates for J0053a/b are similar to the highly star-forming HzRGs and the SPT2349-56 C and G components. The ALESS-scaled SFRs for J0053a and J0053b agree well with the 68\% upper credible intervals from BEAGLE (${\rm SFR}_{10\,{\rm Myr}}\sim871$ and $\sim640\,{\rm M}_\odot{\rm yr}^{-1}$ respectively), although these are still not constrained very well (given the rest-frame UV depth of DES). We use the upper 68\% credible intervals for the stellar masses estimated by BEAGLE to approximate an upper limit on the total protocluster stellar mass of its currently known members. Deep detections in the rest-frame UV-optical would provide better constraints on the (unobscured) stellar mass and SFRs.

Even with our extremely conservative estimates of the (cold) molecular gas mass, the lower limits for J0053a are at the same order of magnitude as the other reference HzRGs. This largely stems from the factor in Equation \ref{eqn:h2_mass}, $\alpha_\mathrm{CO}/r_{4-1}=0.8$. For J0053a, ignoring the SMGs (including those constituting SPT2349-56), the other radio galaxies' molecular gas masses are within a factor of five, which is reasonable for a more realistic gas mass.

The molecular gas mass of J0053b, with more realistic conversions from CO luminosity, could also be five times higher, bringing it in line with the other SMGs. This also applies when comparing the total in-field estimates, as well as considering that the full extent of the J0053 protocluster is not yet known (at least $1.1\arcmin$ diameter) and both SPT2349-56 and the DRC have more protocluster members (in smaller areas $0.7\arcmin$ and $0.3\arcmin$ respectively).


\subsection{Massive, extremely red objects at \texorpdfstring{$z\sim4$}{z\~4}} \label{sec:j0053d}
An additional discovery via the J0053 protocluster (and implicitly by studying the environment of J0053a) is the very red, very massive galaxy that is J0053d, modelled by BEAGLE as a $M_*\sim10^{12}\,{\rm M}_\odot$ ($M_*\sim4\times10^{11}\,{\rm M}_\odot$ lower limit from mass-to-light ratios in P\'egase), potentially very dusty galaxy.

There is mounting evidence from early \textit{JWST} studies that red, optically-dark systems are predominantly dust-obscured star-forming galaxies. With 138 NIRCam-selected red sources in the Cosmic Evolution Early Release Science \citep[CEERS;][]{CEERS_1_early_galaxy_formation_JWST_Finkelstein_2023} survey, \citet[][]{Natures_of_HST-dark_galaxies_Perez-Gonzalez_2023} use the spatially resolved photometry to find that $\sim71\%$ are dusty star-forming galaxies, with only $\sim18\%$ appearing as quiescent or post-starburst galaxies and $\sim11\%$ as young emission-line-dominated systems. \citet[][]{Extremely_red_galaxies_quiescent_or_dusty_Barrufet_2025} yield a similar result ($13_{-6}^{+9}\%$ quiescent galaxies) but more reliably with PRISM spectroscopy containing H$\alpha$ and H$\beta$ emission lines to yield the Balmer decrement. They report on 23 HST-dark galaxies in the CEERS survey that have high dust attenuations ($A_V>2\;$mag), stellar masses $\log(M/M_\odot)\sim10$, and lie squarely on the star-forming main sequence at $2<z<8$. Both the aforementioned photometry and spectroscopy place the optically-dark, red CEERS sources at $2<z<6$ (aside from the smaller fraction of emission-line-dominated starbursts). J0053d's modelled dust attenuation is similar to that of the galaxies in \citet[][]{Extremely_red_galaxies_quiescent_or_dusty_Barrufet_2025}, supporting the more common DSFG scenario. The molecular gas reservoir in J0053d could serve as an indication of the potential obscured SFR given an assumed star formation efficiency and depletion timescale, but these assumptions could vary as much as the estimated SFR by BEAGLE. In the scenario that J0053d is in the minority of post-starburst or passive galaxies, one would need deep NIR spectroscopy as photometry alone is not sufficient to untangle it from the DSFG population, as mentioned by \citet[][]{Extremely_red_galaxies_at_high_redshift_Williams_2024}. Another angle which suggests J0053d is more likely a DSFG is the size of the object, whereby \citet[][]{Extremely_red_galaxies_quiescent_or_dusty_Barrufet_2025} shows that the quiescent galaxies, where one has a confirmed AGN, are far more compact than the other extremely red DSFGs which are extended (J0053d resembling the latter in the HAWK-I imaging). Meanwhile, \citet[][]{LRDs_EROs_Barro_2024} reports that the point-like nature of their CEERS sample is more indicative of AGN-dominated emission despite SED modelling of the photometry suggesting either massive, dusty galaxies (including quiescent templates) or low-mass obscured QSOs.
\citet[][]{Extremely_red_galaxies_quiescent_or_dusty_Barrufet_2025} appears to provide the most reliable metrics so far for assessing the nature of these extremely red objects, with J0053d aligning more with high redshift DSFGs, although the true nature can only be speculated in lieu of deep IR spectroscopy. Whilst some of these \textit{JWST} results pertain to massive DSFGs, J0053d appears to be far more massive than the aforementioned red galaxies, so we will touch upon how rare such an object is relative to the narrow fields probed by \textit{JWST}.

We explore how rare such an object would be in hydrodynamical simulations and utilise Illustris TNG300 \citep[][]{IllustrisTNG_galaxy_groups_clusters_Pilepich_2018} to search for equivalent objects in a 300$^3$ Mpc$^3$ volume, which would correspond to a $\sim135\,\deg^2$ field of view at $z=3.9$. In the $z=4$ bin, the most massive galaxy was $M_*=10^{11.7}\,{\rm M}_\odot$ and resided in a $M_{\rm DM}=10^{13.4}\,{\rm M}_\odot$ dark matter (DM) halo with a co-moving $R_{200}$ size of $890\,$kpc, growing to be $M_{\rm DM}=10^{15.2}\,{\rm M}_\odot$ by $z=0$. If J0053d is truly as massive as the most massive galaxy in TNG300, it must be very rare to stumble upon in narrow fields of view (e.g., with ALMA and {\it JWST}). The chances of finding such systems with the most massive galaxies in the Universe can be enhanced by investigating protocluster environments of HzRGs that can be easily selected in the radio. The indicative DM halo from TNG300 is comparable to the derived DM halo masses for the DRC and SPT2349-56. The DRC's DM halo is estimated to be $M_{\rm DM}\sim10^{13.6}\,{\rm M}_\odot$ \citep[][]{DRC_z_4_proto-cluster_DSFG_Oteo_2018} using the relation between total IR luminosity and halo mass at $z=4$ \citep[][]{BH_galaxy_coevolution_DM_IR_abundance_matching_Aversa_2015}, whilst SPT2349-56's DM halo is estimated to have an upper limit (if not yet virialised) of $M_{\rm DM}\sim10^{13.1}\,{\rm M}_\odot$ \citep[][]{SPT2349-56_galaxy_cluster_core_z_4.3_Miller_2018} using the mass-dispersion relation for galaxy clusters \citep[][]{Virial_scaling_dark_matter_halos_clusters_Evrard_2008}. With these examples, $z\sim4$ SMG protoclusters seem to be progenitors of the rarest, most massive galaxy clusters, with DM halos projected to evolve as massive as that of the Coma cluster by $z=0$.


\subsection{Future Work}
More accurately characterising the known protocluster members and surveying the extent of this overdensity requires additional, deeper, photometry and would benefit from a SED fitting code that can meaningfully incorporate the 100-GHz ALMA data into the energy budget.
Whilst there are examples of HzRGs around cosmic noon that are present in mergers and clusters seen in the optical and NIR, there may be younger analogues of such HzRGs and systems at $z\gtrsim4$ that are too faint in the optical and NIR, but accessible at MIR and millimetre/sub-millimetre wavelengths. Building up the population of HzRGs at $z\gtrsim4$ may require deep follow-up of candidate radio sources between MIR and millimetre wavelengths. The full sample of candidate UHzRGs followed up with ALMA will be presented in a future paper (Hedge et al., in prep) which will also include the other candidates from B22 using photometry from public surveys not yet considered.

\section{Conclusions} \label{sec:conclusions}
We present the discovery of a galaxy over-density identified by clouds of molecular gas detected in CO(4-3) and pinpointed by a powerful radio galaxy at $z=3.879$ with ALMA. Our key findings and interpretations are as follows:

\begin{itemize}
    \item We identify five likely protocluster members and a further two tentative members via CO(4-3) emission, supported by dual detection of \cione{} in both a known powerful radio galaxy and a new sub-millimetre galaxy detected in these ALMA Band 3 observations. This source over-density is $12-20$ times that of the CANDELS fields, with up to eight sources per square arcminute per $0.05$ redshift bin, if both the two tentative and one continuum-only candidate protocluster members are part of the structure.
    
    \item The molecular gas surrounding the radio galaxy demonstrates multiple kinematic components, indicating a recent interaction or merger with another galaxy. This falls in-line with existing literature on radio emission correlated with observed mergers and merger-induced starbursts.
    
    \item We model the hosts of each candidate protocluster member using aperture photometry fed into BEAGLE and recover constraints for members with marginal or better detections in DES $g,\,r,\,i,\,z,\,Y$ and HAWK-I $K_s$ bands (J0053a/d/f). Their masses range from $M_*\sim10^{10.5}-10^{12.1}\,{\rm M}_\odot$, but without deeper observations of the rest-frame UV, their modelled SFRs are not robust.
    
    \item We calculate lower limits to the total molecular gas mass of each CO(4-3) emitter and the total observed cluster molecular gas $M_{\rm gas}>5.5\times10^{10}$.
    
    \item The molecular gas fuelling the radio galaxy is comparable to other notable massive HzRGs rich in CO, with minimum depletion timescales for both the radio galaxy and the protocluster that are similar to these HzRGs and their environments. The approximate $M_{\rm H_2}<M_*$ from current limits is also more similar to these HzRGs, but assuming a less conservative $\alpha_{\rm CO}$ and $r_{4-1}$ would raise the uncertain molecular gas mass estimate to be more similar to literature SMGs presented. 
    
    \item We also identify a rare, optically-dark very massive $M_*\sim10^{12}\,{\rm M}_\odot$ galaxy from a $K_s$-band detection associated with one of the CO(4-3) emitting protocluster members (J0053d).
    
    \item The equivalent dark matter halo containing a galaxy almost the mass of J0053d in Illustris TNG300 is projected to grow from $M_{\rm DM}\sim10^{13}\,{\rm M}_\odot$ to be as massive as the largest halos in the local Universe $M_{\rm DM}\sim10^{15}\,{\rm M}_\odot$, similar to the Coma cluster.
\end{itemize}

It is clear from the wide-bandwidth scans of this one field that molecular gas traced by CO can be both widespread in the environment around powerful radio galaxies (and SMGs) and relatively easy to detect with ALMA, quite often with $\lesssim30$ minutes on-source for a given spectral window \citep[as has been reported repeatedly in the literature;][]{ALMA_quenching_HzRGs_Falkendal_2019,ALMA_NOEMA_molecular_gas_high_z_SFG_Birkin_2021,COS_z6.8_ALMA_Endsley_2023,Dragonfly_2_z_2_ALMA_CO_6-5_merger_ULIRG_HzRG_Emonts_2015,ALMA_HzRG_CO_survey_Emonts_2023}. Future work will provide stronger constraints on the hosts of these molecular gas clouds, e.g. properties such as dust temperature, SFR and stellar mass. Higher frequency ALMA observations (Bands 8 and 9) to probe the peak rest-frame FIR continuum would place the best constraints on dust temperature and inferred SFR of the host. With modern, advanced SED fitting codes, improved constraints on host galaxy SFR and stellar mass would be achievable if continuum detections of protocluster members (both current and yet-to-be-discovered) existed. {\it JWST} or {\it HST} observations would be the most efficient for obtaining detections in the rest-optical or rest-UV continuum respectively, given the non-detections in our deep ground-based imaging with HAWK-I.

\section{Acknowledgements} \label{sec:acknowledgements}
We acknowledge the Noongar people as the traditional owners and custodians of Whadjuk Boodjar, the land on which the majority of this work was completed.

AH acknowledges Tim J. Galvin for the technical knowledge bestowed prior to and during this project.

The authors would like to thank the referee for their positive and constructive comments and suggestions that ultimately strengthened this work.

This paper makes use of the following ALMA data: ADS/JAO.ALMA\#2022.1.01657.S. ALMA is a partnership of ESO (representing its member states), NSF (USA) and NINS (Japan), together with NRC (Canada), MOST and ASIAA (Taiwan), and KASI (Republic of Korea), in cooperation with the Republic of Chile. The Joint ALMA Observatory is operated by ESO, AUI/NRAO and NAOJ.
This paper is based on observations collected at the European Southern Observatory under ESO programme 108.22HY.001.

The Australia Telescope Compact Array is part of the Australia Telescope National Facility (\url{https://ror.org/05qajvd42}) which is funded by the Australian Government for operation as a National Facility managed by CSIRO. We acknowledge the Gomeroi people as the Traditional Owners of the Observatory site.
This research is supported by an Australian Government Research Training Program (RTP) Scholarship.
AH thanks the Astronomical Society of Australia for the support of a Student Travel Grant that assisted in funding conference travel to present portions of this work.

The work of DS was carried out at the Jet Propulsion Laboratory, California Institute of Technology, under a contract with the National Aeronautics and Space Administration (80NM0018D0004).

JA acknowledges financial support from the Science and Technology Foundation (FCT, Portugal) through research grants UIDB/04434/2020 (DOI: 10.54499/UIDB/04434/2020) and UIDP/04434/2020 (DOI: 10.54499/UIDP/04434/2020).

This project used public archival data from the Dark Energy Survey (DES). Funding for the DES Projects has been provided by the U.S. Department of Energy, the U.S. National Science Foundation, the Ministry of Science and Education of Spain, the Science and Technology Facilities Council of the United Kingdom, the Higher Education Funding Council for England, the National Center for Supercomputing Applications at the University of Illinois at Urbana-Champaign, the Kavli Institute of Cosmological Physics at the University of Chicago, the Center for Cosmology and Astro-Particle Physics at the Ohio State University, the Mitchell Institute for Fundamental Physics and Astronomy at Texas A\&M University, Financiadora de Estudos e Projetos, Funda{\c c}{\~a}o Carlos Chagas Filho de Amparo {\`a} Pesquisa do Estado do Rio de Janeiro, Conselho Nacional de Desenvolvimento Cient{\'i}fico e Tecnol{\'o}gico and the Minist{\'e}rio da Ci{\^e}ncia, Tecnologia e Inova{\c c}{\~a}o, the Deutsche Forschungsgemeinschaft, and the Collaborating Institutions in the Dark Energy Survey.
The Collaborating Institutions are Argonne National Laboratory, the University of California at Santa Cruz, the University of Cambridge, Centro de Investigaciones Energ{\'e}ticas, Medioambientales y Tecnol{\'o}gicas-Madrid, the University of Chicago, University College London, the DES-Brazil Consortium, the University of Edinburgh, the Eidgen{\"o}ssische Technische Hochschule (ETH) Z{\"u}rich,  Fermi National Accelerator Laboratory, the University of Illinois at Urbana-Champaign, the Institut de Ci{\`e}ncies de l'Espai (IEEC/CSIC), the Institut de F{\'i}sica d'Altes Energies, Lawrence Berkeley National Laboratory, the Ludwig-Maximilians Universit{\"a}t M{\"u}nchen and the associated Excellence Cluster Universe, the University of Michigan, the National Optical Astronomy Observatory, the University of Nottingham, The Ohio State University, the OzDES Membership Consortium, the University of Pennsylvania, the University of Portsmouth, SLAC National Accelerator Laboratory, Stanford University, the University of Sussex, and Texas A\&M University.
Based in part on observations at Cerro Tololo Inter-American Observatory, National Optical Astronomy Observatory, which is operated by the Association of Universities for Research in Astronomy (AURA) under a cooperative agreement with the National Science Foundation.

This paper makes use of services or code that have been provided by AAO Data Central  (datacentral.org.au).
This research has made use of NASA’s Astrophysics Data System Bibliographic Services.
This research made use of {\sc astropy}:\footnote{\url{http://www.astropy.org}} a community-developed core Python package and an ecosystem of tools and resources for astronomy \citep{astropy_1_2013, astropy_2_2018, astropy_3_2022}.
This research made use of {\sc photutils}, an {\sc astropy} package for detection and photometry of astronomical sources \citep[][]{photutils_1.6.0_Bradley_2022}.
This research also made use of {\sc numpy} \citep[][]{Numpy_Harris_2020}, {\sc matplotlib} \citep[][]{Matplotlib_Hunter_2007}, {\sc scipy} \citep[][]{SciPy_Virtanen_2020}, {\sc lmfit} \citep[][]{lmfit_1.2.2_Newville_2023} and {\sc topcat} \citep[][]{TOPCAT_Taylor_2005}.

\bibliography{references}{}

\begin{thebibliography}{}
\expandafter\ifx\csname natexlab\endcsname\relax\def\natexlab#1{#1}\fi
\providecommand{\url}[1]{\href{#1}{#1}}
\providecommand{\dodoi}[1]{doi:~\href{http://doi.org/#1}{\nolinkurl{#1}}}
\providecommand{\doeprint}[1]{\href{http://ascl.net/#1}{\nolinkurl{http://ascl.net/#1}}}
\providecommand{\doarXiv}[1]{\href{https://arxiv.org/abs/#1}{\nolinkurl{https://arxiv.org/abs/#1}}}

\bibitem[{{Abbott} {et~al.}(2021){Abbott}, {Adam{\'o}w}, {Aguena}, {Allam},
  {Amon}, {Annis}, {Avila}, {Bacon}, {Banerji}, {Bechtol}, {Becker},
  {Bernstein}, {Bertin}, {Bhargava}, {Bridle}, {Brooks}, {Burke}, {Carnero
  Rosell}, {Carrasco Kind}, {Carretero}, {Castander}, {Cawthon}, {Chang},
  {Choi}, {Conselice}, {Costanzi}, {Crocce}, {da Costa}, {Davis}, {De Vicente},
  {DeRose}, {Desai}, {Diehl}, {Dietrich}, {Drlica-Wagner}, {Eckert},
  {Elvin-Poole}, {Everett}, {Evrard}, {Ferrero}, {Fert{\'e}}, {Flaugher},
  {Fosalba}, {Friedel}, {Frieman}, {Garc{\'\i}a-Bellido}, {Gaztanaga},
  {Gelman}, {Gerdes}, {Giannantonio}, {Gill}, {Gruen}, {Gruendl}, {Gschwend},
  {Gutierrez}, {Hartley}, {Hinton}, {Hollowood}, {Honscheid}, {Huterer},
  {James}, {Jeltema}, {Johnson}, {Kent}, {Kron}, {Kuehn}, {Kuropatkin},
  {Lahav}, {Li}, {Lidman}, {Lin}, {MacCrann}, {Maia}, {Manning}, {Maloney},
  {March}, {Marshall}, {Martini}, {Melchior}, {Menanteau}, {Miquel}, {Morgan},
  {Myles}, {Neilsen}, {Ogando}, {Palmese}, {Paz-Chinch{\'o}n}, {Petravick},
  {Pieres}, {Plazas}, {Pond}, {Rodriguez-Monroy}, {Romer}, {Roodman}, {Rykoff},
  {Sako}, {Sanchez}, {Santiago}, {Scarpine}, {Serrano}, {Sevilla-Noarbe},
  {Smith}, {Smith}, {Soares-Santos}, {Suchyta}, {Swanson}, {Tarle}, {Thomas},
  {To}, {Tremblay}, {Troxel}, {Tucker}, {Turner}, {Varga}, {Walker},
  {Wechsler}, {Weller}, {Wester}, {Wilkinson}, {Yanny}, {Zhang}, {Nikutta},
  {Fitzpatrick}, {Jacques}, {Scott}, {Olsen}, {Huang}, {Herrera}, {Juneau},
  {Nidever}, {Weaver}, {Adean}, {Correia}, {de Freitas}, {Freitas},
  {Singulani}, {Vila-Verde}, \& {Linea Science Server}}]{DES_DR2_Abbott_2021}
{Abbott}, T.~M.~C., {Adam{\'o}w}, M., {Aguena}, M., {et~al.} 2021, \apjs, 255,
  20, \dodoi{10.3847/1538-4365/ac00b3}

\bibitem[{{Aravena} {et~al.}(2016){Aravena}, {Decarli}, {Walter}, {Bouwens},
  {Oesch}, {Carilli}, {Bauer}, {Da Cunha}, {Daddi}, {G{\'o}nzalez-L{\'o}pez},
  {Ivison}, {Riechers}, {Smail}, {Swinbank}, {Weiss}, {Anguita}, {Bacon},
  {Bell}, {Bertoldi}, {Cortes}, {Cox}, {Hodge}, {Ibar}, {Inami}, {Infante},
  {Karim}, {Magnelli}, {Ota}, {Popping}, {van der Werf}, {Wagg}, \&
  {Fudamoto}}]{ALMA_CII_spectroscopic_survey_Hubble_UDF_Aravena_2016}
{Aravena}, M., {Decarli}, R., {Walter}, F., {et~al.} 2016, \apj, 833, 71,
  \dodoi{10.3847/1538-4357/833/1/71}

\bibitem[{{Astropy Collaboration} {et~al.}(2013){Astropy Collaboration},
  {Robitaille}, {Tollerud}, {Greenfield}, {Droettboom}, {Bray}, {Aldcroft},
  {Davis}, {Ginsburg}, {Price-Whelan}, {Kerzendorf}, {Conley}, {Crighton},
  {Barbary}, {Muna}, {Ferguson}, {Grollier}, {Parikh}, {Nair}, {Unther},
  {Deil}, {Woillez}, {Conseil}, {Kramer}, {Turner}, {Singer}, {Fox}, {Weaver},
  {Zabalza}, {Edwards}, {Azalee Bostroem}, {Burke}, {Casey}, {Crawford},
  {Dencheva}, {Ely}, {Jenness}, {Labrie}, {Lim}, {Pierfederici}, {Pontzen},
  {Ptak}, {Refsdal}, {Servillat}, \& {Streicher}}]{astropy_1_2013}
{Astropy Collaboration}, {Robitaille}, T.~P., {Tollerud}, E.~J., {et~al.} 2013,
  \aap, 558, A33, \dodoi{10.1051/0004-6361/201322068}

\bibitem[{{Astropy Collaboration} {et~al.}(2018){Astropy Collaboration},
  {Price-Whelan}, {Sip{\H{o}}cz}, {G{\"u}nther}, {Lim}, {Crawford}, {Conseil},
  {Shupe}, {Craig}, {Dencheva}, {Ginsburg}, {Vand erPlas}, {Bradley},
  {P{\'e}rez-Su{\'a}rez}, {de Val-Borro}, {Aldcroft}, {Cruz}, {Robitaille},
  {Tollerud}, {Ardelean}, {Babej}, {Bach}, {Bachetti}, {Bakanov}, {Bamford},
  {Barentsen}, {Barmby}, {Baumbach}, {Berry}, {Biscani}, {Boquien}, {Bostroem},
  {Bouma}, {Brammer}, {Bray}, {Breytenbach}, {Buddelmeijer}, {Burke},
  {Calderone}, {Cano Rodr{\'\i}guez}, {Cara}, {Cardoso}, {Cheedella}, {Copin},
  {Corrales}, {Crichton}, {D'Avella}, {Deil}, {Depagne}, {Dietrich}, {Donath},
  {Droettboom}, {Earl}, {Erben}, {Fabbro}, {Ferreira}, {Finethy}, {Fox},
  {Garrison}, {Gibbons}, {Goldstein}, {Gommers}, {Greco}, {Greenfield},
  {Groener}, {Grollier}, {Hagen}, {Hirst}, {Homeier}, {Horton}, {Hosseinzadeh},
  {Hu}, {Hunkeler}, {Ivezi{\'c}}, {Jain}, {Jenness}, {Kanarek}, {Kendrew},
  {Kern}, {Kerzendorf}, {Khvalko}, {King}, {Kirkby}, {Kulkarni}, {Kumar},
  {Lee}, {Lenz}, {Littlefair}, {Ma}, {Macleod}, {Mastropietro}, {McCully},
  {Montagnac}, {Morris}, {Mueller}, {Mumford}, {Muna}, {Murphy}, {Nelson},
  {Nguyen}, {Ninan}, {N{\"o}the}, {Ogaz}, {Oh}, {Parejko}, {Parley}, {Pascual},
  {Patil}, {Patil}, {Plunkett}, {Prochaska}, {Rastogi}, {Reddy Janga},
  {Sabater}, {Sakurikar}, {Seifert}, {Sherbert}, {Sherwood-Taylor}, {Shih},
  {Sick}, {Silbiger}, {Singanamalla}, {Singer}, {Sladen}, {Sooley},
  {Sornarajah}, {Streicher}, {Teuben}, {Thomas}, {Tremblay}, {Turner},
  {Terr{\'o}n}, {van Kerkwijk}, {de la Vega}, {Watkins}, {Weaver}, {Whitmore},
  {Woillez}, {Zabalza}, \& {Astropy Contributors}}]{astropy_2_2018}
{Astropy Collaboration}, {Price-Whelan}, A.~M., {Sip{\H{o}}cz}, B.~M., {et~al.}
  2018, \aj, 156, 123, \dodoi{10.3847/1538-3881/aabc4f}

\bibitem[{{Astropy Collaboration} {et~al.}(2022){Astropy Collaboration},
  {Price-Whelan}, {Lim}, {Earl}, {Starkman}, {Bradley}, {Shupe}, {Patil},
  {Corrales}, {Brasseur}, {N{"o}the}, {Donath}, {Tollerud}, {Morris},
  {Ginsburg}, {Vaher}, {Weaver}, {Tocknell}, {Jamieson}, {van Kerkwijk},
  {Robitaille}, {Merry}, {Bachetti}, {G{"u}nther}, {Aldcroft},
  {Alvarado-Montes}, {Archibald}, {B{'o}di}, {Bapat}, {Barentsen}, {Baz{'a}n},
  {Biswas}, {Boquien}, {Burke}, {Cara}, {Cara}, {Conroy}, {Conseil}, {Craig},
  {Cross}, {Cruz}, {D'Eugenio}, {Dencheva}, {Devillepoix}, {Dietrich},
  {Eigenbrot}, {Erben}, {Ferreira}, {Foreman-Mackey}, {Fox}, {Freij}, {Garg},
  {Geda}, {Glattly}, {Gondhalekar}, {Gordon}, {Grant}, {Greenfield}, {Groener},
  {Guest}, {Gurovich}, {Handberg}, {Hart}, {Hatfield-Dodds}, {Homeier},
  {Hosseinzadeh}, {Jenness}, {Jones}, {Joseph}, {Kalmbach}, {Karamehmetoglu},
  {Ka{l}uszy{'n}ski}, {Kelley}, {Kern}, {Kerzendorf}, {Koch}, {Kulumani},
  {Lee}, {Ly}, {Ma}, {MacBride}, {Maljaars}, {Muna}, {Murphy}, {Norman},
  {O'Steen}, {Oman}, {Pacifici}, {Pascual}, {Pascual-Granado}, {Patil},
  {Perren}, {Pickering}, {Rastogi}, {Roulston}, {Ryan}, {Rykoff}, {Sabater},
  {Sakurikar}, {Salgado}, {Sanghi}, {Saunders}, {Savchenko}, {Schwardt},
  {Seifert-Eckert}, {Shih}, {Jain}, {Shukla}, {Sick}, {Simpson},
  {Singanamalla}, {Singer}, {Singhal}, {Sinha}, {Sip{H{o}}cz}, {Spitler},
  {Stansby}, {Streicher}, {{{S}}umak}, {Swinbank}, {Taranu}, {Tewary},
  {Tremblay}, {Val-Borro}, {Van Kooten}, {Vasovi{'c}}, {Verma}, {de Miranda
  Cardoso}, {Williams}, {Wilson}, {Winkel}, {Wood-Vasey}, {Xue}, {Yoachim},
  {Zhang}, {Zonca}, \& {Astropy Project Contributors}}]{astropy_3_2022}
{Astropy Collaboration}, {Price-Whelan}, A.~M., {Lim}, P.~L., {et~al.} 2022,
  \apj, 935, 167, \dodoi{10.3847/1538-4357/ac7c74}

\bibitem[{{Aversa} {et~al.}(2015){Aversa}, {Lapi}, {de Zotti}, {Shankar}, \&
  {Danese}}]{BH_galaxy_coevolution_DM_IR_abundance_matching_Aversa_2015}
{Aversa}, R., {Lapi}, A., {de Zotti}, G., {Shankar}, F., \& {Danese}, L. 2015,
  \apj, 810, 74, \dodoi{10.1088/0004-637X/810/1/74}

\bibitem[{{Ba{\~n}ados} {et~al.}(2021){Ba{\~n}ados}, {Mazzucchelli}, {Momjian},
  {Eilers}, {Wang}, {Schindler}, {Connor}, {Andika}, {Barth}, {Carilli},
  {Davies}, {Decarli}, {Fan}, {Farina}, {Hennawi}, {Pensabene}, {Stern},
  {Venemans}, {Wenzl}, \& {Yang}}]{RL_QSO_z_6.82_Banados_2021}
{Ba{\~n}ados}, E., {Mazzucchelli}, C., {Momjian}, E., {et~al.} 2021, \apj, 909,
  80, \dodoi{10.3847/1538-4357/abe239}

\bibitem[{{Ba{\~n}ados} {et~al.}(2023){Ba{\~n}ados}, {Schindler}, {Venemans},
  {Connor}, {Decarli}, {Farina}, {Mazzucchelli}, {Meyer}, {Stern}, {Walter},
  {Fan}, {Hennawi}, {Khusanova}, {Morrell}, {Nanni}, {Noirot}, {Pensabene},
  {Rix}, {Simon}, {Verdoes Kleijn}, {Xie}, {Yang}, \&
  {Connor}}]{Pan-STARRS1_z_5.6_quasar_survey_Banados_2023}
{Ba{\~n}ados}, E., {Schindler}, J.-T., {Venemans}, B.~P., {et~al.} 2023, \apjs,
  265, 29, \dodoi{10.3847/1538-4365/acb3c7}

\bibitem[{{Ba{\~n}ados} {et~al.}(2024){Ba{\~n}ados}, {Khusanova}, {Decarli},
  {Momjian}, {Walter}, {Connor}, {Carilli}, {Mazzucchelli}, {Rojas-Ruiz}, \&
  {Venemans}}]{Blazar_z_7_Banados_2024}
{Ba{\~n}ados}, E., {Khusanova}, Y., {Decarli}, R., {et~al.} 2024, \apjl, 977,
  L46, \dodoi{10.3847/2041-8213/ad823b}

\bibitem[{{Barro} {et~al.}(2024){Barro}, {P{\'e}rez-Gonz{\'a}lez}, {Kocevski},
  {McGrath}, {Trump}, {Simons}, {Somerville}, {Yung}, {Arrabal Haro}, {Akins},
  {Bagley}, {Cleri}, {Costantin}, {Davis}, {Dickinson}, {Finkelstein},
  {Giavalisco}, {G{\'o}mez-Guijarro}, {Hathi}, {Hirschmann}, {Holwerda},
  {Huertas-Company}, {Kartaltepe}, {Koekemoer}, {Lucas}, {Papovich}, {Pirzkal},
  {Seill{\'e}}, {Tacchella}, {Wuyts}, {Wilkins}, {de la Vega}, {Yang}, \&
  {Zavala}}]{LRDs_EROs_Barro_2024}
{Barro}, G., {P{\'e}rez-Gonz{\'a}lez}, P.~G., {Kocevski}, D.~D., {et~al.} 2024,
  \apj, 963, 128, \dodoi{10.3847/1538-4357/ad167e}

\bibitem[{{Barrufet} {et~al.}(2025){Barrufet}, {Oesch}, {Marques-Chaves},
  {Arellano-Cordova}, {Baggen}, {Carnall}, {Cullen}, {Dunlop}, {Gottumukkala},
  {Fudamoto}, {Illingworth}, {Magee}, {McLure}, {McLeod}, {Micha{\l}owski},
  {Stefanon}, {van Dokkum}, \&
  {Weibel}}]{Extremely_red_galaxies_quiescent_or_dusty_Barrufet_2025}
{Barrufet}, L., {Oesch}, P.~A., {Marques-Chaves}, R., {et~al.} 2025, \mnras,
  537, 3453, \dodoi{10.1093/mnras/staf013}

\bibitem[{{Belladitta} {et~al.}(2023){Belladitta}, {Moretti}, {Caccianiga},
  {Dallacasa}, {Spingola}, {Pedani}, {Cassar{\`a}}, \&
  {Bisogni}}]{RL_QSO_z_5.3_Belladitta_2023}
{Belladitta}, S., {Moretti}, A., {Caccianiga}, A., {et~al.} 2023, \aap, 669,
  A134, \dodoi{10.1051/0004-6361/202243855}

\bibitem[{{Birkin} {et~al.}(2021){Birkin}, {Weiss}, {Wardlow}, {Smail},
  {Swinbank}, {Dudzevi{\v{c}}i{\={u}}t{\.{e}}}, {An}, {Ao}, {Chapman}, {Chen},
  {da Cunha}, {Dannerbauer}, {Gullberg}, {Hodge}, {Ikarashi}, {Ivison},
  {Matsuda}, {Stach}, {Walter}, {Wang}, \& {van der
  Werf}}]{ALMA_NOEMA_molecular_gas_high_z_SFG_Birkin_2021}
{Birkin}, J.~E., {Weiss}, A., {Wardlow}, J.~L., {et~al.} 2021, \mnras, 501,
  3926, \dodoi{10.1093/mnras/staa3862}

\bibitem[{{Bogd{\'a}n} {et~al.}(2024){Bogd{\'a}n}, {Goulding}, {Natarajan},
  {Kov{\'a}cs}, {Tremblay}, {Chadayammuri}, {Volonteri}, {Kraft}, {Forman},
  {Jones}, {Churazov}, \&
  {Zhuravleva}}]{Heavy_seed_SMBH_z_10_x-ray_quasar_Bogdan_2024}
{Bogd{\'a}n}, {\'A}., {Goulding}, A.~D., {Natarajan}, P., {et~al.} 2024, Nature
  Astronomy, 8, 126, \dodoi{10.1038/s41550-023-02111-9}

\bibitem[{{Boquien} {et~al.}(2019){Boquien}, {Burgarella}, {Roehlly}, {Buat},
  {Ciesla}, {Corre}, {Inoue}, \& {Salas}}]{CIGALE_3_python_port_Boquien_2019}
{Boquien}, M., {Burgarella}, D., {Roehlly}, Y., {et~al.} 2019, \aap, 622, A103,
  \dodoi{10.1051/0004-6361/201834156}

\bibitem[{Bradley {et~al.}(2022)Bradley, Sipőcz, Robitaille, Tollerud,
  Vinícius, Deil, Barbary, Wilson, Busko, Donath, Günther, Cara, Lim,
  Meßlinger, Conseil, Bostroem, Droettboom, Bray, Bratholm, Barentsen, Craig,
  Ginsburg, Rathi, Pascual, Perren, Georgiev, de~Val-Borro, Kerzendorf, Bach,
  \& Quint}]{photutils_1.6.0_Bradley_2022}
Bradley, L., Sipőcz, B., Robitaille, T., {et~al.} 2022, astropy/photutils:
  1.6.0, 1.6.0,  Zenodo, \dodoi{10.5281/zenodo.7419741}

\bibitem[{{Bressan} {et~al.}(2012){Bressan}, {Marigo}, {Girardi}, {Salasnich},
  {Dal Cero}, {Rubele}, \&
  {Nanni}}]{PARSEC_stellar_evolutionary_track_Bressan_2012}
{Bressan}, A., {Marigo}, P., {Girardi}, L., {et~al.} 2012, \mnras, 427, 127,
  \dodoi{10.1111/j.1365-2966.2012.21948.x}

\bibitem[{{Briggs}(1995)}]{Briggs_deconvolution_scheme_resolved_sources_Briggs_1995}
{Briggs}, D.~S. 1995, PhD thesis, New Mexico Institute of Mining and Technology

\bibitem[{Broderick {et~al.}(2022)Broderick, Drouart, Seymour, Galvin, Wright,
  Carnero~Rosell, Chhetri, Dannerbauer, Driver, Morgan, \&
  et~al.}]{GLEAMing_HzRGs_2_Broderick_2022}
Broderick, J.~W., Drouart, G., Seymour, N., {et~al.} 2022, \pasa, 39, e061,
  \dodoi{10.1017/pasa.2022.42}

\bibitem[{{Broderick} {et~al.}(2024){Broderick}, {Seymour}, {Drouart},
  {Knight}, {Afonso}, {De Breuck}, {Galvin}, {Hedge}, {Lehnert}, {Noirot},
  {Shabala}, {Turner}, \& {Vernet}}]{GLEAMing_HzRGs_3_Broderick_2024}
{Broderick}, J.~W., {Seymour}, N., {Drouart}, G., {et~al.} 2024, \pasa, 41,
  e071, \dodoi{10.1017/pasa.2024.55}

\bibitem[{{Bruzual} \&
  {Charlot}(2003)}]{stellar_population_synthesis_BC03_Bruzual_2003}
{Bruzual}, G., \& {Charlot}, S. 2003, \mnras, 344, 1000,
  \dodoi{10.1046/j.1365-8711.2003.06897.x}

\bibitem[{{Burgarella} {et~al.}(2005){Burgarella}, {Buat}, \&
  {Iglesias-P{\'a}ramo}}]{CIGALE_1_original_Burgarella_2005}
{Burgarella}, D., {Buat}, V., \& {Iglesias-P{\'a}ramo}, J. 2005, \mnras, 360,
  1413, \dodoi{10.1111/j.1365-2966.2005.09131.x}

\bibitem[{{Calzetti} {et~al.}(2000){Calzetti}, {Armus}, {Bohlin}, {Kinney},
  {Koornneef}, \& {Storchi-Bergmann}}]{Dust_attenuation_model_Calzetti_2000}
{Calzetti}, D., {Armus}, L., {Bohlin}, R.~C., {et~al.} 2000, \apj, 533, 682,
  \dodoi{10.1086/308692}

\bibitem[{{Cameron} {et~al.}(2024){Cameron}, {Katz}, {Witten}, {Saxena},
  {Laporte}, \&
  {Bunker}}]{Nebular_dominated_galaxies_hot_stars_top-heavy_IMF_Cameron_2024}
{Cameron}, A.~J., {Katz}, H., {Witten}, C., {et~al.} 2024, \mnras, 534, 523,
  \dodoi{10.1093/mnras/stae1547}

\bibitem[{{Casey} {et~al.}(2014){Casey}, {Narayanan}, \&
  {Cooray}}]{High-z_DSFG_Casey_2014}
{Casey}, C.~M., {Narayanan}, D., \& {Cooray}, A. 2014, \physrep, 541, 45,
  \dodoi{10.1016/j.physrep.2014.02.009}

\bibitem[{{Casoli} {et~al.}(1998){Casoli}, {Sauty}, {Gerin}, {Boselli},
  {Fouque}, {Braine}, {Gavazzi}, {Lequeux}, \&
  {Dickey}}]{Molecular_gas_in_spiral_galaxies_Casoli_1998}
{Casoli}, F., {Sauty}, S., {Gerin}, M., {et~al.} 1998, \aap, 331, 451

\bibitem[{{Chabrier}(2003)}]{IMF_Chabrier_2003}
{Chabrier}, G. 2003, \pasp, 115, 763, \dodoi{10.1086/376392}

\bibitem[{{Chapman} {et~al.}(2024){Chapman}, {Hill}, {Aravena}, {Archipley},
  {Babul}, {Burgoyne}, {Canning}, {Deane}, {De Breuck}, {Gonzalez}, {Hayward},
  {Kim}, {Malkan}, {Marrone}, {McIntyre}, {Murphy}, {Pass}, {Perry}, {Phadke},
  {Rennehan}, {Reuter}, {Rotermund}, {Scott}, {Seymour}, {Solimano}, {Spilker},
  {Stark}, {Sulzenauer}, {Tothill}, {Vieira}, {Vizgan}, {Wang}, \&
  {Weiss}}]{SPT2349-56_z_4.3_radio_loud_AGN_Chapman_2024}
{Chapman}, S.~C., {Hill}, R., {Aravena}, M., {et~al.} 2024, \apj, 961, 120,
  \dodoi{10.3847/1538-4357/ad0b77}

\bibitem[{{Chevallard} \& {Charlot}(2016)}]{Beagle_Chevallard_2016}
{Chevallard}, J., \& {Charlot}, S. 2016, \mnras, 462, 1415,
  \dodoi{10.1093/mnras/stw1756}

\bibitem[{{da Cunha} {et~al.}(2015){da Cunha}, {Walter}, {Smail}, {Swinbank},
  {Simpson}, {Decarli}, {Hodge}, {Weiss}, {van der Werf}, {Bertoldi},
  {Chapman}, {Cox}, {Danielson}, {Dannerbauer}, {Greve}, {Ivison}, {Karim}, \&
  {Thomson}}]{ALESS_SMG_UV_to_radio_modelling_da_Cunha_2015}
{da Cunha}, E., {Walter}, F., {Smail}, I.~R., {et~al.} 2015, \apj, 806, 110,
  \dodoi{10.1088/0004-637X/806/1/110}

\bibitem[{{Dannerbauer} {et~al.}(2014){Dannerbauer}, {Kurk}, {De Breuck},
  {Wylezalek}, {Santos}, {Koyama}, {Seymour}, {Tanaka}, {Hatch}, {Altieri},
  {Coia}, {Galametz}, {Kodama}, {Miley}, {R{\"o}ttgering}, {Sanchez-Portal},
  {Valtchanov}, {Venemans}, \&
  {Ziegler}}]{Spiderweb_galaxy_excess_dusty_starbursts_Dannerbauer_2014}
{Dannerbauer}, H., {Kurk}, J.~D., {De Breuck}, C., {et~al.} 2014, \aap, 570,
  A55, \dodoi{10.1051/0004-6361/201423771}

\bibitem[{{De Breuck} {et~al.}(2010){De Breuck}, Seymour, Stern, Willner,
  Eisenhardt, Fazio, Galametz, Lacy, Rettura, Rocca-Volmerange, \&
  Vernet}]{Spitzer_HzRG_survey_DeBreuck_2010}
{De Breuck}, C., Seymour, N., Stern, D., {et~al.} 2010, \apj, 725, 36,
  \dodoi{10.1088/0004-637X/725/1/36}

\bibitem[{{Downes} \&
  {Solomon}(1998)}]{CO_H2_molecular_gas_mass_conversion_nuclear_rings_starbursts_ULIRGs_Downes_1998}
{Downes}, D., \& {Solomon}, P.~M. 1998, \apj, 507, 615, \dodoi{10.1086/306339}

\bibitem[{{Drouart} {et~al.}(2016){Drouart}, {Rocca-Volmerange}, {De Breuck},
  {Fioc}, {Lehnert}, {Seymour}, {Stern}, \&
  {Vernet}}]{Disentangling_SFR_AGN_in_HzRG_ULIRGS_z_1-4_Drouart_2016}
{Drouart}, G., {Rocca-Volmerange}, B., {De Breuck}, C., {et~al.} 2016, \aap,
  593, A109, \dodoi{10.1051/0004-6361/201526880}

\bibitem[{{Drouart} {et~al.}(2020){Drouart}, {Seymour}, {Galvin}, {Afonso},
  {Callingham}, {De Breuck}, {Johnston-Hollitt}, {Kapi{\'n}ska}, {Lehnert}, \&
  {Vernet}}]{GLEAMing_HzRGs_1_Drouart_2020}
{Drouart}, G., {Seymour}, N., {Galvin}, T.~J., {et~al.} 2020, \pasa, 37, e026,
  \dodoi{10.1017/pasa.2020.6}

\bibitem[{{Emonts} {et~al.}(2014){Emonts}, {Piqueras-L{\'o}pez}, {Colina},
  {Arribas}, {Villar-Mart{\'\i}n}, {Pereira-Santaella}, {Garcia-Burillo}, \&
  {Alonso-Herrero}}]{Hot_cold_molecular_gas_outflow_NGC_3256_secondary_Emonts_2014}
{Emonts}, B.~H.~C., {Piqueras-L{\'o}pez}, J., {Colina}, L., {et~al.} 2014,
  \aap, 572, A40, \dodoi{10.1051/0004-6361/201423805}

\bibitem[{{Emonts} {et~al.}(2013){Emonts}, {Feain}, {R{\"o}ttgering}, {Miley},
  {Seymour}, {Norris}, {Carilli}, {Villar-Mart{\'\i}n}, {Mao}, {Sadler},
  {Ekers}, {van Moorsel}, {Ivison}, {Pentericci}, {Tadhunter}, \&
  {Saikia}}]{Spiderweb_galaxy_CO_Emonts_2013}
{Emonts}, B.~H.~C., {Feain}, I., {R{\"o}ttgering}, H.~J.~A., {et~al.} 2013,
  \mnras, 430, 3465, \dodoi{10.1093/mnras/stt147}

\bibitem[{{Emonts} {et~al.}(2015{\natexlab{a}}){Emonts}, {De Breuck},
  {Lehnert}, {Vernet}, {Gullberg}, {Villar-Mart{\'\i}n}, {Nesvadba}, {Drouart},
  {Ivison}, {Seymour}, {Wylezalek}, \&
  {Barthel}}]{Dragonfly_2_z_2_ALMA_CO_6-5_merger_ULIRG_HzRG_Emonts_2015}
{Emonts}, B.~H.~C., {De Breuck}, C., {Lehnert}, M.~D., {et~al.}
  2015{\natexlab{a}}, \aap, 584, A99, \dodoi{10.1051/0004-6361/201526090}

\bibitem[{{Emonts} {et~al.}(2015{\natexlab{b}}){Emonts}, {Mao}, {Stroe},
  {Pentericci}, {Villar-Mart{\'\i}n}, {Norris}, {Miley}, {De Breuck}, {van
  Moorsel}, {Lehnert}, {Carilli}, {R{\"o}ttgering}, {Seymour}, {Sadler},
  {Ekers}, {Drouart}, {Feain}, {Colina}, {Stevens}, \&
  {Holt}}]{Dragonfly_1_z_2_ATCA_CO_1-0_merger_ULIRG_HzRG_Emonts_2015}
{Emonts}, B.~H.~C., {Mao}, M.~Y., {Stroe}, A., {et~al.} 2015{\natexlab{b}},
  \mnras, 451, 1025, \dodoi{10.1093/mnras/stv930}

\bibitem[{{Emonts} {et~al.}(2016){Emonts}, {Lehnert}, {Villar-Mart{\'\i}n},
  {Norris}, {Ekers}, {van Moorsel}, {Dannerbauer}, {Pentericci}, {Miley},
  {Allison}, {Sadler}, {Guillard}, {Carilli}, {Mao}, {R{\"o}ttgering}, {De
  Breuck}, {Seymour}, {Gullberg}, {Ceverino}, {Jagannathan}, {Vernet}, \&
  {Indermuehle}}]{Molecular_gas_fuels_Spiderweb_galaxy_Emonts_2016}
{Emonts}, B.~H.~C., {Lehnert}, M.~D., {Villar-Mart{\'\i}n}, M., {et~al.} 2016,
  Science, 354, 1128, \dodoi{10.1126/science.aag0512}

\bibitem[{{Emonts} {et~al.}(2018){Emonts}, {Lehnert}, {Dannerbauer}, {De
  Breuck}, {Villar-Mart{\'\i}n}, {Miley}, {Allison}, {Gullberg}, {Hatch},
  {Guillard}, {Mao}, \&
  {Norris}}]{Spiderweb_HzRG_recyled_gas_CGM_CI_Emonts_2018}
{Emonts}, B.~H.~C., {Lehnert}, M.~D., {Dannerbauer}, H., {et~al.} 2018, \mnras,
  477, L60, \dodoi{10.1093/mnrasl/sly034}

\bibitem[{{Emonts} {et~al.}(2023){Emonts}, {Lehnert}, {Lebowitz}, {Miley},
  {Villar-Mart{\'\i}n}, {Norris}, {De Breuck}, {Carilli}, \&
  {Feain}}]{ALMA_HzRG_CO_survey_Emonts_2023}
{Emonts}, B. H.~C., {Lehnert}, M.~D., {Lebowitz}, S., {et~al.} 2023, \apj, 952,
  148, \dodoi{10.3847/1538-4357/acde53}

\bibitem[{Endsley {et~al.}(2023)Endsley, Stark, Lyu, Wang, Yang, Fan, Smit,
  Bouwens, Hainline, \& Schouws}]{COS_z6.8_ALMA_Endsley_2023}
Endsley, R., Stark, D.~P., Lyu, J., {et~al.} 2023, \mnras, 520, 4609,
  \dodoi{10.1093/mnras/stad266}

\bibitem[{{European Southern Observatory}(1998)}]{VLT_white_book_ESO_1998}
{European Southern Observatory}. 1998, {The VLT White Book}

\bibitem[{{Evrard} {et~al.}(2008){Evrard}, {Bialek}, {Busha}, {White}, {Habib},
  {Heitmann}, {Warren}, {Rasia}, {Tormen}, {Moscardini}, {Power}, {Jenkins},
  {Gao}, {Frenk}, {Springel}, {White}, \&
  {Diemand}}]{Virial_scaling_dark_matter_halos_clusters_Evrard_2008}
{Evrard}, A.~E., {Bialek}, J., {Busha}, M., {et~al.} 2008, \apj, 672, 122,
  \dodoi{10.1086/521616}

\bibitem[{{Falkendal} {et~al.}(2019){Falkendal}, {De Breuck}, {Lehnert},
  {Drouart}, {Vernet}, {Emonts}, {Lee}, {Nesvadba}, {Seymour}, {B{\'e}thermin},
  {Kolwa}, {Gullberg}, \& {Wylezalek}}]{ALMA_quenching_HzRGs_Falkendal_2019}
{Falkendal}, T., {De Breuck}, C., {Lehnert}, M.~D., {et~al.} 2019, \aap, 621,
  A27, \dodoi{10.1051/0004-6361/201732485}

\bibitem[{{Finkelstein} {et~al.}(2023){Finkelstein}, {Bagley}, {Ferguson},
  {Wilkins}, {Kartaltepe}, {Papovich}, {Yung}, {Arrabal Haro}, {Behroozi},
  {Dickinson}, {Kocevski}, {Koekemoer}, {Larson}, {Le Bail}, {Morales},
  {P{\'e}rez-Gonz{\'a}lez}, {Burgarella}, {Dav{\'e}}, {Hirschmann},
  {Somerville}, {Wuyts}, {Bromm}, {Casey}, {Fontana}, {Fujimoto}, {Gardner},
  {Giavalisco}, {Grazian}, {Grogin}, {Hathi}, {Hutchison}, {Jha}, {Jogee},
  {Kewley}, {Kirkpatrick}, {Long}, {Lotz}, {Pentericci}, {Pierel}, {Pirzkal},
  {Ravindranath}, {Ryan}, {Trump}, {Yang}, {Bhatawdekar}, {Bisigello}, {Buat},
  {Calabr{\`o}}, {Castellano}, {Cleri}, {Cooper}, {Croton}, {Daddi}, {Dekel},
  {Elbaz}, {Franco}, {Gawiser}, {Holwerda}, {Huertas-Company}, {Jaskot},
  {Leung}, {Lucas}, {Mobasher}, {Pandya}, {Tacchella}, {Weiner}, \&
  {Zavala}}]{CEERS_1_early_galaxy_formation_JWST_Finkelstein_2023}
{Finkelstein}, S.~L., {Bagley}, M.~B., {Ferguson}, H.~C., {et~al.} 2023, \apjl,
  946, L13, \dodoi{10.3847/2041-8213/acade4}

\bibitem[{{Fioc} \&
  {Rocca-Volmerange}(2019)}]{Pegase_spectral_evolution_galaxies_with_dust_Fioc_2019}
{Fioc}, M., \& {Rocca-Volmerange}, B. 2019, \aap, 623, A143,
  \dodoi{10.1051/0004-6361/201833556}

\bibitem[{{Flaugher} {et~al.}(2015){Flaugher}, {Diehl}, {Honscheid}, {Abbott},
  {Alvarez}, {Angstadt}, {Annis}, {Antonik}, {Ballester}, {Beaufore},
  {Bernstein}, {Bernstein}, {Bigelow}, {Bonati}, {Boprie}, {Brooks},
  {Buckley-Geer}, {Campa}, {Cardiel-Sas}, {Castander}, {Castilla}, {Cease},
  {Cela-Ruiz}, {Chappa}, {Chi}, {Cooper}, {da Costa}, {Dede}, {Derylo},
  {DePoy}, {de Vicente}, {Doel}, {Drlica-Wagner}, {Eiting}, {Elliott}, {Emes},
  {Estrada}, {Fausti Neto}, {Finley}, {Flores}, {Frieman}, {Gerdes},
  {Gladders}, {Gregory}, {Gutierrez}, {Hao}, {Holland}, {Holm}, {Huffman},
  {Jackson}, {James}, {Jonas}, {Karcher}, {Karliner}, {Kent}, {Kessler},
  {Kozlovsky}, {Kron}, {Kubik}, {Kuehn}, {Kuhlmann}, {Kuk}, {Lahav}, {Lathrop},
  {Lee}, {Levi}, {Lewis}, {Li}, {Mandrichenko}, {Marshall}, {Martinez},
  {Merritt}, {Miquel}, {Mu{\~n}oz}, {Neilsen}, {Nichol}, {Nord}, {Ogando},
  {Olsen}, {Palaio}, {Patton}, {Peoples}, {Plazas}, {Rauch}, {Reil}, {Rheault},
  {Roe}, {Rogers}, {Roodman}, {Sanchez}, {Scarpine}, {Schindler}, {Schmidt},
  {Schmitt}, {Schubnell}, {Schultz}, {Schurter}, {Scott}, {Serrano}, {Shaw},
  {Smith}, {Soares-Santos}, {Stefanik}, {Stuermer}, {Suchyta}, {Sypniewski},
  {Tarle}, {Thaler}, {Tighe}, {Tran}, {Tucker}, {Walker}, {Wang}, {Watson},
  {Weaverdyck}, {Wester}, {Woods}, {Yanny}, \& {DES
  Collaboration}}]{DECam_Flaugher_2015}
{Flaugher}, B., {Diehl}, H.~T., {Honscheid}, K., {et~al.} 2015, \aj, 150, 150,
  \dodoi{10.1088/0004-6256/150/5/150}

\bibitem[{{Frater} {et~al.}(1992){Frater}, {Brooks}, \&
  {Whiteoak}}]{ATCA_overview_Frater_1992}
{Frater}, R.~H., {Brooks}, J.~W., \& {Whiteoak}, J.~B. 1992, Journal of
  Electrical and Electronics Engineering Australia, 12, 103

\bibitem[{{Ginsburg} {et~al.}(2016){Ginsburg}, {Robitaille}, \&
  {Beaumont}}]{pvextractor_Ginsburg_2016}
{Ginsburg}, A., {Robitaille}, T., \& {Beaumont}, C. 2016, {pvextractor:
  Position-Velocity Diagram Extractor}, Astrophysics Source Code Library,
  record ascl:1608.010

\bibitem[{{Ginsburg} {et~al.}(2015){Ginsburg}, {Robitaille}, {Beaumont},
  {Rosolowsky}, {Leroy}, {Brogan}, {Hunter}, {Teuben}, \&
  {Brisbin}}]{radio_astronomy_tools_python_Ginsburg_2015}
{Ginsburg}, A., {Robitaille}, T., {Beaumont}, C., {et~al.} 2015, in
  Astronomical Society of the Pacific Conference Series, Vol. 499, Revolution
  in Astronomy with ALMA: The Third Year, ed. D.~{Iono}, K.~{Tatematsu},
  A.~{Wootten}, \& L.~{Testi}, 363--364

\bibitem[{{Gloudemans} {et~al.}(2022){Gloudemans}, {Duncan}, {Saxena},
  {Harikane}, {Hill}, {Zeimann}, {R{\"o}ttgering}, {Yang}, {Best},
  {Ba{\~n}ados}, {Drabent}, {Hardcastle}, {Hennawi}, {Lansbury},
  {Magliocchetti}, {Miley}, {Nanni}, {Shimwell}, {Smith}, {Venemans}, \&
  {Wagenveld}}]{24_RL_quasars_z_4.9_to_6.6_Gloudemans_2022}
{Gloudemans}, A.~J., {Duncan}, K.~J., {Saxena}, A., {et~al.} 2022, \aap, 668,
  A27, \dodoi{10.1051/0004-6361/202244763}

\bibitem[{{Goulding} {et~al.}(2023){Goulding}, {Greene}, {Setton}, {Labbe},
  {Bezanson}, {Miller}, {Atek}, {Bogd{\'a}n}, {Brammer}, {Chemerynska},
  {Cutler}, {Dayal}, {Fudamoto}, {Fujimoto}, {Furtak}, {Kokorev}, {Khullar},
  {Leja}, {Marchesini}, {Natarajan}, {Nelson}, {Oesch}, {Pan}, {Papovich},
  {Price}, {van Dokkum}, {Wang}, {Weaver}, {Whitaker}, \&
  {Zitrin}}]{UNCOVER_SMBH_growth_UHZ1_JWST_spectroscopy_Goulding_2023}
{Goulding}, A.~D., {Greene}, J.~E., {Setton}, D.~J., {et~al.} 2023, \apjl, 955,
  L24, \dodoi{10.3847/2041-8213/acf7c5}

\bibitem[{{Grogin} {et~al.}(2011){Grogin}, {Kocevski}, {Faber}, {Ferguson},
  {Koekemoer}, {Riess}, {Acquaviva}, {Alexander}, {Almaini}, {Ashby}, {Barden},
  {Bell}, {Bournaud}, {Brown}, {Caputi}, {Casertano}, {Cassata}, {Castellano},
  {Challis}, {Chary}, {Cheung}, {Cirasuolo}, {Conselice}, {Roshan Cooray},
  {Croton}, {Daddi}, {Dahlen}, {Dav{\'e}}, {de Mello}, {Dekel}, {Dickinson},
  {Dolch}, {Donley}, {Dunlop}, {Dutton}, {Elbaz}, {Fazio}, {Filippenko},
  {Finkelstein}, {Fontana}, {Gardner}, {Garnavich}, {Gawiser}, {Giavalisco},
  {Grazian}, {Guo}, {Hathi}, {H{\"a}ussler}, {Hopkins}, {Huang}, {Huang},
  {Jha}, {Kartaltepe}, {Kirshner}, {Koo}, {Lai}, {Lee}, {Li}, {Lotz}, {Lucas},
  {Madau}, {McCarthy}, {McGrath}, {McIntosh}, {McLure}, {Mobasher},
  {Moustakas}, {Mozena}, {Nandra}, {Newman}, {Niemi}, {Noeske}, {Papovich},
  {Pentericci}, {Pope}, {Primack}, {Rajan}, {Ravindranath}, {Reddy}, {Renzini},
  {Rix}, {Robaina}, {Rodney}, {Rosario}, {Rosati}, {Salimbeni}, {Scarlata},
  {Siana}, {Simard}, {Smidt}, {Somerville}, {Spinrad}, {Straughn}, {Strolger},
  {Telford}, {Teplitz}, {Trump}, {van der Wel}, {Villforth}, {Wechsler},
  {Weiner}, {Wiklind}, {Wild}, {Wilson}, {Wuyts}, {Yan}, \&
  {Yun}}]{CANDELS_Grogin_2011}
{Grogin}, N.~A., {Kocevski}, D.~D., {Faber}, S.~M., {et~al.} 2011, \apjs, 197,
  35, \dodoi{10.1088/0067-0049/197/2/35}

\bibitem[{{G{\"u}sten} {et~al.}(2006){G{\"u}sten}, {Nyman}, {Schilke},
  {Menten}, {Cesarsky}, \& {Booth}}]{APEX_ALMA_pathfinder_Gusten_2006}
{G{\"u}sten}, R., {Nyman}, L.~{\r{A}}., {Schilke}, P., {et~al.} 2006, \aap,
  454, L13, \dodoi{10.1051/0004-6361:20065420}

\bibitem[{Harris {et~al.}(2020)Harris, Millman, van~der Walt, Gommers,
  Virtanen, Cournapeau, Wieser, Taylor, Berg, Smith, Kern, Picus, Hoyer, van
  Kerkwijk, Brett, Haldane, del R{\'{i}}o, Wiebe, Peterson,
  G{\'{e}}rard-Marchant, Sheppard, Reddy, Weckesser, Abbasi, Gohlke, \&
  Oliphant}]{Numpy_Harris_2020}
Harris, C.~R., Millman, K.~J., van~der Walt, S.~J., {et~al.} 2020, Nature, 585,
  357, \dodoi{10.1038/s41586-020-2649-2}

\bibitem[{{Hodge} {et~al.}(2013){Hodge}, {Karim}, {Smail}, {Swinbank},
  {Walter}, {Biggs}, {Ivison}, {Weiss}, {Alexander}, {Bertoldi}, {Brandt},
  {Chapman}, {Coppin}, {Cox}, {Danielson}, {Dannerbauer}, {De Breuck},
  {Decarli}, {Edge}, {Greve}, {Knudsen}, {Menten}, {Rix}, {Schinnerer},
  {Simpson}, {Wardlow}, \& {van der
  Werf}}]{ALESS_SMG_catalogue_multiplicity_Hodge_2013}
{Hodge}, J.~A., {Karim}, A., {Smail}, I., {et~al.} 2013, \apj, 768, 91,
  \dodoi{10.1088/0004-637X/768/1/91}

\bibitem[{Hunter(2007)}]{Matplotlib_Hunter_2007}
Hunter, J.~D. 2007, Computing in Science \& Engineering, 9, 90,
  \dodoi{10.1109/MCSE.2007.55}

\bibitem[{{Ighina} {et~al.}(2021){Ighina}, {Belladitta}, {Caccianiga},
  {Broderick}, {Drouart}, {Moretti}, \& {Seymour}}]{RL_QSO_z_6.44_Ighina_2021}
{Ighina}, L., {Belladitta}, S., {Caccianiga}, A., {et~al.} 2021, \aap, 647,
  L11, \dodoi{10.1051/0004-6361/202140362}

\bibitem[{{Ighina} {et~al.}(2023){Ighina}, {Caccianiga}, {Moretti},
  {Belladitta}, {Broderick}, {Drouart}, {Leung}, \&
  {Seymour}}]{EoR_RL_QSOs_Ighina_2023}
{Ighina}, L., {Caccianiga}, A., {Moretti}, A., {et~al.} 2023, \mnras, 519,
  2060, \dodoi{10.1093/mnras/stac3668}

\bibitem[{{Ighina} {et~al.}(2022){Ighina}, {Leung}, {Broderick}, {Drouart},
  {Seymour}, {Belladitta}, {Caccianiga}, {Lenc}, {Moretti}, {An}, {Galvin},
  {Heald}, {Huynh}, {McConnell}, {Murphy}, {Pritchard}, {Quici}, {Shabala},
  {Tingay}, {Turner}, {Wang}, \&
  {White}}]{RL_QSO_z_6.44_properties_Ighina_2022}
{Ighina}, L., {Leung}, J.~K., {Broderick}, J.~W., {et~al.} 2022, \aap, 663,
  A73, \dodoi{10.1051/0004-6361/202142733}

\bibitem[{{Ighina} {et~al.}(2024){Ighina}, {Caccianiga}, {Moretti},
  {Broderick}, {Leung}, {Paterson}, {Rigamonti}, {Seymour}, {Belladitta},
  {Drouart}, {Galvin}, \& {Hurley-Walker}}]{RL_QSO_z_6.5_multi-wl_Ighina_2024}
{Ighina}, L., {Caccianiga}, A., {Moretti}, A., {et~al.} 2024, \aap, 687, A242,
  \dodoi{10.1051/0004-6361/202449369}

\bibitem[{{Inoue} {et~al.}(2014){Inoue}, {Shimizu}, {Iwata}, \&
  {Tanaka}}]{IGM_absorption_model_Inoue_2014}
{Inoue}, A.~K., {Shimizu}, I., {Iwata}, I., \& {Tanaka}, M. 2014, \mnras, 442,
  1805, \dodoi{10.1093/mnras/stu936}

\bibitem[{{Ivison} {et~al.}(2012){Ivison}, {Smail}, {Amblard}, {Arumugam}, {De
  Breuck}, {Emonts}, {Feain}, {Greve}, {Haas}, {Ibar}, {Jarvis}, {Kov{\'a}cs},
  {Lehnert}, {Nesvadba}, {R{\"o}ttgering}, {Seymour}, \&
  {Wylezalek}}]{Gas-rich_mergers_starbursting_radio_galaxies_Ivison_2012}
{Ivison}, R.~J., {Smail}, I., {Amblard}, A., {et~al.} 2012, \mnras, 425, 1320,
  \dodoi{10.1111/j.1365-2966.2012.21544.x}

\bibitem[{{Jeong} {et~al.}(2025){Jeong}, {Jeon}, {Song}, \&
  {Bromm}}]{Simulating_top-heavy_IMF_high_SFE_enhanced_UV_luminosity_Jeong_2025}
{Jeong}, T.~B., {Jeon}, M., {Song}, H., \& {Bromm}, V. 2025, \apj, 980, 10,
  \dodoi{10.3847/1538-4357/ada27d}

\bibitem[{{Kissler-Patig} {et~al.}(2008){Kissler-Patig}, {Pirard}, {Casali},
  {Moorwood}, {Ageorges}, {Alves de Oliveira}, {Baksai}, {Bedin}, {Bendek},
  {Biereichel}, {Delabre}, {Dorn}, {Esteves}, {Finger}, {Gojak}, {Huster},
  {Jung}, {Kiekebush}, {Klein}, {Koch}, {Lizon}, {Mehrgan}, {Petr-Gotzens},
  {Pritchard}, {Selman}, \& {Stegmeier}}]{HAWK-I_Kissler-Patig_2008}
{Kissler-Patig}, M., {Pirard}, J.~F., {Casali}, M., {et~al.} 2008, \aap, 491,
  941, \dodoi{10.1051/0004-6361:200809910}

\bibitem[{{Kodra} {et~al.}(2023){Kodra}, {Andrews}, {Newman}, {Finkelstein},
  {Fontana}, {Hathi}, {Salvato}, {Wiklind}, {Wuyts}, {Broussard}, {Chartab},
  {Conselice}, {Cooper}, {Dekel}, {Dickinson}, {Ferguson}, {Gawiser}, {Grogin},
  {Iyer}, {Kartaltepe}, {Kassin}, {Koekemoer}, {Koo}, {Lucas}, {Mantha},
  {McIntosh}, {Mobasher}, {Pacifici}, {P{\'e}rez-Gonz{\'a}lez}, \&
  {Santini}}]{CANDELS_photz_catalogue_Kodra_2023}
{Kodra}, D., {Andrews}, B.~H., {Newman}, J.~A., {et~al.} 2023, \apj, 942, 36,
  \dodoi{10.3847/1538-4357/ac9f12}

\bibitem[{{Kolwa} {et~al.}(2023){Kolwa}, {De Breuck}, {Vernet}, {Wylezalek},
  {Wang}, {Popping}, {Man}, {Harrison}, \&
  {Andreani}}]{Faint_Ci_z_3.5_HzRG_Kolwa_2023}
{Kolwa}, S., {De Breuck}, C., {Vernet}, J., {et~al.} 2023, \mnras, 525, 5831,
  \dodoi{10.1093/mnras/stad2647}

\bibitem[{{Koopmans} {et~al.}(2015){Koopmans}, {Pritchard}, {Mellema},
  {Aguirre}, {Ahn}, {Barkana}, {van Bemmel}, {Bernardi}, {Bonaldi}, {Briggs},
  {de Bruyn}, {Chang}, {Chapman}, {Chen}, {Ciardi}, {Dayal}, {Ferrara},
  {Fialkov}, {Fiore}, {Ichiki}, {Illiev}, {Inoue}, {Jelic}, {Jones}, {Lazio},
  {Maio}, {Majumdar}, {Mack}, {Mesinger}, {Morales}, {Parsons}, {Pen},
  {Santos}, {Schneider}, {Semelin}, {de Souza}, {Subrahmanyan}, {Takeuchi},
  {Vedantham}, {Wagg}, {Webster}, {Wyithe}, {Datta}, \&
  {Trott}}]{Cosmic_Dawn_EoR_SKA_Koopmans_2015}
{Koopmans}, L., {Pritchard}, J., {Mellema}, G., {et~al.} 2015, in Advancing
  Astrophysics with the Square Kilometre Array (AASKA14), 1,
  \dodoi{10.22323/1.215.0001}

\bibitem[{{Kurk} {et~al.}(2000){Kurk}, {R{\"o}ttgering}, {Pentericci}, {Miley},
  {van Breugel}, {Carilli}, {Ford}, {Heckman}, {McCarthy}, \&
  {Moorwood}}]{Spiderweb_galaxy_z_2.2_Lyman_alpha_emitters_Kurk_2000}
{Kurk}, J.~D., {R{\"o}ttgering}, H.~J.~A., {Pentericci}, L., {et~al.} 2000,
  \aap, 358, L1, \dodoi{10.48550/arXiv.astro-ph/0005058}

\bibitem[{{Lang}(2014)}]{unWISE_coadds_Lang_2014}
{Lang}, D. 2014, \aj, 147, 108, \dodoi{10.1088/0004-6256/147/5/108}

\bibitem[{{Lebowitz} {et~al.}(2023){Lebowitz}, {Emonts}, {Terndrup},
  {Burchett}, {Prochaska}, {Drouart}, {Villar-Mart{\'\i}n}, {Lehnert}, {De
  Breuck}, {Vernet}, \&
  {Alatalo}}]{Dragonfly_3_z_2_HzRG_jet_brightening_disk_merger_ULIRG_Lebowitz_2023}
{Lebowitz}, S., {Emonts}, B., {Terndrup}, D.~M., {et~al.} 2023, \apj, 951, 73,
  \dodoi{10.3847/1538-4357/acd3ed}

\bibitem[{{Liu} {et~al.}(2021){Liu}, {Daddi}, {Schinnerer}, {Saito}, {Leroy},
  {Silverman}, {Valentino}, {Magdis}, {Gao}, {Jin}, {Puglisi}, \&
  {Groves}}]{CO_molecular_gas_state_local_high_z_galaxies_Liu_2021}
{Liu}, D., {Daddi}, E., {Schinnerer}, E., {et~al.} 2021, \apj, 909, 56,
  \dodoi{10.3847/1538-4357/abd801}

\bibitem[{{Long} {et~al.}(2020){Long}, {Cooray}, {Ma}, {Casey}, {Wardlow},
  {Nayyeri}, {Ivison}, {Farrah}, \&
  {Dannerbauer}}]{DRC_z_4_detailed_evolution_Long_2020}
{Long}, A.~S., {Cooray}, A., {Ma}, J., {et~al.} 2020, \apj, 898, 133,
  \dodoi{10.3847/1538-4357/ab9d1f}

\bibitem[{{Marigo} {et~al.}(2013){Marigo}, {Bressan}, {Nanni}, {Girardi}, \&
  {Pumo}}]{COLIBRI_stellar_evolutionary_track_Marigo_2013}
{Marigo}, P., {Bressan}, A., {Nanni}, A., {Girardi}, L., \& {Pumo}, M.~L. 2013,
  \mnras, 434, 488, \dodoi{10.1093/mnras/stt1034}

\bibitem[{{McMullin} {et~al.}(2007){McMullin}, {Waters}, {Schiebel}, {Young},
  \& {Golap}}]{CASA_McMullin_2007}
{McMullin}, J.~P., {Waters}, B., {Schiebel}, D., {Young}, W., \& {Golap}, K.
  2007, in Astronomical Society of the Pacific Conference Series, Vol. 376,
  Astronomical Data Analysis Software and Systems XVI, ed. R.~A. {Shaw},
  F.~{Hill}, \& D.~J. {Bell}, 127

\bibitem[{{Miley} \& {De Breuck}(2008)}]{HzRG_review_paper_Miley_DeBreuck_2008}
{Miley}, G., \& {De Breuck}, C. 2008, \aapr, 15, 67,
  \dodoi{10.1007/s00159-007-0008-z}

\bibitem[{{Miley} {et~al.}(2006){Miley}, {Overzier}, {Zirm}, {Ford}, {Kurk},
  {Pentericci}, {Blakeslee}, {Franx}, {Illingworth}, {Postman}, {Rosati},
  {R{\"o}ttgering}, {Venemans}, \& {Helder}}]{Spiderweb_galaxy_Miley_2006}
{Miley}, G.~K., {Overzier}, R.~A., {Zirm}, A.~W., {et~al.} 2006, \apjl, 650,
  L29, \dodoi{10.1086/508534}

\bibitem[{{Miller} {et~al.}(2018){Miller}, {Chapman}, {Aravena}, {Ashby},
  {Hayward}, {Vieira}, {Wei{\ss}}, {Babul}, {B{\'e}thermin}, {Bradford},
  {Brodwin}, {Carlstrom}, {Chen}, {Cunningham}, {De Breuck}, {Gonzalez},
  {Greve}, {Harnett}, {Hezaveh}, {Lacaille}, {Litke}, {Ma}, {Malkan},
  {Marrone}, {Morningstar}, {Murphy}, {Narayanan}, {Pass}, {Perry}, {Phadke},
  {Rennehan}, {Rotermund}, {Simpson}, {Spilker}, {Sreevani}, {Stark},
  {Strandet}, \& {Strom}}]{SPT2349-56_galaxy_cluster_core_z_4.3_Miller_2018}
{Miller}, T.~B., {Chapman}, S.~C., {Aravena}, M., {et~al.} 2018, \nat, 556,
  469, \dodoi{10.1038/s41586-018-0025-2}

\bibitem[{{Miszalski} {et~al.}(2022){Miszalski}, {O'Toole}, {Tocknell},
  {Marnoch}, \& {Ryder}}]{Data_central_data_aggregation_service_Miszalski_2022}
{Miszalski}, B., {O'Toole}, S.~J., {Tocknell}, J., {Marnoch}, L., \& {Ryder},
  S.~D. 2022, in Society of Photo-Optical Instrumentation Engineers (SPIE)
  Conference Series, Vol. 12189, Software and Cyberinfrastructure for Astronomy
  VII, 121892S, \dodoi{10.1117/12.2642065}

\bibitem[{{Momjian} {et~al.}(2021){Momjian}, {Ba{\~n}ados}, {Carilli},
  {Walter}, \& {Mazzucchelli}}]{RL_QSO_z_6.82_resolved_radio_Momjian_2021}
{Momjian}, E., {Ba{\~n}ados}, E., {Carilli}, C.~L., {Walter}, F., \&
  {Mazzucchelli}, C. 2021, \aj, 161, 207, \dodoi{10.3847/1538-3881/abe6ae}

\bibitem[{{Newville} {et~al.}(2023){Newville}, {Otten}, {Nelson}, {Stensitzki},
  {Ingargiola}, {Allan}, {Fox}, {Carter}, {Micha{\l}}, {Osborn}, {Pustakhod},
  {Lneuhaus}, {Weigand}, {Aristov}, {Glenn}, {Deil}, {Mgunyho}, {Mark},
  {Hansen}, {Pasquevich}, {Foks}, {Zobrist}, {Frost}, {Stuermer}, {Azelcer},
  {Polloreno}, {Persaud}, {Hedegaard Nielsen}, {Pompili}, \&
  {Eendebak}}]{lmfit_1.2.2_Newville_2023}
{Newville}, M., {Otten}, R., {Nelson}, A., {et~al.} 2023, {lmfit/lmfit-py:
  1.2.2}, 1.2.2,  Zenodo, \dodoi{10.5281/zenodo.8145703}

\bibitem[{{Noirot} {et~al.}(2016){Noirot}, {Vernet}, {De Breuck}, {Wylezalek},
  {Galametz}, {Stern}, {Mei}, {Brodwin}, {Cooke}, {Gonzalez}, {Hatch},
  {Rettura}, \&
  {Stanford}}]{CARLA_clusters_around_radio_loud_AGN_z_2_structures_Noirot_2016}
{Noirot}, G., {Vernet}, J., {De Breuck}, C., {et~al.} 2016, \apj, 830, 90,
  \dodoi{10.3847/0004-637X/830/2/90}

\bibitem[{{Noirot} {et~al.}(2018){Noirot}, {Stern}, {Mei}, {Wylezalek},
  {Cooke}, {De Breuck}, {Galametz}, {Hatch}, {Vernet}, {Brodwin}, {Eisenhardt},
  {Gonzalez}, {Jarvis}, {Rettura}, {Seymour}, \&
  {Stanford}}]{CARLA_clusters_around_radio_loud_AGN_z_1_3_structures_Noirot_2018}
{Noirot}, G., {Stern}, D., {Mei}, S., {et~al.} 2018, \apj, 859, 38,
  \dodoi{10.3847/1538-4357/aabadb}

\bibitem[{{Noll} {et~al.}(2009){Noll}, {Burgarella}, {Giovannoli}, {Buat},
  {Marcillac}, \& {Mu{\~n}oz-Mateos}}]{CIGALE_2_SINGS_test_sample_Noll_2009}
{Noll}, S., {Burgarella}, D., {Giovannoli}, E., {et~al.} 2009, \aap, 507, 1793,
  \dodoi{10.1051/0004-6361/200912497}

\bibitem[{{Oke}(1974)}]{AB_magnitude_system_absolute_SED_white_dwarfs_Oke_1974}
{Oke}, J.~B. 1974, \apjs, 27, 21, \dodoi{10.1086/190287}

\bibitem[{{Oteo} {et~al.}(2018){Oteo}, {Ivison}, {Dunne}, {Manilla-Robles},
  {Maddox}, {Lewis}, {de Zotti}, {Bremer}, {Clements}, {Cooray}, {Dannerbauer},
  {Eales}, {Greenslade}, {Omont}, {Perez{\textendash}Fourn{\'o}n}, {Riechers},
  {Scott}, {van der Werf}, {Weiss}, \&
  {Zhang}}]{DRC_z_4_proto-cluster_DSFG_Oteo_2018}
{Oteo}, I., {Ivison}, R.~J., {Dunne}, L., {et~al.} 2018, \apj, 856, 72,
  \dodoi{10.3847/1538-4357/aaa1f1}

\bibitem[{{P{\'e}rez-Gonz{\'a}lez} {et~al.}(2023){P{\'e}rez-Gonz{\'a}lez},
  {Barro}, {Annunziatella}, {Costantin}, {Garc{\'\i}a-Argum{\'a}nez},
  {McGrath}, {M{\'e}rida}, {Zavala}, {Arrabal Haro}, {Bagley}, {Backhaus},
  {Behroozi}, {Bell}, {Bisigello}, {Buat}, {Calabr{\`o}}, {Casey}, {Cleri},
  {Coogan}, {Cooper}, {Cooray}, {Dekel}, {Dickinson}, {Elbaz}, {Ferguson},
  {Finkelstein}, {Fontana}, {Franco}, {Gardner}, {Giavalisco},
  {G{\'o}mez-Guijarro}, {Grazian}, {Grogin}, {Guo}, {Huertas-Company}, {Jogee},
  {Kartaltepe}, {Kewley}, {Kirkpatrick}, {Kocevski}, {Koekemoer}, {Long},
  {Lotz}, {Lucas}, {Papovich}, {Pirzkal}, {Ravindranath}, {Somerville},
  {Tacchella}, {Trump}, {Wang}, {Wilkins}, {Wuyts}, {Yang}, \&
  {Yung}}]{Natures_of_HST-dark_galaxies_Perez-Gonzalez_2023}
{P{\'e}rez-Gonz{\'a}lez}, P.~G., {Barro}, G., {Annunziatella}, M., {et~al.}
  2023, \apjl, 946, L16, \dodoi{10.3847/2041-8213/acb3a5}

\bibitem[{{Pillepich} {et~al.}(2018){Pillepich}, {Nelson}, {Hernquist},
  {Springel}, {Pakmor}, {Torrey}, {Weinberger}, {Genel}, {Naiman}, {Marinacci},
  \& {Vogelsberger}}]{IllustrisTNG_galaxy_groups_clusters_Pilepich_2018}
{Pillepich}, A., {Nelson}, D., {Hernquist}, L., {et~al.} 2018, \mnras, 475,
  648, \dodoi{10.1093/mnras/stx3112}

\bibitem[{{Riechers} {et~al.}(2019){Riechers}, {Pavesi}, {Sharon}, {Hodge},
  {Decarli}, {Walter}, {Carilli}, {Aravena}, {da Cunha}, {Daddi}, {Dickinson},
  {Smail}, {Capak}, {Ivison}, {Sargent}, {Scoville}, \&
  {Wagg}}]{COLDz_CO_luminosity_function_high_redshift_Riechers_2019}
{Riechers}, D.~A., {Pavesi}, R., {Sharon}, C.~E., {et~al.} 2019, \apj, 872, 7,
  \dodoi{10.3847/1538-4357/aafc27}

\bibitem[{{Rocca-Volmerange} {et~al.}(2004){Rocca-Volmerange}, {Le Borgne}, {De
  Breuck}, {Fioc}, \& {Moy}}]{K_z_relation_for_RG_Rocca-Volmerange_2004}
{Rocca-Volmerange}, B., {Le Borgne}, D., {De Breuck}, C., {Fioc}, M., \& {Moy},
  E. 2004, \aap, 415, 931, \dodoi{10.1051/0004-6361:20031717}

\bibitem[{{Salpeter}(1955)}]{IMF_Salpeter_1955}
{Salpeter}, E.~E. 1955, \apj, 121, 161, \dodoi{10.1086/145971}

\bibitem[{{S{\'a}nchez} \& {DES Collaboration}(2010)}]{DES_design_Sanchez_2010}
{S{\'a}nchez}, E., \& {DES Collaboration}. 2010, in Journal of Physics
  Conference Series, Vol. 259, Journal of Physics Conference Series (IOP),
  012080, \dodoi{10.1088/1742-6596/259/1/012080}

\bibitem[{{S{\'a}nchez-Bl{\'a}zquez} {et~al.}(2006){S{\'a}nchez-Bl{\'a}zquez},
  {Peletier}, {Jim{\'e}nez-Vicente}, {Cardiel}, {Cenarro},
  {Falc{\'o}n-Barroso}, {Gorgas}, {Selam}, \&
  {Vazdekis}}]{MILES_spectral_templates_Sanchez-Blazquez_2006}
{S{\'a}nchez-Bl{\'a}zquez}, P., {Peletier}, R.~F., {Jim{\'e}nez-Vicente}, J.,
  {et~al.} 2006, \mnras, 371, 703, \dodoi{10.1111/j.1365-2966.2006.10699.x}

\bibitem[{{Scoville} {et~al.}(2016){Scoville}, {Sheth}, {Aussel}, {Vanden
  Bout}, {Capak}, {Bongiorno}, {Casey}, {Murchikova}, {Koda},
  {{\'A}lvarez-M{\'a}rquez}, {Lee}, {Laigle}, {McCracken}, {Ilbert}, {Pope},
  {Sanders}, {Chu}, {Toft}, {Ivison}, \&
  {Manohar}}]{Mass_and_SFR_law_dust_continuum_ALMA_COSMOS_Scoville_2016}
{Scoville}, N., {Sheth}, K., {Aussel}, H., {et~al.} 2016, \apj, 820, 83,
  \dodoi{10.3847/0004-637X/820/2/83}

\bibitem[{{Serra} {et~al.}(2015){Serra}, {Westmeier}, {Giese}, {Jurek},
  {Fl{\"o}er}, {Popping}, {Winkel}, {van der Hulst}, {Meyer}, {Koribalski},
  {Staveley-Smith}, \& {Courtois}}]{SoFiA_description_Serra_2015}
{Serra}, P., {Westmeier}, T., {Giese}, N., {et~al.} 2015, \mnras, 448, 1922,
  \dodoi{10.1093/mnras/stv079}

\bibitem[{{Seymour} {et~al.}(2007){Seymour}, {Stern}, {De Breuck}, {Vernet},
  {Rettura}, {Dickinson}, {Dey}, {Eisenhardt}, {Fosbury}, {Lacy}, {McCarthy},
  {Miley}, {Rocca-Volmerange}, {R{\"o}ttgering}, {Stanford}, {Teplitz}, {van
  Breugel}, \& {Zirm}}]{RG_Hosts_across_cosmic_time_Seymour_2007}
{Seymour}, N., {Stern}, D., {De Breuck}, C., {et~al.} 2007, \apjs, 171, 353,
  \dodoi{10.1086/517887}

\bibitem[{{Seymour} {et~al.}(2012){Seymour}, {Altieri}, {De Breuck}, {Barthel},
  {Coia}, {Conversi}, {Dannerbauer}, {Dey}, {Dickinson}, {Drouart}, {Galametz},
  {Greve}, {Haas}, {Hatch}, {Ibar}, {Ivison}, {Jarvis}, {Kov{\'a}cs}, {Kurk},
  {Lehnert}, {Miley}, {Nesvadba}, {Rawlings}, {Rettura}, {R{\"o}ttgering},
  {Rocca-Volmerange}, {S{\'a}nchez-Portal}, {Santos}, {Stern}, {Stevens},
  {Valtchanov}, {Vernet}, \&
  {Wylezalek}}]{MRC1138-262_HzRG_z_2.2_Spiderweb_galaxy_coeval_growth_Seymour_2012}
{Seymour}, N., {Altieri}, B., {De Breuck}, C., {et~al.} 2012, \apj, 755, 146,
  \dodoi{10.1088/0004-637X/755/2/146}

\bibitem[{{Siringo} {et~al.}(2009){Siringo}, {Kreysa}, {Kov{\'a}cs},
  {Schuller}, {Wei{\ss}}, {Esch}, {Gem{\"u}nd}, {Jethava}, {Lundershausen},
  {Colin}, {G{\"u}sten}, {Menten}, {Beelen}, {Bertoldi}, {Beeman}, \&
  {Haller}}]{LABOCA_Large_APEX_Bolometric_Camera_Siringo_2009}
{Siringo}, G., {Kreysa}, E., {Kov{\'a}cs}, A., {et~al.} 2009, \aap, 497, 945,
  \dodoi{10.1051/0004-6361/200811454}

\bibitem[{{Smith} \& {Bromm}(2019)}]{SMBHs_in_early_universe_Smith_2019}
{Smith}, A., \& {Bromm}, V. 2019, Contemporary Physics, 60, 111,
  \dodoi{10.1080/00107514.2019.1615715}

\bibitem[{{Smith} {et~al.}(2017){Smith}, {Bromm}, \&
  {Loeb}}]{First_SMBHs_Smith_2017}
{Smith}, A., {Bromm}, V., \& {Loeb}, A. 2017, Astronomy and Geophysics, 58,
  3.22, \dodoi{10.1093/astrogeo/atx099}

\bibitem[{{Solomon} \& {Vanden Bout}(2005)}]{high_z_molecular_gas_Solomon_2005}
{Solomon}, P.~M., \& {Vanden Bout}, P.~A. 2005, \araa, 43, 677,
  \dodoi{10.1146/annurev.astro.43.051804.102221}

\bibitem[{{Spilker} {et~al.}(2014){Spilker}, {Marrone}, {Aguirre}, {Aravena},
  {Ashby}, {B{\'e}thermin}, {Bradford}, {Bothwell}, {Brodwin}, {Carlstrom},
  {Chapman}, {Crawford}, {de Breuck}, {Fassnacht}, {Gonzalez}, {Greve},
  {Gullberg}, {Hezaveh}, {Holzapfel}, {Husband}, {Ma}, {Malkan}, {Murphy},
  {Reichardt}, {Rotermund}, {Stalder}, {Stark}, {Strandet}, {Vieira},
  {Wei{\ss}}, \& {Welikala}}]{sub-mm_SED_high_z_DSFG_Spilker_2014}
{Spilker}, J.~S., {Marrone}, D.~P., {Aguirre}, J.~E., {et~al.} 2014, \apj, 785,
  149, \dodoi{10.1088/0004-637X/785/2/149}

\bibitem[{{Taylor}(2005)}]{TOPCAT_Taylor_2005}
{Taylor}, M.~B. 2005, in Astronomical Society of the Pacific Conference Series,
  Vol. 347, Astronomical Data Analysis Software and Systems XIV, ed.
  P.~{Shopbell}, M.~{Britton}, \& R.~{Ebert}, 29

\bibitem[{{Venemans} {et~al.}(2002){Venemans}, {Kurk}, {Miley},
  {R{\"o}ttgering}, {van Breugel}, {Carilli}, {De Breuck}, {Ford}, {Heckman},
  {McCarthy}, \& {Pentericci}}]{HzRG_protocluster_BCG_ancestor_Venemans_2002}
{Venemans}, B.~P., {Kurk}, J.~D., {Miley}, G.~K., {et~al.} 2002, \apjl, 569,
  L11, \dodoi{10.1086/340563}

\bibitem[{{Venemans} {et~al.}(2007){Venemans}, {R{\"o}ttgering}, {Miley}, {van
  Breugel}, {de Breuck}, {Kurk}, {Pentericci}, {Stanford}, {Overzier}, {Croft},
  \& {Ford}}]{HzRG_protocluster_BCG_population_2007}
{Venemans}, B.~P., {R{\"o}ttgering}, H.~J.~A., {Miley}, G.~K., {et~al.} 2007,
  \aap, 461, 823, \dodoi{10.1051/0004-6361:20053941}

\bibitem[{Virtanen {et~al.}(2020)Virtanen, Gommers, Oliphant, Haberland, Reddy,
  Cournapeau, Burovski, Peterson, Weckesser, Bright, {van der Walt}, Brett,
  Wilson, Millman, Mayorov, Nelson, Jones, Kern, Larson, Carey, Polat, Feng,
  Moore, {VanderPlas}, Laxalde, Perktold, Cimrman, Henriksen, Quintero, Harris,
  Archibald, Ribeiro, Pedregosa, {van Mulbregt}, \& {SciPy 1.0
  Contributors}}]{SciPy_Virtanen_2020}
Virtanen, P., Gommers, R., Oliphant, T.~E., {et~al.} 2020, Nature Methods, 17,
  261, \dodoi{10.1038/s41592-019-0686-2}

\bibitem[{{Volonteri}(2012)}]{SMBH_formation_evolution_Volonteri_2012}
{Volonteri}, M. 2012, Science, 337, 544, \dodoi{10.1126/science.1220843}

\bibitem[{{Wang} {et~al.}(2021){Wang}, {Yang}, {Fan}, {Hennawi}, {Barth},
  {Banados}, {Bian}, {Boutsia}, {Connor}, {Davies}, {Decarli}, {Eilers},
  {Farina}, {Green}, {Jiang}, {Li}, {Mazzucchelli}, {Nanni}, {Schindler},
  {Venemans}, {Walter}, {Wu}, \& {Yue}}]{QSO_z_7.642_Wang_2021}
{Wang}, F., {Yang}, J., {Fan}, X., {et~al.} 2021, \apjl, 907, L1,
  \dodoi{10.3847/2041-8213/abd8c6}

\bibitem[{{Wei{\ss}} {et~al.}(2009){Wei{\ss}}, {Kov{\'a}cs}, {Coppin}, {Greve},
  {Walter}, {Smail}, {Dunlop}, {Knudsen}, {Alexander}, {Bertoldi}, {Brandt},
  {Chapman}, {Cox}, {Dannerbauer}, {De Breuck}, {Gawiser}, {Ivison}, {Lutz},
  {Menten}, {Koekemoer}, {Kreysa}, {Kurczynski}, {Rix}, {Schinnerer}, \& {van
  der Werf}}]{LESS_LABOCA_ECDFS_Survey_South_Weiss_2009}
{Wei{\ss}}, A., {Kov{\'a}cs}, A., {Coppin}, K., {et~al.} 2009, \apj, 707, 1201,
  \dodoi{10.1088/0004-637X/707/2/1201}

\bibitem[{{Wei{\ss}} {et~al.}(2013){Wei{\ss}}, {De Breuck}, {Marrone},
  {Vieira}, {Aguirre}, {Aird}, {Aravena}, {Ashby}, {Bayliss}, {Benson},
  {B{\'e}thermin}, {Biggs}, {Bleem}, {Bock}, {Bothwell}, {Bradford}, {Brodwin},
  {Carlstrom}, {Chang}, {Chapman}, {Crawford}, {Crites}, {de Haan}, {Dobbs},
  {Downes}, {Fassnacht}, {George}, {Gladders}, {Gonzalez}, {Greve},
  {Halverson}, {Hezaveh}, {High}, {Holder}, {Holzapfel}, {Hoover}, {Hrubes},
  {Husband}, {Keisler}, {Lee}, {Leitch}, {Lueker}, {Luong-Van}, {Malkan},
  {McIntyre}, {McMahon}, {Mehl}, {Menten}, {Meyer}, {Murphy}, {Padin},
  {Plagge}, {Reichardt}, {Rest}, {Rosenman}, {Ruel}, {Ruhl}, {Schaffer},
  {Shirokoff}, {Spilker}, {Stalder}, {Staniszewski}, {Stark}, {Story},
  {Vanderlinde}, {Welikala}, \&
  {Williamson}}]{ALMA_redshifts_SPT_survey_Weiss_2013}
{Wei{\ss}}, A., {De Breuck}, C., {Marrone}, D.~P., {et~al.} 2013, \apj, 767,
  88, \dodoi{10.1088/0004-637X/767/1/88}

\bibitem[{{Westmeier} {et~al.}(2021){Westmeier}, {Kitaeff}, {Pallot}, {Serra},
  {van der Hulst}, {Jurek}, {Elagali}, {For}, {Kleiner}, {Koribalski},
  {Lee-Waddell}, {Mould}, {Reynolds}, {Rhee}, \&
  {Staveley-Smith}}]{SoFiA2_description_Westmeier_2021}
{Westmeier}, T., {Kitaeff}, S., {Pallot}, D., {et~al.} 2021, \mnras, 506, 3962,
  \dodoi{10.1093/mnras/stab1881}

\bibitem[{{Williams} {et~al.}(2024){Williams}, {Alberts}, {Ji}, {Hainline},
  {Lyu}, {Rieke}, {Endsley}, {Suess}, {Sun}, {Johnson}, {Florian}, {Shivaei},
  {Rujopakarn}, {Baker}, {Bhatawdekar}, {Boyett}, {Bunker}, {Cameron},
  {Carniani}, {Charlot}, {Curtis-Lake}, {DeCoursey}, {de Graaff}, {Egami},
  {Eisenstein}, {Gibson}, {Hausen}, {Helton}, {Maiolino}, {Maseda}, {Nelson},
  {P{\'e}rez-Gonz{\'a}lez}, {Rieke}, {Robertson}, {Saxena}, {Tacchella},
  {Willmer}, \&
  {Willott}}]{Extremely_red_galaxies_at_high_redshift_Williams_2024}
{Williams}, C.~C., {Alberts}, S., {Ji}, Z., {et~al.} 2024, \apj, 968, 34,
  \dodoi{10.3847/1538-4357/ad3f17}

\bibitem[{{Wootten} \& {Thompson}(2009)}]{ALMA_Wootten_Thompson_2009}
{Wootten}, A., \& {Thompson}, A.~R. 2009, IEEE Proceedings, 97, 1463,
  \dodoi{10.1109/JPROC.2009.2020572}

\bibitem[{{Wright} {et~al.}(2010){Wright}, {Eisenhardt}, {Mainzer}, {Ressler},
  {Cutri}, {Jarrett}, {Kirkpatrick}, {Padgett}, {McMillan}, {Skrutskie},
  {Stanford}, {Cohen}, {Walker}, {Mather}, {Leisawitz}, {Gautier}, {McLean},
  {Benford}, {Lonsdale}, {Blain}, {Mendez}, {Irace}, {Duval}, {Liu}, {Royer},
  {Heinrichsen}, {Howard}, {Shannon}, {Kendall}, {Walsh}, {Larsen}, {Cardon},
  {Schick}, {Schwalm}, {Abid}, {Fabinsky}, {Naes}, \&
  {Tsai}}]{WISE_Wright_2010}
{Wright}, E.~L., {Eisenhardt}, P. R.~M., {Mainzer}, A.~K., {et~al.} 2010, \aj,
  140, 1868, \dodoi{10.1088/0004-6256/140/6/1868}

\bibitem[{{Wylezalek} {et~al.}(2013){Wylezalek}, {Galametz}, {Stern}, {Vernet},
  {De Breuck}, {Seymour}, {Brodwin}, {Eisenhardt}, {Gonzalez}, {Hatch},
  {Jarvis}, {Rettura}, {Stanford}, \&
  {Stevens}}]{Clusters_around_radio_loud_AGN_z_1_3_Wylezalek_2013}
{Wylezalek}, D., {Galametz}, A., {Stern}, D., {et~al.} 2013, \apj, 769, 79,
  \dodoi{10.1088/0004-637X/769/1/79}

\bibitem[{{Wylezalek} {et~al.}(2014){Wylezalek}, {Vernet}, {De Breuck},
  {Stern}, {Brodwin}, {Galametz}, {Gonzalez}, {Jarvis}, {Hatch}, {Seymour}, \&
  {Stanford}}]{Cluster_MIR_luminosity_function_z_1_3_Wylezalek_2014}
{Wylezalek}, D., {Vernet}, J., {De Breuck}, C., {et~al.} 2014, \apj, 786, 17,
  \dodoi{10.1088/0004-637X/786/1/17}

\bibitem[{{Young} \&
  {Knezek}(1989)}]{Molecular_atomic_gas_ratio_in_spiral_galaxies_Young_1989}
{Young}, J.~S., \& {Knezek}, P.~M. 1989, \apjl, 347, L55,
  \dodoi{10.1086/185606}

\bibitem[{{Young} \&
  {Scoville}(1991)}]{Molecular_gas_in_spiral_galaxies_Young_1991}
{Young}, J.~S., \& {Scoville}, N.~Z. 1991, \araa, 29, 581,
  \dodoi{10.1146/annurev.aa.29.090191.003053}

\end{thebibliography}
\bibliographystyle{aasjournal}

\newpage

\appendix
\section{Supplementary Tables \& Figures}\label{app:supplementary}

\renewcommand\thefigure{\thesection\arabic{figure}}    
\setcounter{figure}{0}

\begin{figure*}[ht]
    \includegraphics[trim={0.9cm 0.4cm 1.5cm 1.5cm},clip,width=\linewidth]{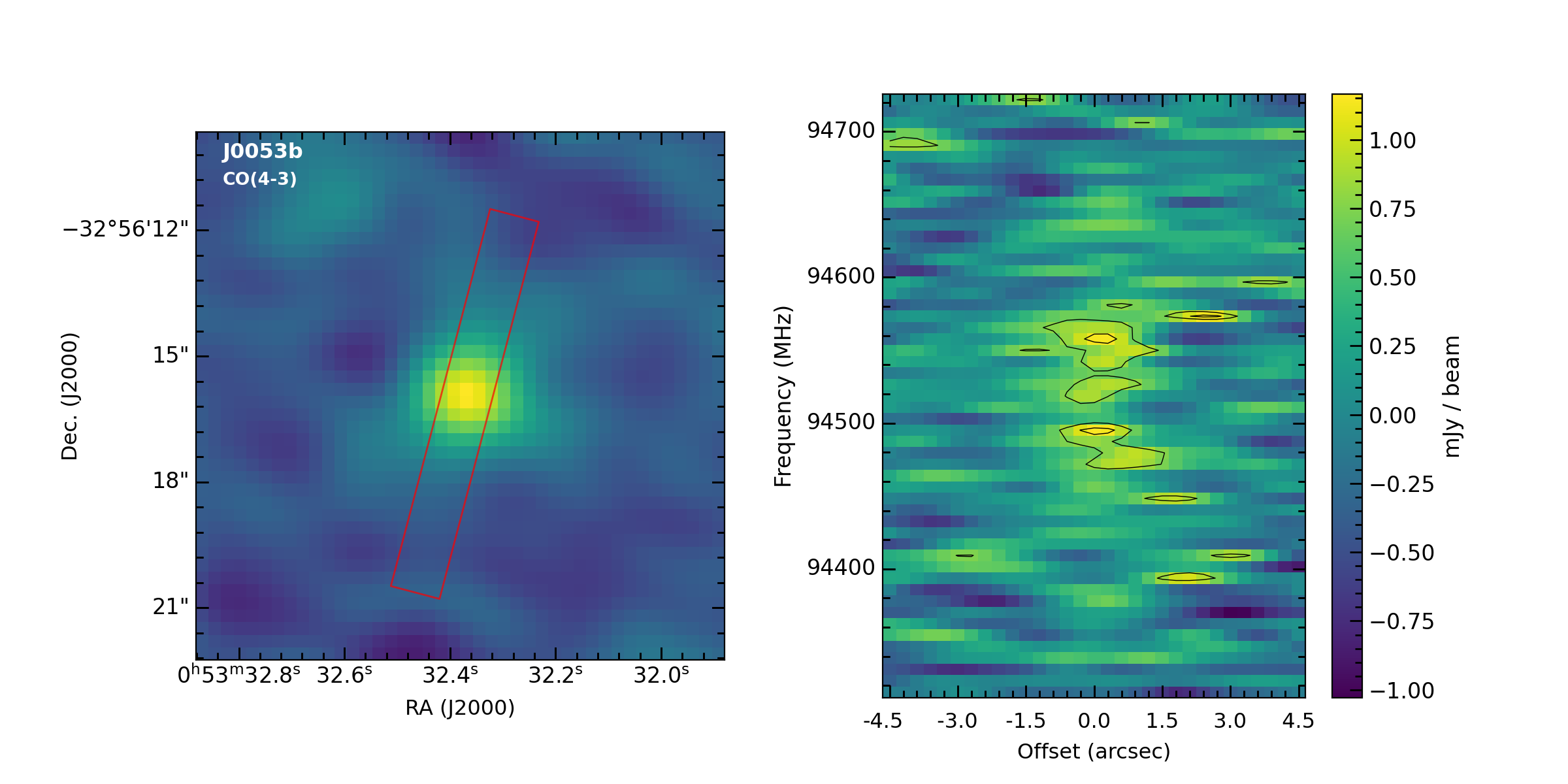}
    \caption{Position-velocity diagram (PVD) of the CO(4-3) associated with J0053b in the cube, after smoothing to the smallest common resolution across the five tunings. The PVD ({\it right panel}) is made using \texttt{pvextractor} \citep[][]{radio_astronomy_tools_python_Ginsburg_2015,pvextractor_Ginsburg_2016} with a 9\arcsec{} length slit of 1.2\arcsec{} width centred on the centroid of the CO(4-3) segment and along the position angle of the segment, both estimated by \texttt{photutils} on the FWZI linemap ({\it left panel}).}
    \label{fig:pvd_J0053b}
\end{figure*}

\begin{figure}
\figurenum{A2}
\plotone{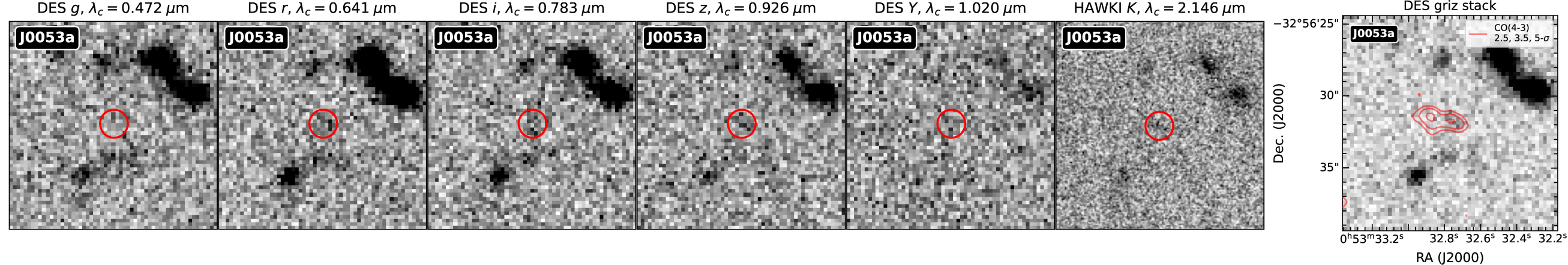}
\plotone{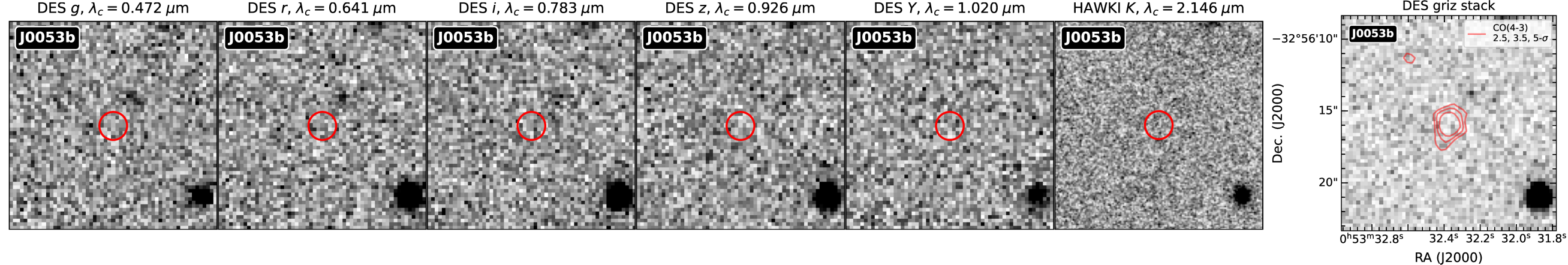}
{\includegraphics[width=0.7\textwidth]{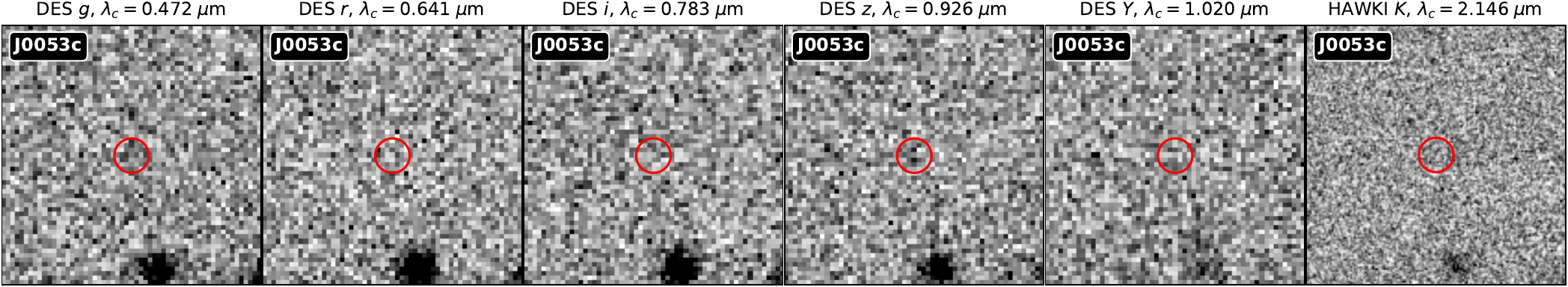} \hspace{8cm}}
\plotone{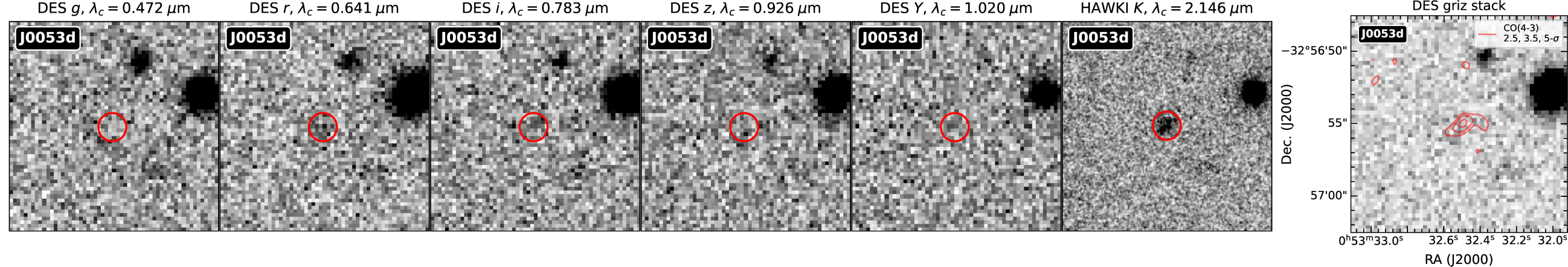}
\plotone{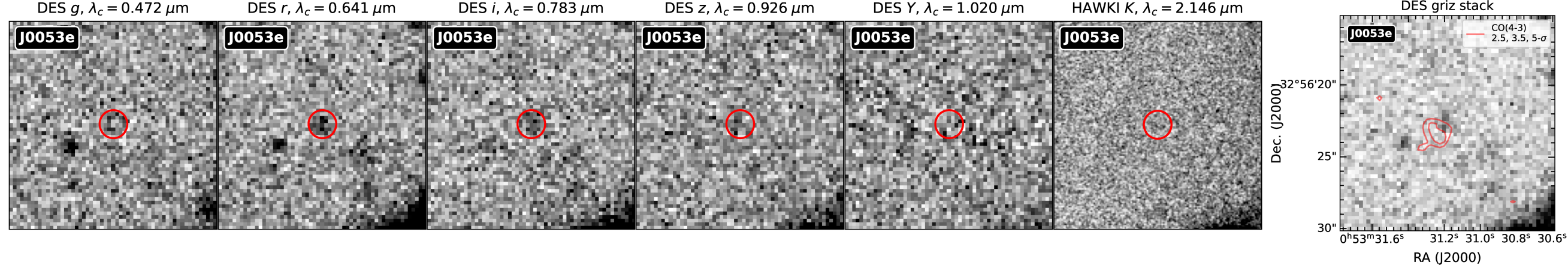}
\plotone{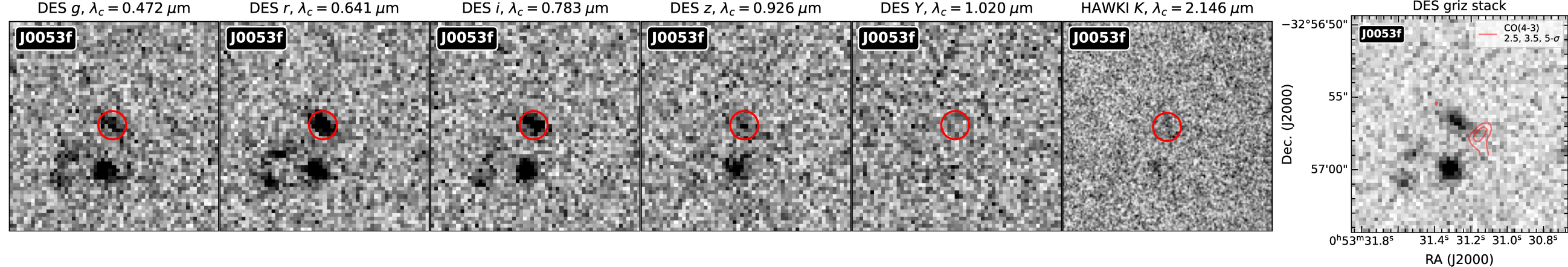}
\plotone{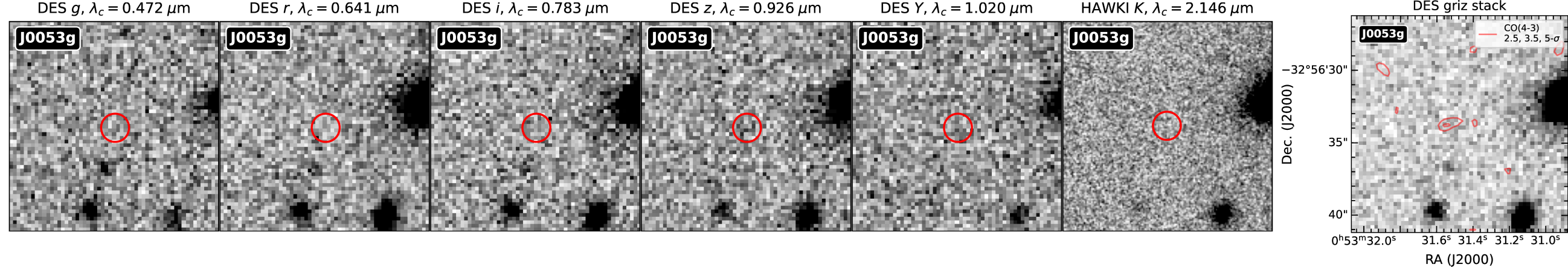}
\plotone{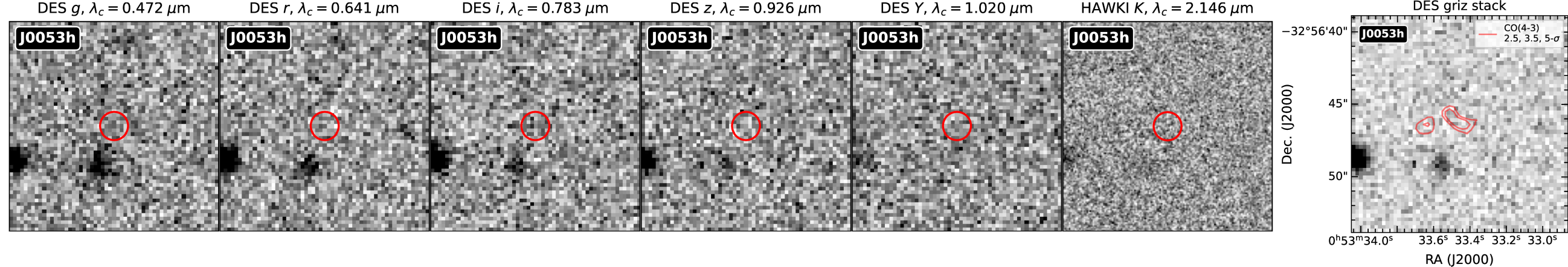}
\caption{Multi-wavelength cutouts, $15\hbox{$^{\prime\prime}$}\times15\hbox{$^{\prime\prime}$}$, in the DES bands and HAWK-I $K_s$-band with a red circle overlaid indicating the aperture used to extract the aperture photometry at the nominal host galaxy/molecular gas position. An additional cutout is provided which shows a stack of the DES $g,\,r,\,i$ and $z$ bands overlaid with contours (red) of the CO(4-3) emission, starting at $2.5\sigma_{\rm linemap}$ and increasing by factors of $\sqrt{2}$, where $\sigma_{\rm linemap}$ is the RMS noise in the FWZI CO(4-3) linemap.}
\label{fig:photometry_cutouts}
\end{figure}

\renewcommand{\thefigure}{\thesection\arabic{figure}}

\end{document}